\documentclass[structabstract]{aa}  
\usepackage{graphicx}
\usepackage{txfonts}
\usepackage{longtable}
\usepackage{natbib}
\bibpunct{(}{)}{;}{a}{}{,}

\begin{document}

\title{The HARPS search for southern extra-solar planets\thanks{Based on observations made with the HARPS instrument on the ESO 3.6-m telescope at La Silla Observatory (Chile), under program IDs 072.C-0488, 183.C-0972, 083.C-1001 and following periods.}}

\subtitle{XXXI. Magnetic activity cycles in solar-type stars: statistics and impact on precise radial velocities}

\author{C. Lovis\inst{1}
\and X. Dumusque\inst{1,2}
\and N. C. Santos\inst{2,3,1}
\and F. Bouchy\inst{4,5}
\and M. Mayor\inst{1}
\and F. Pepe\inst{1}
\and D. Queloz\inst{1}
\and \\ D. S\'egransan\inst{1}
\and S. Udry\inst{1}
}

\institute{Observatoire de Gen\`eve, Universit\'e de Gen\`eve, 51 ch. des Maillettes, CH-1290 Versoix, Switzerland \\
\email{christophe.lovis@unige.ch}
\and Centro de Astrof\'isica, Universidade do Porto, Rua das Estrelas, P4150-762 Porto, Portugal
\and Departamento de F\'{\i}sica e Astronomia, Faculdade de Ci\^encias, Universidade do Porto, Portugal
\and Institut d'Astrophysique de Paris, UMR7095 CNRS, Universit\'e Pierre \& Marie Curie, 98bis Bd Arago, F-75014 Paris, France
\and Observatoire de Haute-Provence, CNRS/OAMP, F-04870 St. Michel l'Observatoire, France
}

\date{Received 26 July 2011 / Accepted ...}

\abstract
{Searching for extrasolar planets through radial velocity measurements relies on the stability of stellar photospheres. Several phenomena are known to affect line profiles in solar-type stars, among which stellar oscillations, granulation and magnetic activity through spots, plages and activity cycles.}
{We aim at characterizing the statistical properties of magnetic activity cycles, and studying their impact on spectroscopic measurements such as radial velocities, line bisectors and line shapes.}
{We use data from the HARPS high-precision planet-search sample comprising 304 FGK stars followed over about 7 years. We obtain high-precision Ca II H\&K chromospheric activity measurements and convert them to $R'_{\mathrm{HK}}$ indices using an updated calibration taking into account stellar metallicity. We study $R'_{\mathrm{HK}}$ variability as a function of time and search for possible correlations with radial velocities and line shape parameters.}
{The obtained long-term precision of $\sim$0.35\% on S-index measurements is about 3 times better than the canonical Mt Wilson survey, which opens new possibilities to characterize stellar activity. We classify stars according to the magnitude and timescale of the Ca II H\&K variability, and identify activity cycles whenever possible. We find that 39$\pm$8\% of old solar-type stars in the solar neighborhood do not show any activity cycles (or only very weak ones), while 61$\pm$8\% do have one. Non-cycling stars are almost only found among G dwarfs and at mean activity levels $\log{R'_{\mathrm{HK}}}$ $<$ -4.95. Magnetic cycle amplitude generally decreases with decreasing activity level. A significant fraction of stars exhibit small variations in radial velocities and line shape parameters that are correlated with activity cycles. The sensitivity of radial velocities to magnetic cycles increases towards hotter stars, while late K dwarfs are almost insensitive.}
{Activity cycles do induce long-period, low-amplitude radial velocity variations, at levels up to $\sim$25 m\,s$^{-1}$. Caution is therefore mandatory when searching for long-period exoplanets. However, these effects can be corrected to high precision by detrending the radial velocity data using simultaneous measurements of Ca II H\&K flux and line shape parameters.}

\keywords{Planetary systems -- Stars: activity -- Stars: chromospheres -- Line: profiles -- Techniques: radial velocities -- Techniques: spectroscopic}

\maketitle

\section{Introduction}

The advent of large-scale Doppler surveys to search for extrasolar planets around FGKM stars in the solar neighborhood has produced an impressive body of high-resolution spectroscopic data extending over the past 20 years or so. Precise radial velocities are obviously the main products that are derived from these data, but more generally the existing libraries of spectra make it possible to study in details many properties of solar-type stars. The knowledge of extrasolar planets has always been intimately related to the knowledge of their parent stars. An important example is the need for precise fundamental stellar parameters such as effective temperature, metallicity, mass and radius to properly derive exoplanet parameters. The behavior of stellar photospheres also plays a crucial role: since the discovery of 51 Peg b \citep{mayor95}, questions have arisen about the effects of stellar photospheric "jitter" on the detection and characterization of exoplanets \citep[e.g.][]{Baliunas97}. The outer convective envelope that is present in solar-type stars gives rise to several phenomena that can have an impact on derived exoplanet properties, through variations in measured disk-averaged radial velocity and photometric flux. In order of increasing timescales, one can mention p-mode oscillations \citep{christensen04}, granulation and supergranulation \citep{harvey85}, magnetic activity inducing surface inhomogeneities rotating with the star \citep{saar97}, and finally magnetic activity cycles over timescales of years and decades \citep{baliunas95}. From a stellar physics point of view, all these phenomena are obviously interesting in their own right since they can reveal the details of stellar internal structure and dynamo mechanisms at play in solar-type stars. Moreover, the statistical study of stellar activity in a large number solar-type stars can also inform us about the past and future evolution of our Sun, and its potential impact on Earth climate.

With the ever-increasing precision of radial velocity (RV) and transit observations of exoplanets, the need to better understand the host stars has become more acute in recent years. In particular, radial velocity surveys are now probing the sub-m\,s$^{-1}$ regime \citep{pepe11}, a level at which stellar activity perturbations become non-negligible even for quiet stars and must be monitored. The study of the effects of star spots on radial velocities has a long history, see e.g. \citet{saar97,saar98,santos00,saar00,queloz01}. More recently, the field has received closer attention again following the discovery of transiting planets around relatively active stars like CoRoT-7 \citep{queloz09,boisse11,hatzes11} or M dwarfs like GJ~436 \citep{knutson11}. At the same time, efforts to characterize RV jitter in quiet solar-type stars, in particular the Sun itself, have also intensified. For example, \citet{dumusque11a,dumusque11b} simulate the impact of p-mode oscillations, granulation, supergranulation and active regions on RV data, based on asteroseismic observations of a few solar-type stars. \citet{lagrange10} and \citet{meunier10} use solar irradiance and Ca II archival data to infer the level of RV jitter caused by spots and plages. While these studies offer important insights into the causes of RV jitter, they rely at least partly on unverified assumptions to relate a given physical phenomenon at the stellar surface to the corresponding RV perturbation, as actually measured by RV planet-search techniques. The ultimate "truth" in this domain must necessarily come from actual measurements of precise RVs, photometry and/or spectroscopic diagnostics in the form of densely-sampled time series covering the relevant timescales. The present paper is an attempt is this direction, focusing on the longer timescales.

Over years and decades, the main cause of variability in the Sun is the occurrence of a magnetic cycle with a quasi-periodicity of 11 years (or 22 years depending on definition). The study of this cycle has a long history, starting with its discovery by S. H. Schwabe in 1843. The question whether a similar phenomenon also occurs in other stars was addressed in the 1960s by O. Wilson, who started what would become the famous Mt Wilson program for measuring chromospheric emission in stars of the solar neighborhood \citep{wilson68,wilson78}. The Mt Wilson spectrophotometers regularly monitored the flux in the core of the Ca II H \& K lines in the sample stars for more than 30 years, yielding what is still today the reference database for magnetic activity in solar-type stars. Long-term results from this program were published in several papers \citep[e.g.][]{wilson78,duncan91,baliunas95}, which demonstrated that activity cycles similar to the solar one are indeed common in other stars.

In the meantime, other stellar activity surveys, and the development of large-scale Doppler exoplanet programs, have also produced a large quantity of Ca II H \& K observations, and several studies have been conducted to investigate magnetic activity levels in solar-type stars \citep[e.g.][]{henry96,santos00,tinney02,wright04a,hall07,isaacson10,santos10}. All these "second-generation" projects tried to reproduce as closely as possible the method used by the Mt Wilson survey to measure Ca II H \& K emission, i.e. by computing the so-called S index which normalizes the measured Ca II core flux by the flux in two "continuum" bandpasses on the blue and red sides of the Ca II lines. The Mt Wilson scale, as defined by the S index, has therefore become the standard way of measuring chromospheric activity in stars.

Most projects have actually focused on the mean activity level of stars, and not on the time variability of the activity. To the best of our knowledge, only \citet{baliunas95}, using the Mt Wilson data, have actually performed a systematic search for magnetic cycles and derived their main parameters such as cycle period. In the present paper we attempt to do a similar study to better characterize overall cycle properties in older solar-type stars, and compare them to the Mt Wilson survey. Then we search for the effects of magnetic cycles on the shapes of photospheric lines, with the ultimate goal of disentangling stellar RV "jitter" from real barycentric motions of the star.

The idea that magnetic cycles may influence spectral lines dates back at least to \citet{dravins85}. The rationale behind this is that the convective patterns at the surface of solar-type stars may change along the magnetic cycle under the influence of changing magnetic field strength. Indeed, the convection is greatly reduced in active regions, which causes a decrease in the convective blueshift usually exhibited by photospheric lines \citep[see][and references therein for more detailed studies of these effects]{dravins82,livingston82,brandt90,gray92,lindegren03,meunier10}.

A few attempts to measure such an effect in the Sun yielded somewhat contradictory results: indeed, \citet{deming94} find a peak-to-peak RV variation of 28 m\,s$^{-1}$ over the solar cycle, while \citet{mcmillan93} obtain constant RVs within $\sim$4 m\,s$^{-1}$. We note here that the two studies used very different spectral lines for their measurements (CO lines in the IR vs. atomic UV lines). Measurements of similar RV effects in other stars have remained inconclusive to date \citep[e.g.][]{campbell88,saar00, santos10}, although \citet{santos10} find clear correlations in several stars between magnetic cycles and line shape parameters like bisector and line width.

In the present paper we use high-precision data from the HARPS spectrograph \citep{mayor03}, obtained in the context of a planet-search survey around FGK dwarfs in the solar neighborhood. This survey focuses on the search for low-mass planets, i.e. Neptunes and super-Earths, which induce RV variations of only a few m\,s$^{-1}$ on their parent star \citep[see e.g.][]{lovis06a,mayor09b,lovis11,pepe11}. Potential effects of magnetic cycles at this level could induce spurious long-term drifts in the data that may perturb the detection of low-mass objects at shorter periods due to limited sampling, or even mimic the signal of giant planets orbiting at several AUs from the star. It is therefore crucial to understand the effects magnetic cycles may have on our sample stars. In an accompanying paper to the present study, \citet{dumusque11c} show several examples of planetary systems around stars showing magnetic cycles, and discuss how to correct the perturbing effects of these.

\section{Observations and data reduction}

\subsection{The stellar sample}
\label{SectSample}

\begin{figure}
\centering
\includegraphics[width=\columnwidth]{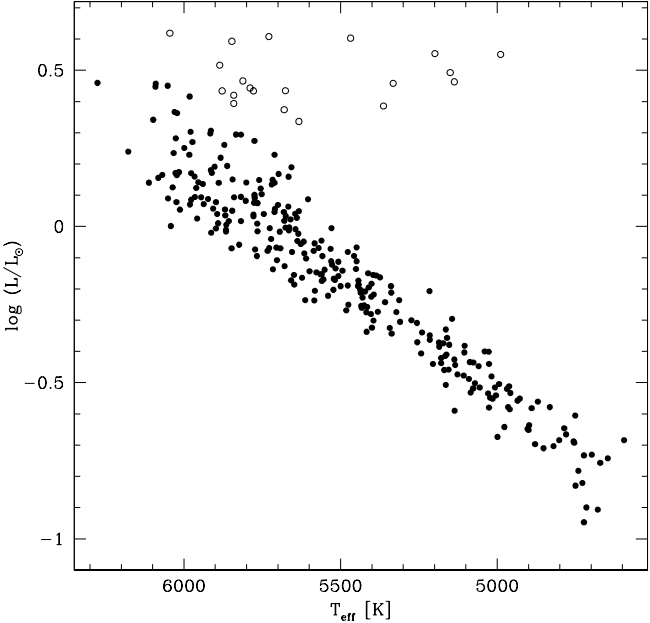}
\caption{H-R diagram of the 304 stars in the sample. Subgiants are shown as open dots.}
\label{FigPlotHR}
\end{figure}

The original HARPS FGK high-precision sample comprises about 400 FGK stars chosen from the volume-limited (50 pc) CORALIE sample of $\sim$1600 stars \citep{udry00}. The selection has been made based on criteria to minimize stellar radial velocity jitter and thus maximize the detectability of low-mass exoplanets. The main selection criteria were: 1) $v \sin{i} <$ 3-4 km\,s$^{-1}$, and 2) exclude all kinds of binary or multiple stars (based on CORALIE measurements). The main effect of this is to generate a stellar sample that is biased against young stars, due to the $v \sin{i}$ cut. Otherwise, we expect that this sample is still representative of old (single) solar-type stars in the solar neighborhood.

The sample stars have been regularly monitored with HARPS since 2003. Single observations usually reach SNR $>$ 100 per pixel at 550 nm, in order to achieve a radial velocity precision of 1\,m\,s$^{-1}$ or better. Observing cadence varies significantly from star to star, depending on radial velocity variability, the presence of planetary systems, and activity level. Active stars with $\log{R'_{\mathrm{HK}}}$ $\gtrsim$ -4.7 that could not be eliminated based on CORALIE measurements were screened early by HARPS and only rarely measured. As of 2011, the number of data points per star varies between 1 and more than 200, which makes any global analysis of the sample challenging because of the very irregular sampling.

For the purpose of this paper, we further selected stars with at least 6 measurements as of May 2011, and with a total time span of observations of at least 1000 days. We finally obtain a sample of 304 stars, still representative of old FGK stars in the solar neighborhood. Fundamental properties of these stars have been obtained by \citet{sousa08} using the same HARPS spectra that we use here. Their analysis yields effective temperature, metallicity and surface gravity, while the Hipparcos catalogue provides absolute magnitude and luminosity through a bolometric correction. Fig.~\ref{FigPlotHR} shows a H-R diagram of the 304 stars that we selected for this study. They span a range in effective temperature between 4600 and 6200 K, and a range in metallicity [Fe/H] between -0.84 and +0.39.

We immediately see from Fig.~\ref{FigPlotHR} that some stars have already evolved off the main sequence, i.e. have become subgiants. Because of their lower gravities and higher luminosities, chromospheric activity measurements in these stars cannot be calibrated in the same way as for main-sequence stars, and we will therefore consider them separately in the remainder of this paper. To separate subgiants from main-sequence stars, we fitted the main sequence in Fig.~\ref{FigPlotHR} and place a limit at +0.25 dex in luminosity (0.63 mag) above the main sequence. According to this criterion, there are 20 subgiants among the 304 sample stars, shown as open dots in the figure.

\subsection{HARPS Ca II H \& K index}
\label{SectHK}

We compute the HARPS S index by following the original Mt Wilson procedure as closely as possible \citep[see in particular][]{duncan91}. We work on extracted 2D spectra in (order, pixel) space, as produced by the standard HARPS pipeline. The spectra are wavelength-calibrated, flat-fielded and de-blazed using calibration frames acquired at the beginning of each night. Their wavelength scale is then transformed to the rest frame of the star using the measured stellar radial velocity and the barycentric correction. Besides this standard treatment, there are two other key steps in the data reduction that are particularly important for S index computation and other spectrophotometric measurements:

\begin{itemize}

\item Background subtraction: there is a small but non-negligible amount of diffuse light in the spectrograph that must be subtracted to preserve spectrophotometric accuracy and precision. In HARPS the diffuse light has two origins: stellar light (fiber A) and ThAr light (fiber B) when observing in simultaneous reference mode. Scattered stellar light does not affect precision to first order since its level is directly proportional to the total stellar flux entering the spectrograph. More problematic is the diffuse light from ThAr because it has a constant flux level, independent of stellar flux. As a consequence, a clear dependence of the S index on SNR is produced, particularly towards low SNR. Correcting the background is therefore essential for spectrophotometric accuracy and precision. We measure the background locally on the detector using the inter-order pixels and taking the mode of the pixel value distribution as the background value. Then the local values are interpolated with splines in the two directions to generate a global background frame that is subtracted from the science frame.

\item Contamination by saturated ThAr lines: besides producing diffuse light, the ThAr spectrum also contains a number of saturated lines that contaminate neighboring stellar spectral orders on the detector. To correct this, we developed a new spectrum extraction algorithm based on the usual Horne optimal extraction algorithm \citep{horne86}. We modified the algorithm to simultaneously fit the order profile in cross-dispersion direction {\em and} the contaminating flux coming from the neighboring fiber. We obtained a reference ThAr spectrum with fiber B only (no light on fiber A), and use it as a template for contamination on fiber A. Only one extra free parameter is required in the extraction procedure (a global scaling of the contamination level).

\end{itemize}

We verified that these improvements removed most of the strong systematics as a function of SNR that were affecting Ca II H \& K measurements before any background subtraction was applied. S-index measurements on standard stars show that we can now still get reliable results at SNR values of only a few per pixel in the continuum at 4000 \AA. Residual instrumental systematics as a function of SNR are still visible on some stars, and their potential causes should be investigated further (e.g. CCD non-linearity). They are however negligible for the purpose of this paper.

To compute the S index, we define exactly the same spectral passbands as for the Mt Wilson HKP-2 spectrophotometer \citep{duncan91}. For the Ca II H \& K line cores, the bands are centered at 3933.664 \AA\ ($K$) and 3968.470 \AA\ ($H$), and have a triangular shape with a FWHM of 1.09 \AA. For the continuum passbands, the $V$ band has a width of 20 \AA\ and is centered at 3901.070 \AA, while the $R$ band has the same width and is centered at 4001.070 \AA. The Mt Wilson S index is then defined as:

\begin{equation}
S = \alpha \, \frac{H + K}{R + V} .
\end{equation}

In this equation, $H$, $K$, $R$ and $V$ represent the total flux in each passband, while $\alpha$ is a calibration constant. The value of $\alpha$ is reported as being either fixed at 2.40 or 2.30 \citep{duncan91}, or regularly adjusted following calibration \citep{baliunas95}. We note here two important facts:

\begin{itemize}

\item The Mt Wilson spectrophotometer is designed in such a way that a rapidly rotating slit mask sequentially observes the four different channels at high frequency, and exposes the $H$ and $K$ channels 8 times longer than the continuum passbands.

\item When working with 'normal' spectrographs like HARPS, we will then get 8 times less flux in the $H$ and $K$ passbands than the Mt Wilson spectrophotometer. To be on the Mt Wilson S-index scale, we have to multiply the $H$ and $K$ fluxes by $\alpha$ and then by 8.

\end{itemize}

Now, instead of working with integrated fluxes in each passband, we could also consider working with mean fluxes per wavelength interval, i.e. $\tilde{X} = X / \Delta\lambda_X$, where $X$ represents any bandpass and $\Delta\lambda_X$ is the effective bandpass width. The Mt Wilson S index can then be written as:

\begin{eqnarray}
\nonumber S &=& \alpha \cdot 8 \cdot \frac{\Delta\lambda_{HK}}{\Delta\lambda_{RV}} \cdot \frac{\tilde{H} + \tilde{K}}{\tilde{R} + \tilde{V}} 
= \alpha \cdot 8 \cdot \frac{1.09\,\AA}{20\,\AA} \cdot \frac{\tilde{H} + \tilde{K}}{\tilde{R} + \tilde{V}} \\
&\approx& \frac{\tilde{H} + \tilde{K}}{\tilde{R} + \tilde{V}} .
\end{eqnarray}

We see that, accidentally or not, S is actually equal (or very close) to the ratio of the mean fluxes per wavelength interval. Indeed, with $\alpha = 2.3$, the numerical factor is just 1.0028. The simplicity of this expression led us to actually use it, i.e. work with mean fluxes per wavelength interval instead of integrated fluxes. One advantage of this approach is that it helps minimize potential edge effects at band boundaries due to finite pixel size and errors in wavelength calibration.

Together with the S value, we also compute the uncertainty on S due to photon shot noise through error propagation.

\subsection{Calibration to Mt Wilson scale}

As explained above, we expect that the HARPS S index will already be very close to the Mt Wilson scale, so that in principle no further calibration should be needed. However, at least two aspects make it still necessary: potential scattered light in the Mt Wilson spectrophotometer and/or HARPS, and the exact value of the calibration constant $\alpha$. In any case, a simple linear calibration between the HARPS and Mt Wilson systems should be sufficient.

\begin{table}
\caption{Reference stars for S-index calibration. Mt Wilson values are from \citet{baliunas95}.}
\label{TableCalibrators}
\centering
\begin{tabular}{l c c}
\hline\hline
Star & $S_{\mathrm{MW}}$ & $S_{\mathrm{HARPS}}$ \\
\hline
HD 10700		& 0.171	& 0.142 \\
HD 16160		& 0.226	& 0.194 \\
HD 23249		& 0.137	& 0.108 \\
HD 152391	& 0.393	& 0.341 \\
HD 160346	& 0.300	& 0.250 \\
HD 216385	& 0.142	& 0.119 \\
HD 219834	& 0.155	& 0.121 \\
\hline
\end{tabular}
\end{table}

The problem of calibrating the S scale is that a number of 'standard' stars observed with both instruments are needed. However, many stars show intrinsic variability because of rotational modulations and/or magnetic cycles that will induce noise in the calibration procedure. We choose to limit ourselves to a small sample of stars that have been well observed at Mt Wilson and by HARPS, and whose intrinsic variability remains moderate. Several stars come from a dedicated HARPS program aiming at following Mt Wilson stars on the long-term \citep{santos10}. Table~\ref{TableCalibrators} gives the list of calibrators with their mean Mt Wilson and HARPS values. Mt Wilson values are taken from \citet{baliunas95}. The calibration stars span a range in S values that almost covers the range of values encountered in our sample. A linear fit to Mt Wilson and HARPS values yields the desired calibration:

\begin{equation}
S_{\mathrm{MW}} = 1.111 \cdot S_{\mathrm{HARPS}} + 0.0153 .
\end{equation}

The dispersion of the residuals around the fit is 0.0043, which represents only a few percent of the fitted S values. Considering intrinsic stellar variability, the quality of the fit is probably as good as it can possibly be. As expected, the obtained coefficients are very close to a 1:1 relation. Interestingly, the observed differences between $S_{\mathrm{MW}}$ and $S_{\mathrm{HARPS}}$ (about 0.03) are compatible with the estimated scattered light level in the H \& K passbands of the Mt Wilson spectrophotometer \citep{duncan91}. Actually, before correcting the HARPS background, the calibration relation was even closer to 1:1, suggesting that both instruments roughly have the same amount of scattered light. All these results confirm that we have devised an accurate way of measuring $S_{\mathrm{MW}}$ from HARPS spectra.

\subsection{Conversion to $R'_{\mathrm{HK}}$ and metallicity effects}
\label{SectRHK}

The Mt Wilson S index provides the Ca II H \& K core flux normalized to the neighboring continuum. This flux contains both a photospheric and chromospheric component, whose proportions vary as a function of stellar effective temperature and activity level. The quantity of interest is usually the chromospheric flux alone, which is directly related to the amount of energy that heats the chromosphere through the magnetic field. To be able to usefully compare stars between one another, one then has to subtract the photospheric component and to normalize the chromospheric flux to the total (bolometric) luminosity of the star. With this goal in mind, \citet{noyes84} introduced the well-known quantity $R'_{\mathrm{HK}}$ (and its logarithmic version), which is defined as:

\begin{equation}
R'_{\mathrm{HK}} = C_{\mathrm{cf}}(B-V) \cdot S - R_{\mathrm{phot}}(B-V) ,
\end{equation}

where $C_{\mathrm{cf}}(B-V)$ is a conversion factor that corrects for the varying flux in the continuum passbands as a function of $B-V$ and normalizes to the bolometric luminosity, while $R_{\mathrm{phot}}$ is the photospheric contribution in the H \& K passbands (also dependent on $B-V$).

The main focus of this paper is to study temporal variations in the activity level of our sample stars, and their impact on radial velocity measurements and other quantities. In this context, we have to determine which activity-related quantity is the most appropriate to consider. We assume that, at the stellar surface, the signature of magnetic activity cycles is mainly a variation in the average magnetic field strength and filling factor of active regions. This is tightly related to the amount of magnetically-induced chromospheric heating, measured by $R'_{\mathrm{HK}}$. Further assuming an approximately linear relation between changes in chromospheric emission and photospheric line shapes (for small variations), we expect that $R'_{\mathrm{HK}}$ (and not its logarithm or the $S$ index) is the proper quantity to consider. In the remaining of this paper we will therefore use $R'_{\mathrm{HK}}$, multiplied by $10^5$ for practical purposes. The logarithm $\log{R'_{\mathrm{HK}}}$ remains the preferred quantity (because in widespread use) when considering only the mean activity level of a star.

\begin{figure*}
\centering
\includegraphics[width=\columnwidth]{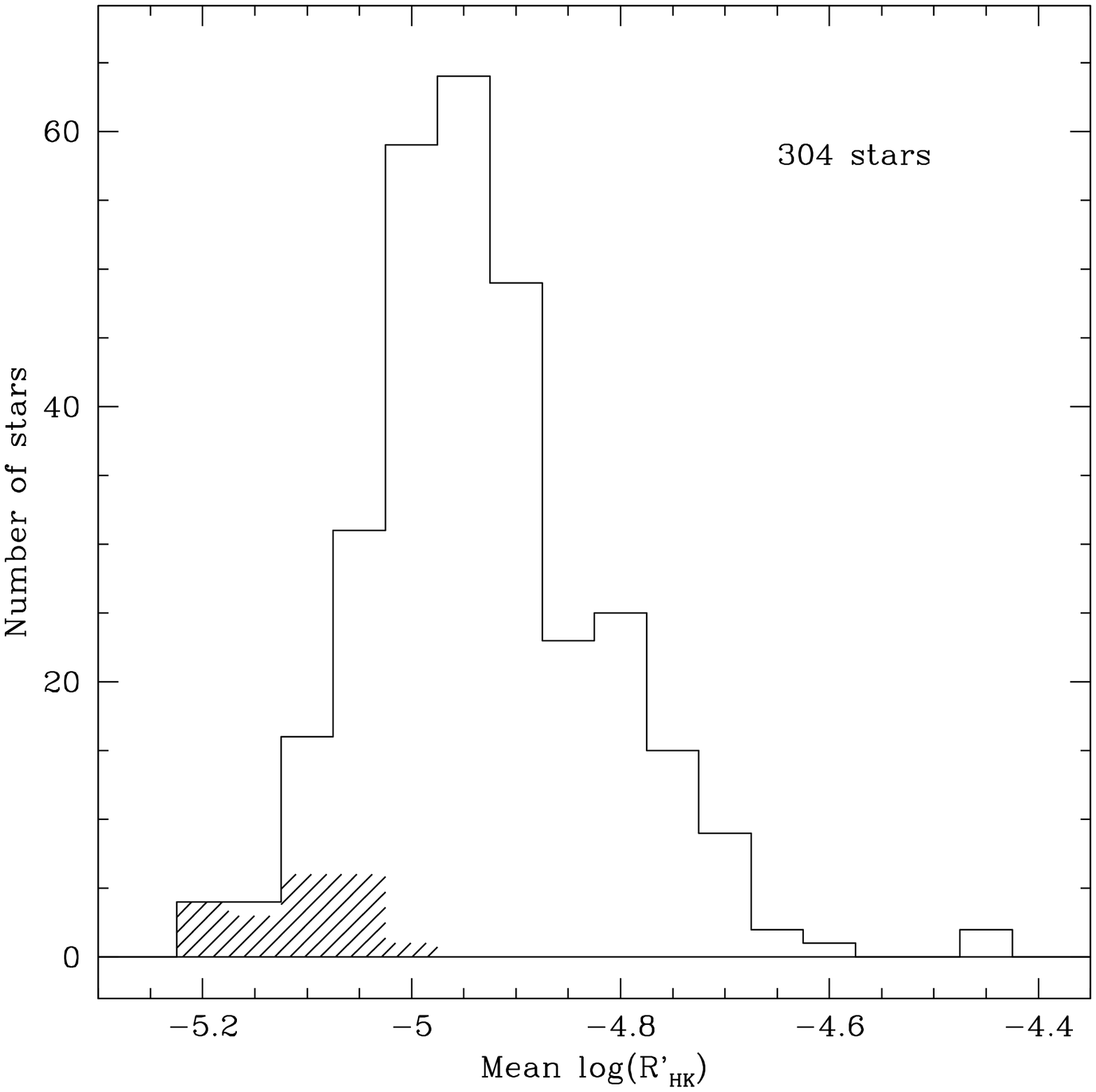}
\includegraphics[width=\columnwidth]{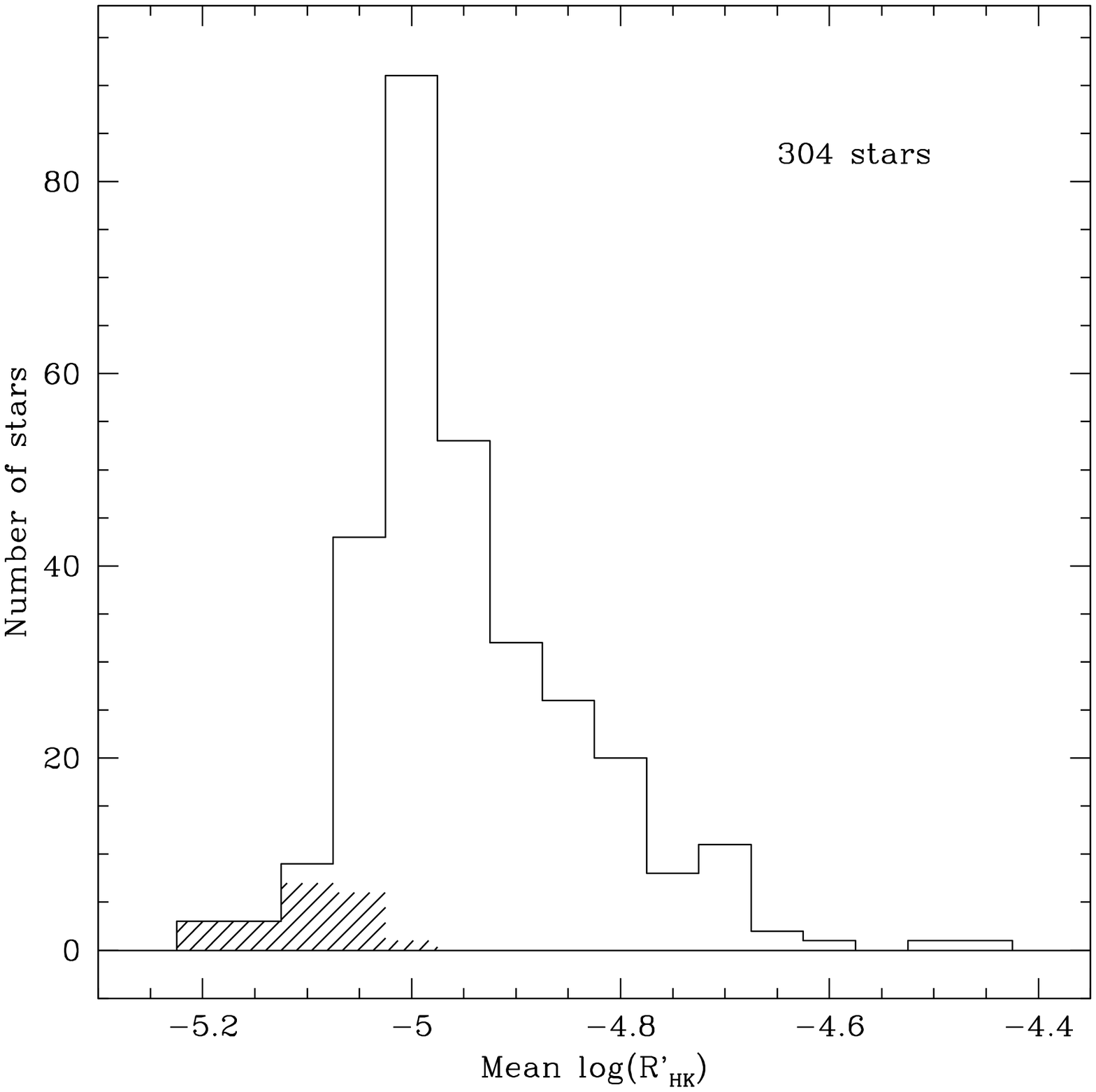}
\caption{{\em Left}. Distribution of the mean $\log{R'_{\mathrm{HK}}}$ values for the 304 stars in the sample, computed with the Noyes et al. $R'_{\mathrm{HK}}$ calibration. The shaded area represents subgiants. {\em Right}. Same figure, but using the updated $R'_{\mathrm{HK}}$ calibration taking into account metallicity. The main peak is significantly narrower.}
\label{FigHistoMean}
\end{figure*}

We show in Fig.~\ref{FigHistoMean} the distribution of mean $\log{R'_{\mathrm{HK}}}$ values for the 304 stars in the sample, as computed with the above formula. A large peak is visible around -4.95. As noted in Sect.~\ref{SectSample}, the sample stars have been selected for being slow rotators based on $v \sin{i}$ measurements. As a consequence, the sample is biased towards old, inactive solar-type stars and cannot be considered as representative of the solar neighborhood. As expected, Fig.~\ref{FigHistoMean} shows a lack of young stars with $\log{R'_{\mathrm{HK}}} \gtrsim -4.70$, and thus does not exhibit the bimodal distribution usually obtained for the solar neighborhood \citep[e.g.][]{vaughan80}. As a side comment, we note that unbiased planet-search samples such as the CORALIE and HARPS volume-limited samples do show a bimodal distribution of $\log{R'_{\mathrm{HK}}}$ values (Marmier et al., in prep.; Lo Curto et al., in prep.).

Before going further, we examine possible systematics of stellar origin in the derivation of $R'_{\mathrm{HK}}$ values. The above formula for $R'_{\mathrm{HK}}$ has been computed by \citet{middelkoop82} and \citet{noyes84} for main-sequence stars only, and taking into account a dependence on $B-V$ only. Post-main-sequence evolution (subgiants) and metallicity are two important factors that have an impact on stellar spectra but are not taken into account in the $R'_{\mathrm{HK}}$ calibration. As already noted by several authors, we can therefore expect systematic differences in $R'_{\mathrm{HK}}$ values between main-sequence stars and subgiants on the one hand, and between stars of different metallicities on the other hand.

It was noted in Sect.~\ref{SectSample} that 20 stars in the sample are actually subgiants based on their position above the main sequence in the H-R diagram. It turns out that all these stars have $\log{R'_{\mathrm{HK}}} < -5.00$, with a mean value at -5.11 and a minimum value of -5.20. Their distribution is shown as a dashed histogram in Fig.~\ref{FigHistoMean}. We therefore confirm here the well-known, abnormally low $\log{R'_{\mathrm{HK}}}$ values measured on subgiants, caused by their higher luminosities and/or lower surface gravities compared to main-sequence stars of the same color. We will consider these stars separately in the analysis below.

\begin{figure*}
\centering
\includegraphics[width=\columnwidth]{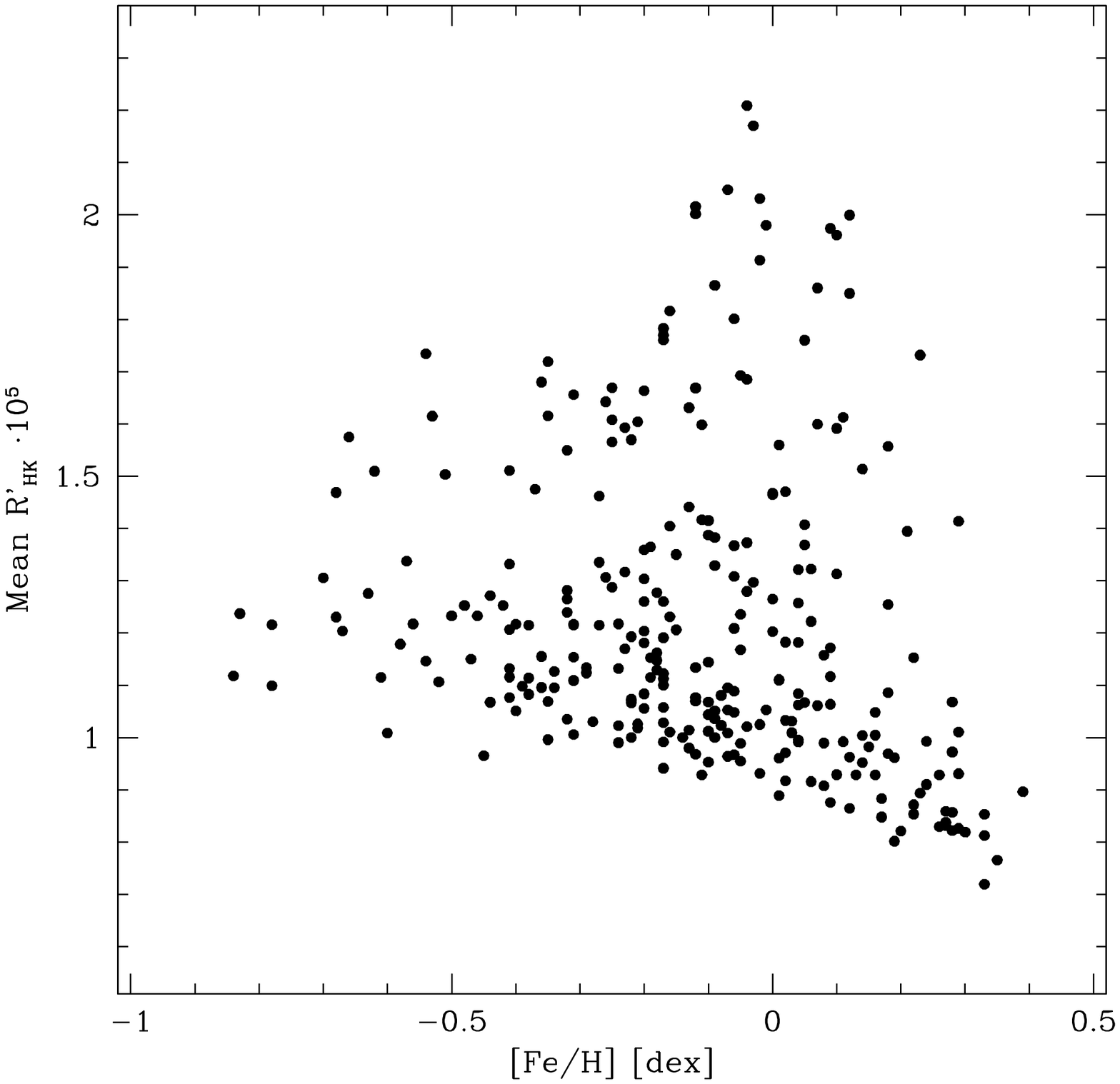}
\includegraphics[width=\columnwidth]{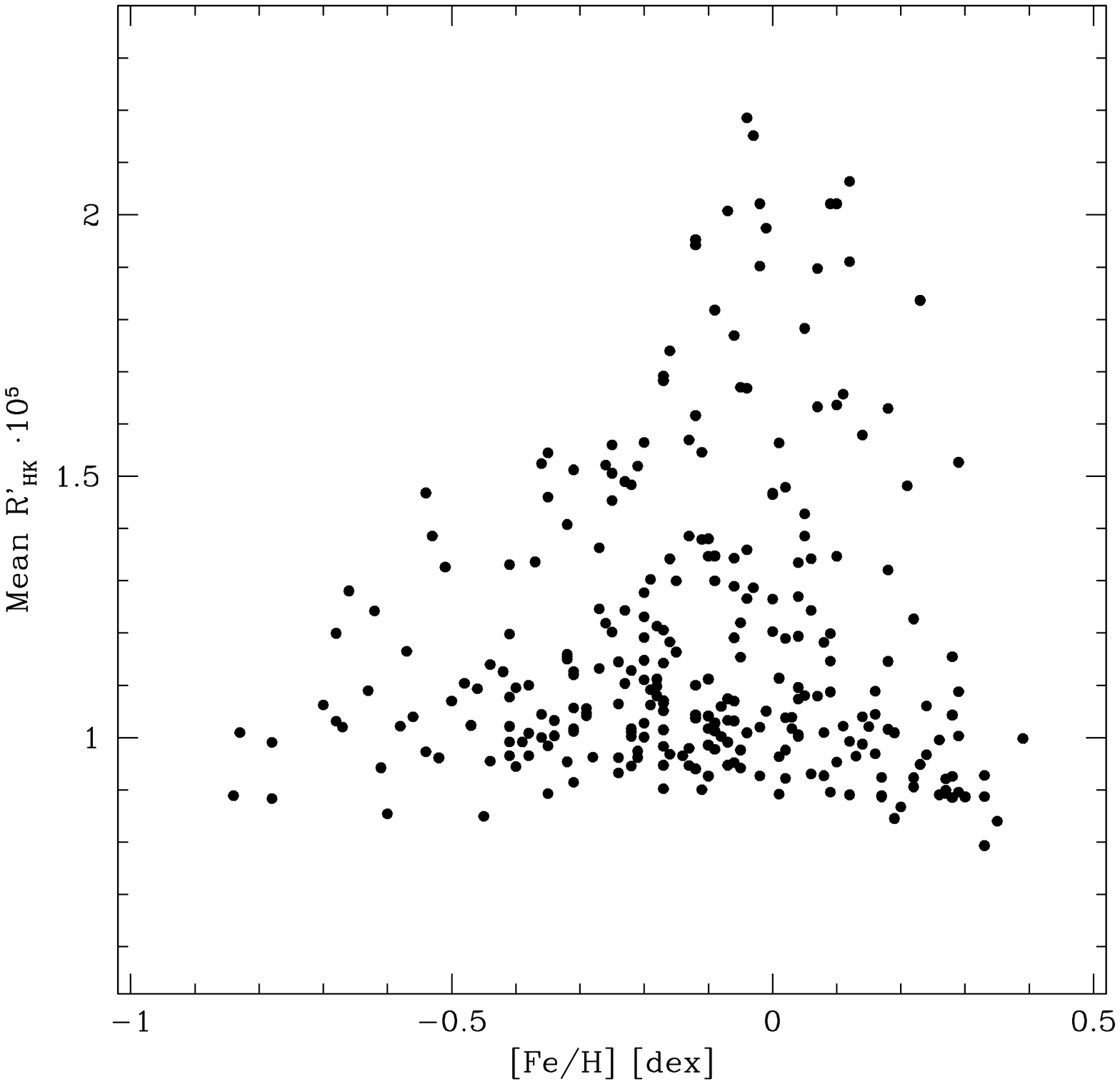}
\caption{{\em Left}. Mean $R'_{\mathrm{HK}}$ as a function of stellar metallicity, obtained with the standard Noyes et al. calibration. Metal-poor stars show a trend towards higher $R'_{\mathrm{HK}}$ values than metal-rich stars. {\em Right}. Corrected mean $R'_{\mathrm{HK}}$. The trend has been successfully suppressed.}
\label{FigMetallicity}
\end{figure*}

We now turn to the effects of metallicity. Fig.~\ref{FigMetallicity} shows the mean $R'_{\mathrm{HK}}$ values as a function of stellar metallicity for all stars except the 20 subgiants. The lower envelope of the points (describing the inactive stars) shows a clear linear trend with metallicity, in the sense that metal-poor stars have higher $R'_{\mathrm{HK}}$ values than metal-rich stars. This can be understood by considering that, at a given $B-V$ color, metal-poor stars are under-luminous compared to metal-rich stars. As a consequence, the continuum passbands are weaker and the measured S value is larger. This leads to a larger $R'_{\mathrm{HK}}$ value since the calibration of \citet{noyes84} assumes that bolometric luminosity (actually $\sigma T_{\mathrm{eff}}^4$) only depends on $B-V$. Assuming that most of the metallicity effect in $R'_{\mathrm{HK}}$ measurements originates in this too simple calibration of luminosity, we can correct for it by multiplying $R'_{\mathrm{HK}}$ with the suitable luminosity ratio:

\begin{equation}
\label{EqNewCal}
R'_{\mathrm{HK,new}} = R'_{\mathrm{HK}} \cdot \left(\frac{T_{\mathrm{eff}}(B-V, \mathrm{[Fe/H]})}{T_{\mathrm{eff}}(B-V, 0)}\right)^4 .
\end{equation}

We now need to express $T_{\mathrm{eff}}$ as a function of $B-V$ and metallicity. Such an expression is actually provided by \citet{sousa08}, obtained from exactly the same data as we are using in this paper:

\begin{equation}
T_{\mathrm{eff}} = 9114 - 6827\,(B-V) + 2638\,(B-V)^2 + 368\,\mathrm{[Fe/H]} .
\end{equation}

The fit is reported to have a standard deviation of the residuals of only 47 K. Since we are using in this work the same effective temperatures, metallicities, $B-V$ colors and even stellar sample as \citet{sousa08}, this expression for $T_{\mathrm{eff}}$ is the ideal one for our needs. Substituting $T_{\mathrm{eff}}$ in Eq.~\ref{EqNewCal}, we finally obtain:

\begin{equation}
R'_{\mathrm{HK,new}} = R'_{\mathrm{HK}} \cdot \left(1 + \frac{368\,\mathrm{[Fe/H]}}{9114 - 6827\,(B-V) + 2638\,(B-V)^2}\right)^4
\end{equation}

The right panel of Fig.~\ref{FigMetallicity} is the same figure as the left panel, but using the updated calibration. The previously existing trend has been successfully corrected, confirming our assumption that it was related to an oversimplified $B-V$ vs. $T_{\mathrm{eff}}$ conversion in the Noyes et al. calibration. Similarly, we show in the right panel of Fig.~\ref{FigHistoMean} the corrected distribution of mean $\log{R'_{\mathrm{HK}}}$ values for our sample. The histogram shows a narrower and higher peak than before, centered around -5.00. We interpret this as the sign that metallicity differences were artificially smearing the $R'_{\mathrm{HK}}$ values of stars that actually have very similar activity levels.

\begin{figure}
\centering
\includegraphics[width=\columnwidth]{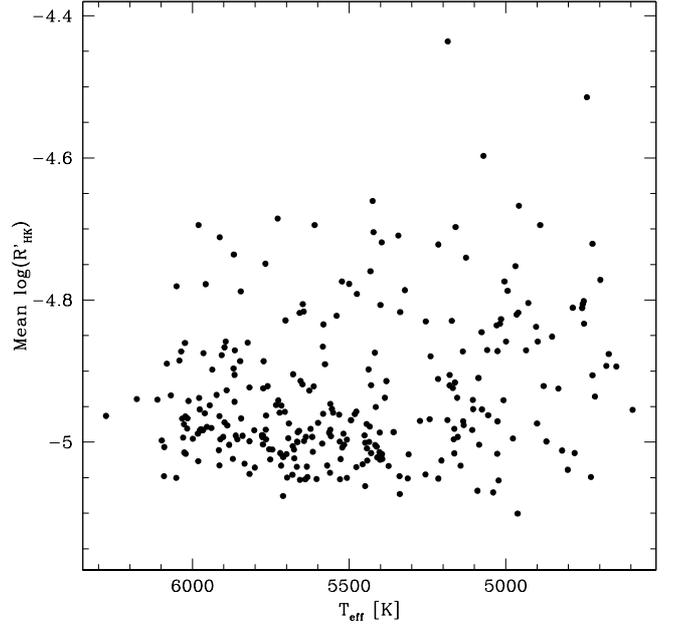}
\caption{Corrected mean $\log{R'_{\mathrm{HK}}}$ values as a function of stellar effective temperature.}
\label{FigTeffMean}
\end{figure}

Finally, we plot in Fig.~\ref{FigTeffMean} the corrected mean $\log{R'_{\mathrm{HK}}}$ values as a function of stellar effective temperature. No systematic trend can be seen, confirming again that the $R'_{\mathrm{HK}}$ calibration takes properly into account the variable stellar properties across the mass and metallicity ranges that we consider here, for stars on the main sequence. We note however a different distribution of $\log{R'_{\mathrm{HK}}}$ values between G and K dwarfs: while G dwarfs are densely packed around -5.0, K dwarfs ($T_{\mathrm{eff}} \lesssim 5300\,\mathrm{K}$) are clearly more regularly spread over the range -4.7 to -5.0. This can possibly be interpreted as the signature of a slower decrease in activity level with age among K dwarfs, as discussed by \citet{mamajek08}. These authors provide, in their Fig.~11, a calibration of the $\log{R'_{\mathrm{HK}}}$ evolution as a function of age and spectral type. Assuming that G and K dwarfs in the solar neighborhood have roughly the same age distribution ($\sim$2-8 Gyr for our sample), we expect that most G dwarfs will have already reached the $\log{R'_{\mathrm{HK}}}$ = -5.0 level, while K dwarfs are still spread between -4.7 and -5.0. The observed accumulation of G dwarfs at -5.0 suggests however that stars do not smoothly evolve towards even lower activity levels once they have reached this floor, but rather remain at -5.0 for an extended period of time.

\subsection{Radial velocities and line shape parameters}

The high-precision radial velocities and line shape parameters used in this paper are those obtained with the standard HARPS pipeline (see Lovis et al., in prep., for a more detailed description). They are all derived from the cross-correlation function (CCF) that is built using a binary template containing several thousands of spectral lines. Radial velocities are then obtained by simply fitting a Gaussian to the CCF profile and measuring its centroid. The fit also yields the FWHM of the CCF and its contrast, which we use here as spectroscopic indicators of stellar activity. Moreover, we also compute the CCF bisector span \citep{queloz01}, which represents the velocity difference between the upper and lower parts of the CCF, and is useful to characterize line shape asymmetries.

HARPS not only delivers precise radial velocities at the level of $\sim$1 m\,s$^{-1}$ and better, but also spectroscopic indicators of the same quality. This is extremely valuable in the context of exoplanet searches since it allows us to monitor stellar activity simultaneously with the radial velocity measurements, and this not only for active stars but also for 'quiet' ones with $\log{R'_{\mathrm{HK}}}$ around -5.0. Several recent papers have already highlighted the use that can be made of these high-precision indicators \citep{queloz09,santos10,lovis11,dumusque11c,pepe11}.

\section{Statistics of Ca II H \& K variability}

\subsection{Raw dispersions and measurement errors}

\begin{figure}
\centering
\includegraphics[width=\columnwidth]{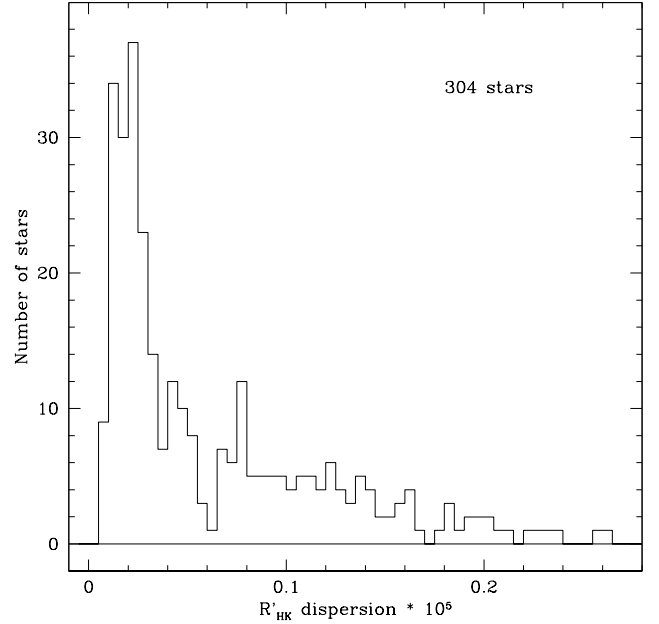}
\caption{Distribution of the $R'_{\mathrm{HK}}$ dispersions for the 304 stars of the sample.}
\label{FigHistoStdev}
\end{figure}

\begin{figure}
\centering
\includegraphics[width=\columnwidth]{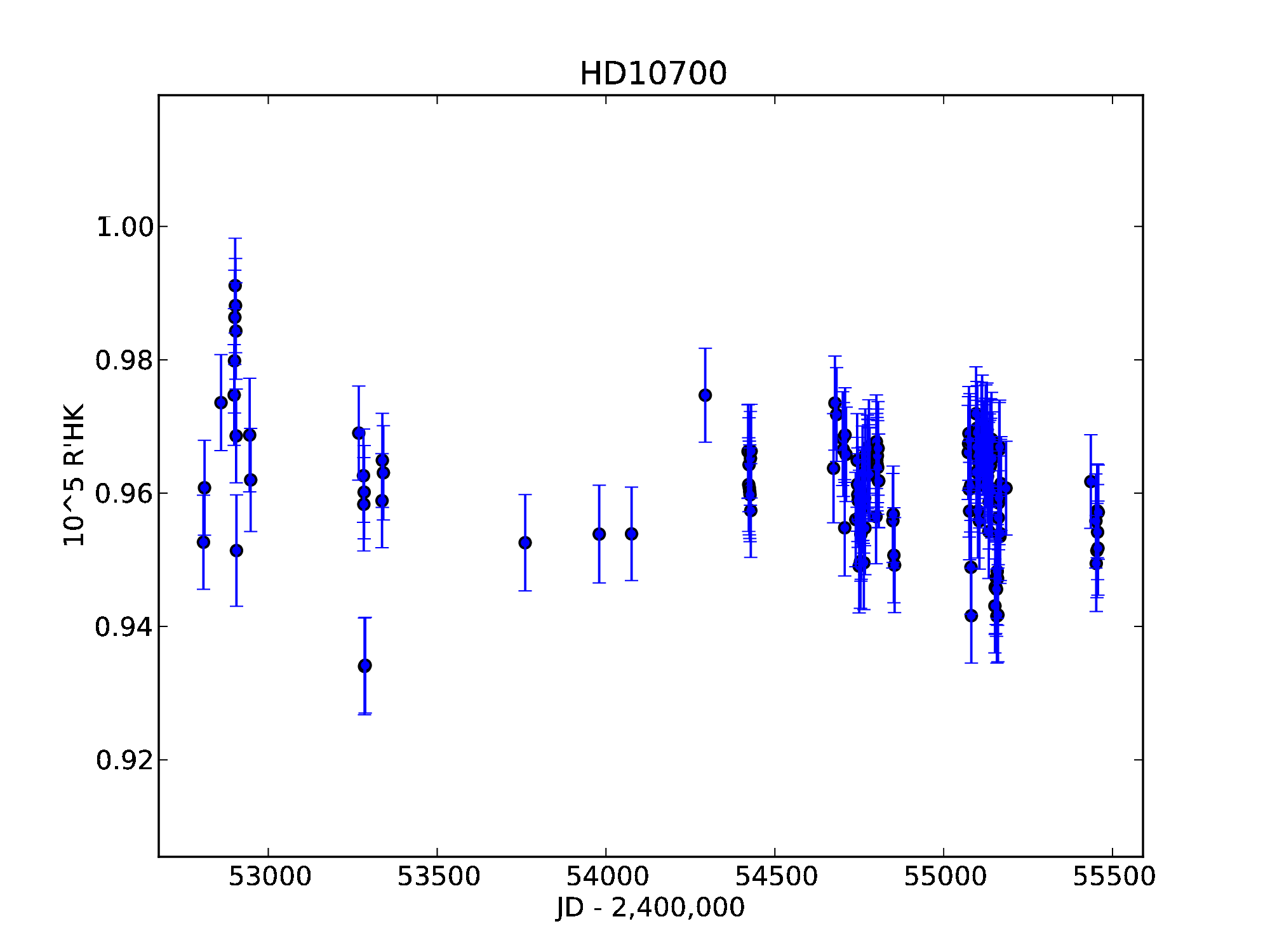}
\caption{Time series of $R'_{\mathrm{HK}}$ measurements for the standard star $\tau$ Ceti (HD 10700).}
\label{FigTimeSeriesHD10700}
\end{figure}

Before searching for signals of stellar origin in chromospheric emission measurements, we first examine the raw dispersions of the $R'_{\mathrm{HK}}$ time series and try to characterize instrumental errors. Fig.~\ref{FigHistoStdev} shows the distribution of the $R'_{\mathrm{HK}}$ standard deviations for the 304 stars in the sample. The peak of the distribution is located at $\sim$0.02. A tail of higher-variability stars is present, either due to short-term variability or magnetic cycles as we will see below. On the other side of the distribution, we note that several well-measured stars exhibit a scatter around 0.01. As an example, Fig.~\ref{FigTimeSeriesHD10700} shows the time series of the well-known standard star $\tau$ Ceti (HD 10700), which comprises 153 data points spread over more than 7 years. The dispersion of the data is only 0.0089 in $R'_{\mathrm{HK}}$, corresponding to a relative variation around the mean of 0.93\%, and to a relative variation of the S index of 0.35\%. This demonstrates both the high instrumental precision reached by HARPS and the remarkable stability of this star. Clearly, $\tau$ Ceti does not show any magnetic activity cycle over the time span of the observations, confirming its previously reported status of very quiet star.

The determination of reliable error bars on the $R'_{\mathrm{HK}}$ measurements is a difficult issue. On the one hand, one has to disentangle instrumental errors from intrinsic stellar variability and on the other hand, instrumental errors may be dominated by systematic effects that are difficult to estimate. Because the SNR in the Ca II H \& K line cores is usually low (less than 5:1 per extracted pixel for the fainter stars in the sample), and because of the large SNR contrast with the continuum, S-index measurements are sensitive to several instrumental effects like non-perfect diffuse light subtraction, residual cosmic hits, CCD defects, CCD charge transfer inefficiency and non-linearity, etc. (see Sect.~\ref{SectHK}).

The observed $R'_{\mathrm{HK}}$ dispersions are typically 2-4 times larger than the estimated photon noise levels for the sample stars, most of which are between 0.002 and 0.020 (median value 0.007). Plotting the normalized standard deviations as a function of photon errors, we note that the lower envelope of the points departs from 1.0 when photon errors become smaller than $\sim$0.007. This suggests that either all stars are intrinsically variable at least at that level, or that we have hit an instrumental noise floor. Given the various effects mentioned above, we are more inclined to believe in the instrumental explanation. Consequently, we quadratically add an additional noise term of 0.007 to all error bars on $R'_{\mathrm{HK}}$ in the following analysis (we note that the precise value of this noise floor is unimportant for the magnetic cycle search). Towards lower SNR, the photon errors seem to correctly describe the lower envelope of standard deviations, at least up to a photon noise level of $\sim$0.03 (the worst case for our sample). It is likely that other effects come into play at even lower SNR (e.g. non-perfect background subtraction), but that regime is beyond the scope of this paper.

In conclusion, we reach a long-term precision of $\sim$0.01 in $R'_{\mathrm{HK}}$ on bright stars, corresponding to a relative precision $\sigma_S / S$ of $\sim$0.35\%. Interestingly, these HARPS measurements are about a factor of 3 more precise than those of the Mt Wilson H \& K project, which reach a long-term precision of about 1\% in $S$ \citep{baliunas95}. To the best of our knowledge, the long-term chromospheric data presented here are likely the most precise ever reported for stars other than the Sun. As a final comment on measurement precision, we note that in the case of the Sun, typical, small active regions induce a modulation in chromospheric emission of $\sim$1\% as they rotate with the star. The precision reached by HARPS should therefore allow us to be sensitive to individual active regions crossing the disk of solar-type stars, provided they are observed at a suitable cadence. Many rotational modulations are indeed readily visible in HARPS $R'_{\mathrm{HK}}$ data, which will be the topic of a subsequent paper.

\subsection{Method and classification}

The large dataset at our disposal allows us to perform a global study of the behavior of chromospheric emission in old solar-type stars as a function of time. In this exercise, the main difficulty arises from the extremely inhomogeneous time series for individual stars, i.e. the varying number of observations per star and the time span of the observations. Indeed, observations were scheduled as a function of the radial velocity behavior of each star, with planet-host stars receiving significantly more attention than RV-stable stars. A global analysis of chromospheric variability has therefore to be robust enough to deal with scarce as well as dense time series in a consistent way.

As a first step, our main goal is to classify stars into broad categories according to their Ca II H \& K variability. We characterize variability in two different ways: 1) a search for significant long-term periodicities in the time series, and 2) a short-term variability criterion based on the standard deviation of the measurements.

\subsubsection{Periodicities}
\label{SectPeriodicities}

\begin{figure}
\centering
\includegraphics[width=\columnwidth]{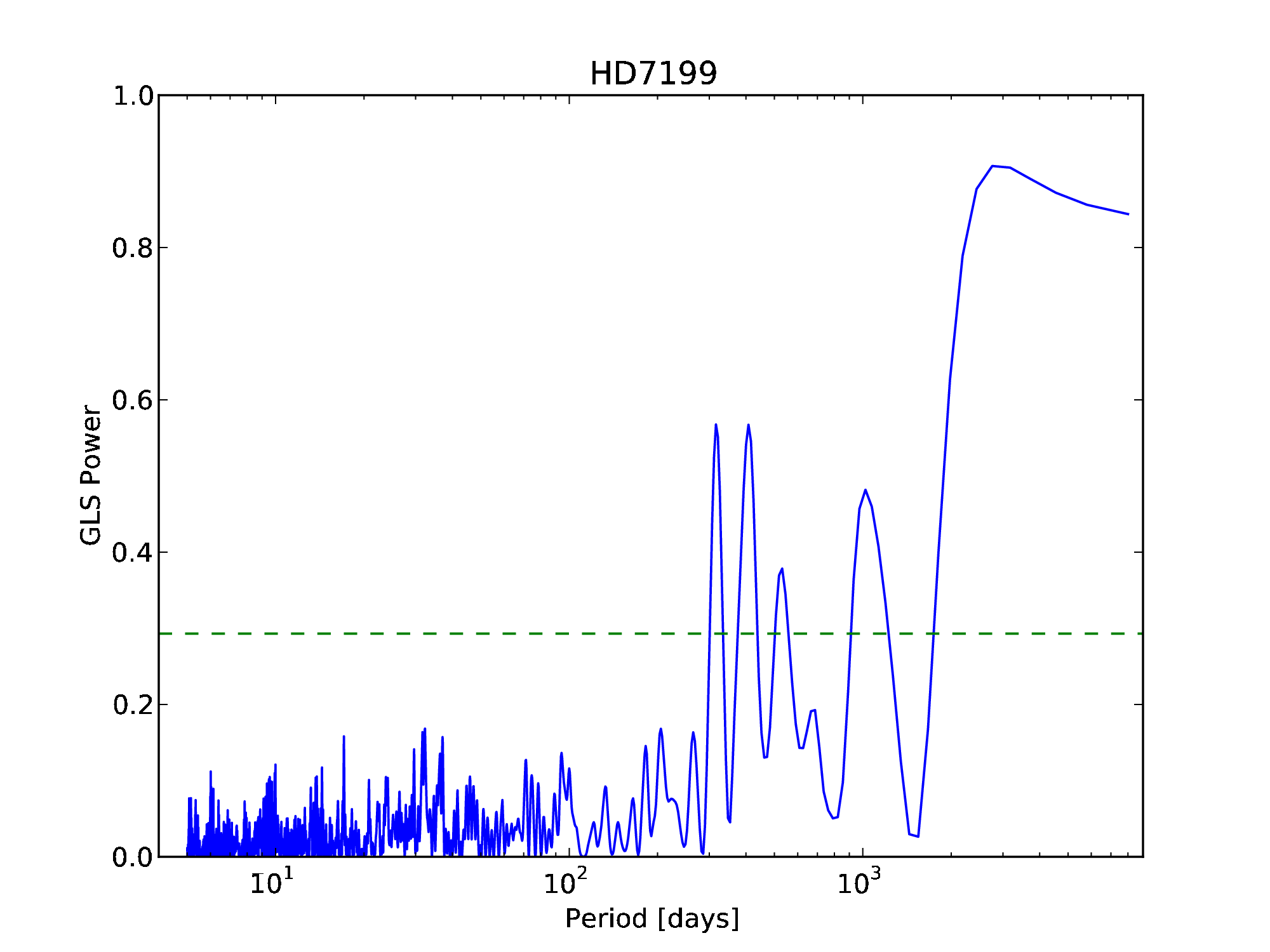}
\includegraphics[width=\columnwidth]{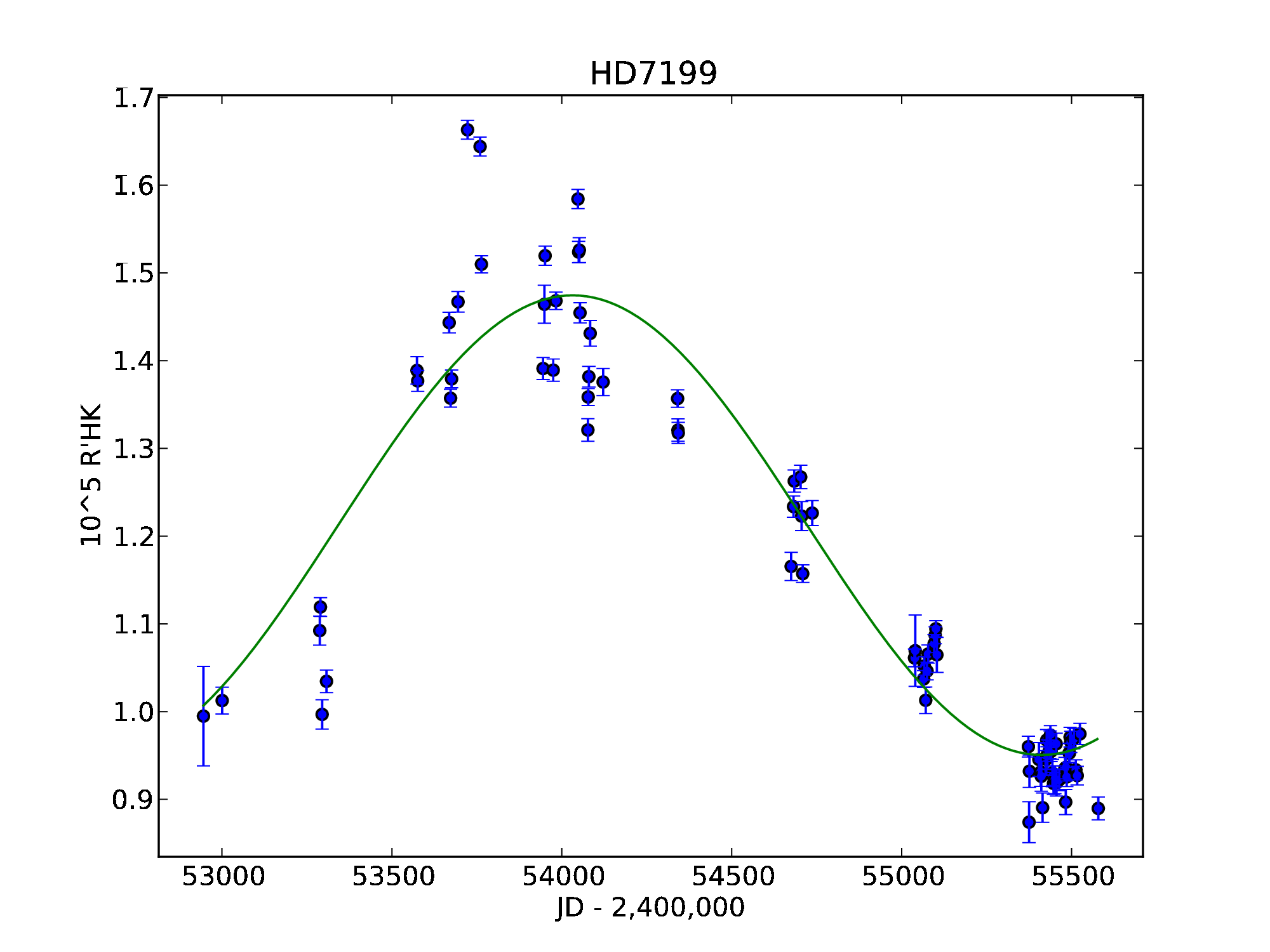}
\caption{Top: GLS periodogram of the $R'_{\mathrm{HK}}$ data for the star HD\,7199. The dashed line denotes the 1\% false-alarm probability threshold. Bottom: Time series of the $R'_{\mathrm{HK}}$ data with the fitted magnetic cycle model overplotted.}
\label{FigHD7199}
\end{figure}

To search for periodicities, we use the generalized Lomb-Scargle periodogram \citep[GLS,][]{zechmeister09} and consider a given periodogram peak as significant when its power exceeds the 1\% false-alarm probability (FAP) threshold. FAP levels are computed by performing random permutations of the data while keeping times of observations fixed. This procedure provides a robust way of finding significant periodicities in a very inhomogeneous dataset in terms of number of observations per star. We fit magnetic cycles with a simple sinusoid at the period of maximum power, as provided by the GLS. As an example, Fig.~\ref{FigHD7199} shows the periodogram and fitted time series for the star HD\,7199, which exhibits a strong magnetic cycle at a period of 2760 days \citep[see][for a detailed study of this star]{dumusque11c}.

We further split stars with a detected signal into two groups: short-term and long-term periodicities, with a limit between the two at $P$ = 2 yr. We choose 2 yr for two main reasons: 1) it is an intermediate timescale separating the stellar rotation timescale on the one hand and stellar magnetic cycles on the other hand, and 2) it minimizes the risk of misclassifying stars because of aliasing problems related to various aliases in the window function of the observations. Putting the limit at shorter periods creates a risk of misclassifying stars that show strong rotationally-induced variability and poor sampling, while putting the limit at longer periods may miss true activity cycles (i.e. not related to rotational modulations) that exist at timescales of a few years. In brief, poor sampling often prevents us from reliably detecting magnetic cycles below $\sim$2 yr (if they at all exist) because it is then difficult to disentangle them from rotationally-induced short-term variability. We therefore focus hereafter on magnetic cycles in the 2-11 yr period range.

For all stars that exhibit no significant long-term modulations, we compute detection limits on a possible sinusoidal signal at long periods. This allows us to use all available information in our dataset, i.e. to put constraints on the existence of magnetic cycles for apparently constant stars and stars with only few measurements. The output of these computations is an upper limit on the $R'_{\mathrm{HK}}$ semi-amplitude that cannot be excluded by the data. In order to obtain a unique limit for all long-period signals, we average the upper limits over periods of 2-11 yr. We choose 11 yr because the typical time span of the observations ($\sim$6-7 yr) allows us to have meaningful upper limits roughly up to this timescale, and because it is the period of the solar magnetic cycle (to be able to compare with the Sun). The Sun exhibits a roughly sinusoidal variation in $R'_{\mathrm{HK}}$ with a semi-amplitude $A$ of $\sim$0.26. As we will see below, the sample stars show magnetic cycles with a variety of semi-amplitudes, including very low ones. It is therefore difficult to define a threshold amplitude in $R'_{\mathrm{HK}}$ that would exclude the existence of a magnetic cycle. In this work we arbitrarily choose $A$ = 0.04 as the minimum semi-amplitude for a 'large-amplitude' magnetic cycle. This corresponds to about 15\% of the solar $R'_{\mathrm{HK}}$ semi-amplitude. This choice is motivated by the average sensitivity of our survey: a much lower threshold would reject too many stars and the global results would be strongly affected by small-number statistics.

In computing detection limits, an artificial signal is injected into the data and its detectability is checked. There are three free parameters in this exercise: the period, amplitude and phase of the signal. While the goal is to derive a limiting amplitude as a function of period, the phase is an extra parameter that must be probed independently, adding an extra dimension to the problem: detection limits should actually be computed for every phase value. At each period, one can then derive the distribution of detection limits for all phases and, assuming all phases are equiprobable, finally assign a probability of detecting a given amplitude at a given period. While this treatment of phase may not be necessary for well-sampled time series with no significant gaps, it becomes important for our sample of irregularly-observed stars because detection limits at a given period often vary widely as a function of phase. In summary, for stars with no significant long-term signals, we compute detection limits for a set of 12 equidistant phases at all periods. From this, and given a semi-amplitude $A_0$, we can compute a lower limit to the probability that no cycle of semi-amplitude $A_0$ or larger exists:

\begin{equation}
\label{EqProb}
Prob\,(A < A_0) \geq \frac{1}{2 \pi} \int_{A_{\mathrm{lim}} < A_0} d\phi ,
\end{equation}

where $A_{\mathrm{lim}}(\phi)$ is the detection limit as a function of phase $\phi$, which takes values between 0 and $2 \pi$.

\begin{figure}
\centering
\includegraphics[width=\columnwidth]{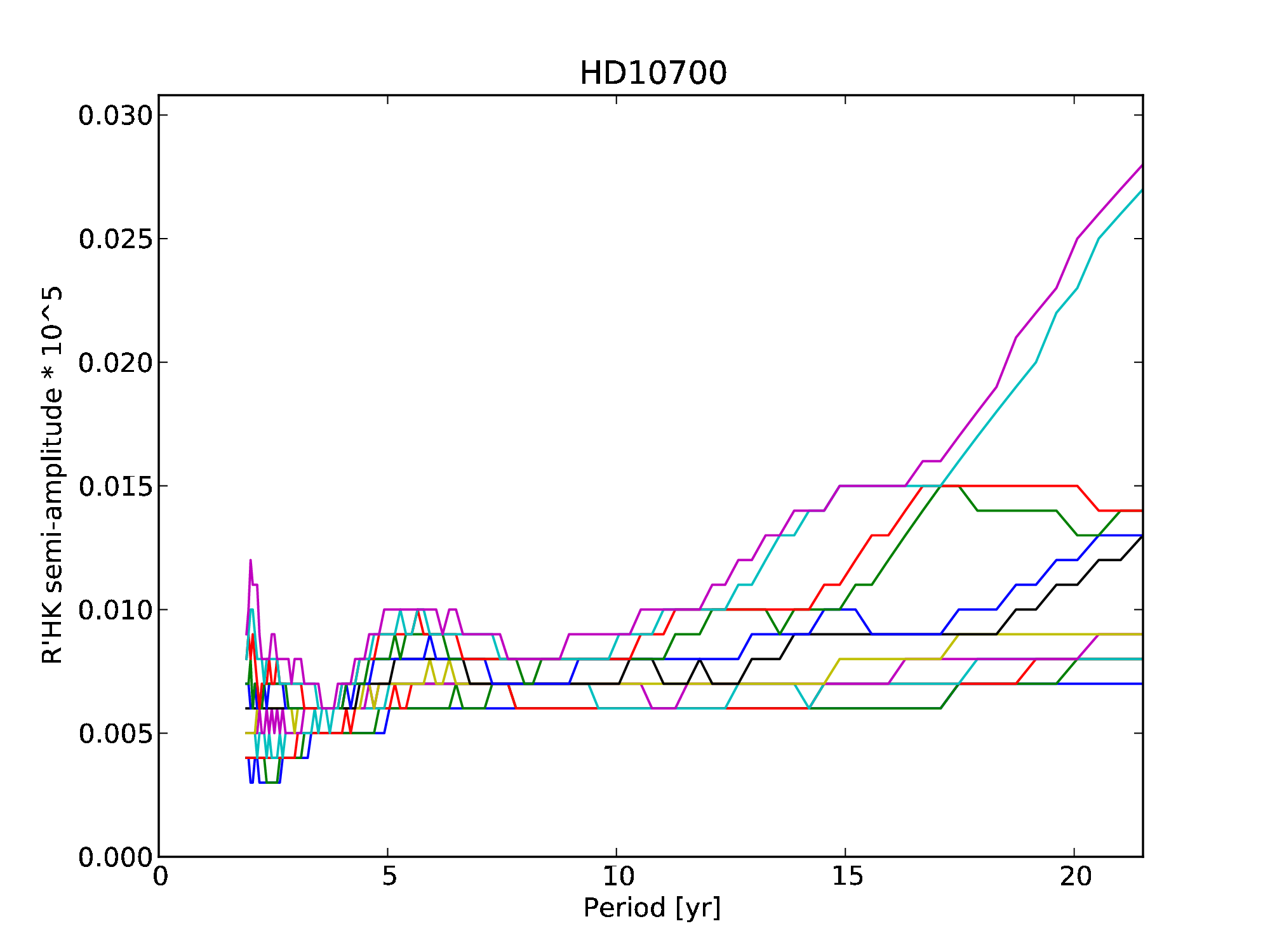}
\caption{Detection limits for sinusoidal signals in $R'_{\mathrm{HK}}$ measurements for the standard star $\tau$ Ceti (HD 10700). The step-like curves are due to the step size of 0.001 used in the computations.}
\label{FigDetectionLimitsHD10700}
\end{figure}

Fig.~\ref{FigDetectionLimitsHD10700} shows the results of the detection limits calculations for $\tau$ Ceti. The different curves denote the 12 equidistant phases simulated at each period. The upper curve represents the detection limit for the 'worst' phase, i.e. the minimum semi-amplitude that can be detected whatever the phase of the signal. The lower curve represents the minimal semi-amplitude that can be detected for the most favorable case, i.e. for the best 8\% (=1/12) of all phases. At a period of 15 yr, we see that a semi-amplitude of 0.01 can be excluded for many phases, but not for all. Counting how many curves are crossed from semi-amplitude 0 to 0.01, we can estimate that a signal with semi-amplitude 0.01 at 15 yr can be excluded with a probability larger than 67\% (=8/12). As another example, we immediately see from Fig.~\ref{FigDetectionLimitsHD10700} that a semi-amplitude of 0.04 (the limit for a 'large-amplitude' magnetic cycle) can be fully excluded for periods between 2-20 yr.

In summary, we define the following variability classes regarding magnetic cycles:
\begin{itemize}
\item Stars showing a magnetic cycle, i.e. significant long-term ($P >$ 2 yr) modulation; among these, 'large-amplitude' magnetic cycles are those with $R'_{\mathrm{HK}}$ semi-amplitude $A > 0.04$.
\item Stars without a magnetic cycle: the detection limit on the $R'_{\mathrm{HK}}$ semi-amplitude is $<$ 0.04 for periods between 2-11 yr.
\end{itemize}

\subsubsection{Short-term variability}

In this work we mainly focus on long-term modulations in chromospheric emission, i.e. magnetic cycles. Therefore, stars exhibiting significant long-term periodicities in $R'_{\mathrm{HK}}$ will receive particular attention in the following sections. However, significant variability also occurs on the short term, and we attempt to quantify it through the standard deviation of the measurements, computed after subtracting the best-fitting long-term signal ($P >$~2~yr), whether significant or not. Short-term variability in $R'_{\mathrm{HK}}$ contains a wealth of information about stellar rotation that deserves a separate study. Again, the inhomogeneity of the time series makes a global analysis more complicated, since the cadence of the observations plays a major role in the ability to characterize rotational modulations. Whereas the best-observed stars may reveal their rotation periods by a simple look at the $R'_{\mathrm{HK}}$ curve, stars with few measurements will only show scatter with undetermined characteristic timescale. In this context, the (short-term) standard deviation of the measurements provides a useful diagnostic that is not very sensitive to the number of data points per star.

\subsection{Results}

\addtocounter{table}{1}
\addtocounter{table}{1}

The global results of the $R'_{\mathrm{HK}}$ analysis are given in Tables~\ref{TableResultsCycle} and \ref{TableResultsNoCycle} for all stars in the sample. Table~\ref{TableResultsCycle} contains the stars showing a significant magnetic cycle. We fit magnetic cycles with a simple sinusoid at the period of maximum power, as provided by the GLS periodogram \citep{zechmeister09}. Confidence intervals for the period are obtained from Monte Carlo simulations: in each trial, different noise realizations on individual data points are drawn from a Gaussian distribution with a standard deviation equal to the measured short-term scatter in the time series, and the GLS periodogram is then recomputed on the modified dataset. The distribution of periods of maximum power yields the desired confidence intervals for the period.

Provided for each star in Table~\ref{TableResultsCycle} are: the number of observations, their time span, the mean and median $\log{R'_{\mathrm{HK}}}$ values, the raw and short-term $R'_{\mathrm{HK}}$ standard deviations, the magnetic cycle period with 68\% confidence intervals, the mean $\log{R'_{\mathrm{HK}}}$ value $\gamma$ as obtained from the magnetic cycle fit, and the cycle semi-amplitude $A$. Finally, estimates of the stellar rotation period and age are also provided, based on the activity-rotation-age calibration of \citet{mamajek08} and using $\gamma$ as the long-term mean $\log{R'_{\mathrm{HK}}}$ value.

Table~\ref{TableResultsNoCycle} provides results for stars with no detected magnetic cycle. The first 7 columns of the table are the same as for cycling stars. Then the detection limit on the $R'_{\mathrm{HK}}$ semi-amplitude $A_{\mathrm{lim}}$ is provided, averaged over periods of 2-11 yr (maximum value over all phases). The following column gives the minimum probability that no cycle exists, taking into account the variable sensitivity as a function of phase (see Eq.~\ref{EqProb}). Finally, estimated rotation periods and ages are also given, based on the mean $\log{R'_{\mathrm{HK}}}$ value.

\subsection{Magnetic cycles}
\label{SectMagneticCycles}

\subsubsection{Occurrence}

We now examine the results of the search for magnetic cycles. Among 284 main-sequence FGK stars, we find 99 stars with a detected magnetic cycle, and 20 stars where a magnetic cycle with $A > 0.04$ over 2-11 yr can be fully excluded. This means there are 165 stars with no detected cycle, but for which the existence of a cycle cannot be completely excluded. Among the 99 stars with a magnetic cycle, 16 of them have $R'_{\mathrm{HK}}$ semi-amplitude smaller than 0.04. Therefore, there are 83 stars with a large-amplitude cycle and 36 stars with no large-amplitude cycle. From this we deduce that about 70\% (83/119) of the sample stars exhibit a large-amplitude cycle, while 30\% (36/119) do not. These estimates assume that there is no observational bias related to the classification of stars in either category, e.g. that stars in either category have not been preferentially observed. Observations have been primarily scheduled as a function of the radial velocity variability of stars, and only indirectly according to their activity behavior. We however need to be cautious about observational biases because it is generally easier to detect an existing cycle than to completely exclude any cycle with amplitude $A > 0.04$. Indeed, the latter case requires more data points on average than the former one because of the need to put constraints at all phases, and the increased cycle detectability as amplitude increases. The fraction of 30\% of non-cycling stars is therefore probably a lower bound. More conservatively, one should first provide the occurrence rates normalized over the whole sample: 29\% (83/284) of the sample stars have a large-amplitude cycle, 13\% (36/284) do not, and 58\% (165/284) have an undetermined status.

We try to further refine these statistics of occurrence by using the information from as many stars as possible, i.e. taking into account the detection limits on $R'_{\mathrm{HK}}$ signals also for stars where a magnetic cycle larger than 0.04 cannot be fully excluded (i.e. at all phases). We follow the methodology described in Sect.~\ref{SectPeriodicities} and compute for each sample star the lower limit to the probability that no cycle larger than 0.04 exists for periods between 2-11 yr (Eq.~\ref{EqProb}). By summing all these probabilities one can derive another estimate of the proportion of quiet stars in the whole sample. We obtain that at least 26\% of sample stars do not have a magnetic cycle, to be compared to at least 29\% which do have one (and 45\% remain unclassified). In this case, the 'classification bias' has shifted in the other direction since we used stars with less data points for the non-cycling category, and the 26-to-29\% ratio can be considered as an upper bound to the fraction of non-cycling stars. We conclude that between 30 and 47\% of older solar-type stars in the solar neighborhood do not have any magnetic cycle with $A > 0.04$, while between 53 and 70\% do have one.

\subsubsection{Periods and amplitudes}

\begin{figure}
\centering
\includegraphics[width=\columnwidth]{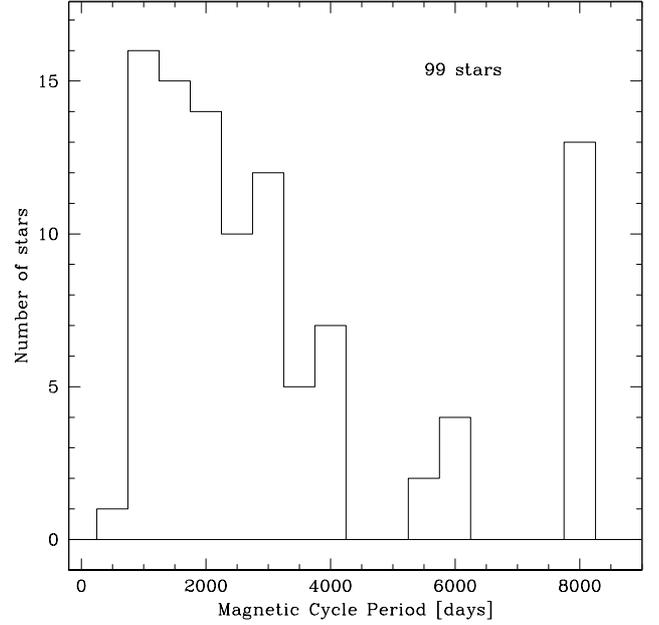}
\caption{Distribution of magnetic cycle periods for the 99 stars exhibiting significant long-term modulations in $R'_{\mathrm{HK}}$. Periods are poorly constrained beyond $\sim$4000 days due to the limited duration of the survey. Unconstrained long-term trends are represented in the last bin at 8000 days.}
\label{FigHistoP}
\end{figure}

\begin{figure}
\centering
\includegraphics[width=\columnwidth]{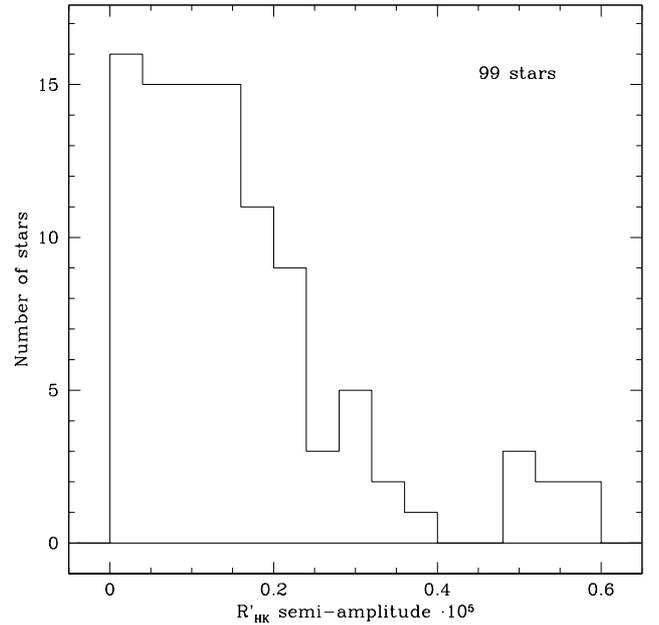}
\caption{Distribution of magnetic cycle semi-amplitudes for the 99 stars exhibiting significant long-term modulations in $R'_{\mathrm{HK}}$.}
\label{FigHistoK}
\end{figure}

Fig.~\ref{FigHistoP} and Fig.~\ref{FigHistoK} show the distributions of magnetic cycle periods and semi-amplitudes, respectively. Perhaps surprisingly, the period distribution shows a peak at relatively short periods ($\sim$3-5 yr) and a sharp decrease at long periods. Our survey is not long enough to properly constrain periods beyond $\sim$4000 days, but long-term trends are nonetheless easily detected if they are of sufficient amplitude. These are shown in the last bin in Fig.~\ref{FigHistoP}. Even if all these unconstrained long-term trends were better characterized, they would not be in sufficient numbers to significantly change the overall shape of the period distribution. The Sun, with its period of 11 yr (4000 d) appears to have a rather long cycle compared to other solar-type stars (80\% of the stars have a period shorter than the Sun). As for cycle semi-amplitudes (Fig.~\ref{FigHistoK}), we note a broad peak at $\sim$0.1 in $R'_{\mathrm{HK}}$ and a large number of low-amplitude cycles. Again, the Sun appears to have a rather large $A$ of 0.26 compared to other solar-type stars (82\% have a lower semi-amplitude). Finally, we note that the first bin in Fig.~\ref{FigHistoK} contains semi-amplitudes smaller than 0.04, i.e. cycles that are not considered 'large-amplitude' in the present work. We nevertheless see that several stars show variations at that level, but these cycles are difficult to measure and require a large number of data points to be detected.

\subsubsection{Correlations with activity level and stellar properties}

\begin{figure}
\centering
\includegraphics[width=\columnwidth]{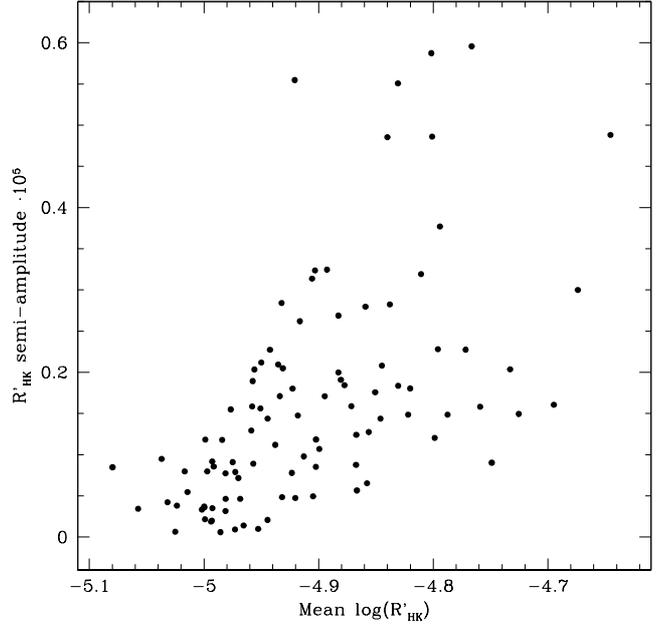}
\caption{Magnetic cycle semi-amplitude as a function of mean activity level.}
\label{FigPlotCK}
\end{figure}

\begin{figure}
\centering
\includegraphics[width=\columnwidth]{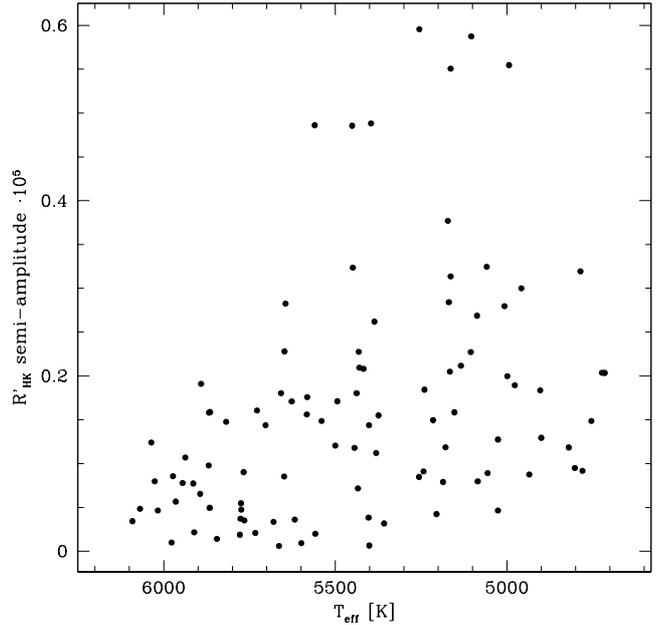}
\caption{Magnetic cycle semi-amplitude as a function of stellar effective temperature.}
\label{FigPlotTeffK}
\end{figure}

We further searched for possible correlations between cycle period, amplitude, mean activity level and effective temperature. We found nothing obvious between period and amplitude, except perhaps a lack of large-amplitude cycles at short periods. When examining period vs. mean activity level, it seems that stars more active than $\log{R'_{\mathrm{HK}}}$ = -4.75 only exhibit short periods ($P <$ 3000 d), although this trend remains uncertain because of small-number statistics. Overall, it is difficult to analyze cycle periods in more details because of the rather large error bars affecting many periods. A clearer picture emerges when plotting cycle amplitude as a function of mean activity level (see Fig.~\ref{FigPlotCK}). Inactive stars with $\log{R'_{\mathrm{HK}}}$ = -5.0 only show low cycle amplitudes, while more active stars tend to show larger amplitudes on average, although there is a rather large scatter around this trend. We also plot in Fig.~\ref{FigPlotTeffK} cycle semi-amplitude vs. effective temperature. We see that the largest amplitudes are concentrated in the range 5000-5600\,K, with a decreasing trend on both sides of the temperature scale. In particular, hotter stars do not show semi-amplitudes larger than $\sim$0.2.

\begin{figure}
\centering
\includegraphics[width=\columnwidth]{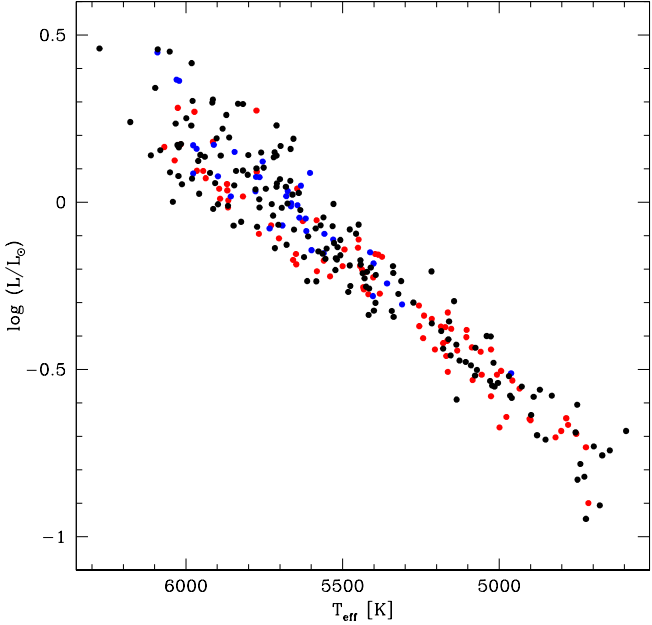}
\caption{H-R diagram showing stars with a large-amplitude magnetic cycle (red), stars with no large-amplitude cycle (blue), and stars of undetermined status (black).}
\label{FigPlotTeffL}
\end{figure}

\begin{figure}
\centering
\includegraphics[width=\columnwidth]{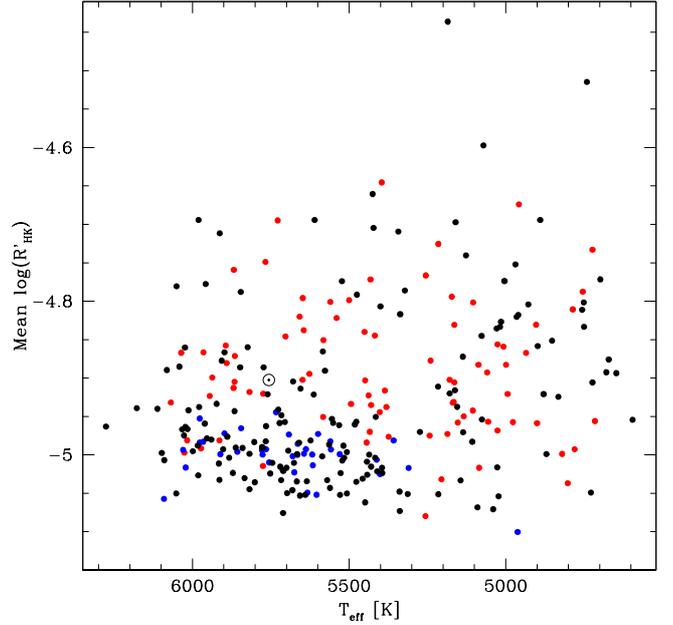}
\caption{$T_{\mathrm{eff}}$ vs. $\log{R'_{\mathrm{HK}}}$ plot showing stars with a large-amplitude magnetic cycle (red), stars with no large-amplitude cycle (blue), and stars of undetermined status (black). The position of the Sun is shown as a $\odot$ symbol.}
\label{FigPlotTeffC}
\end{figure}

We now examine how stars with and without a magnetic cycle are distributed in a H-R diagram and in a $T_{\mathrm{eff}}$ vs. $\log{R'_{\mathrm{HK}}}$ plot. Fig.~\ref{FigPlotTeffL} shows a H-R diagram of the 284 main-sequence stars in the sample, with a different symbol used for a) stars with a large-amplitude cycle, b) stars with no large-amplitude cycle, and c) stars with insufficient observations to be classified. We immediately see that almost all non-cycling stars are found at higher temperatures ($T_{\mathrm{eff}} \gtrsim 5300\, \mathrm{K}$), while stars with large amplitudes seem to exist at all temperatures. In the $T_{\mathrm{eff}}$ vs. $\log{R'_{\mathrm{HK}}}$ plot (Fig.~\ref{FigPlotTeffC}), we see that all non-cycling stars are concentrated in the low-activity, high-temperature regime. In particular, there is no constant star above $\log{R'_{\mathrm{HK}}}$ = -4.94. Stars with a large-amplitude cycle are distributed roughly homogeneously in temperature at larger mean activity levels, although there are some cycling stars below $\log{R'_{\mathrm{HK}}}$ = -5.0.

We can think of two possible interpretations for these results. Firstly, considering the mean $\log{R'_{\mathrm{HK}}}$ value as an age indicator for solar-type stars, these findings would indicate an evolution in magnetic cycles characteristics as stars age, in the sense of a decrease in cycle amplitude with age. This interpretation is also supported by Fig.~\ref{FigPlotCK}, and the existence of many low-amplitude cycles (Fig.~\ref{FigHistoK}). This picture fits well with the known paucity of magnetic cycles in subgiants (see Sect.~\ref{SectSubGiants}). However, a second explanation might also be possible: since we know that the Sun undergoes periods of extended activity minima (e.g. the Maunder minimum), it is possible that the non-cycling stars in our sample are in fact in a temporary state of minimum activity. This is supported by the fact that some stars have been caught in a transition state between cycling and non-cycling behaviors \citep[see e.g.][]{baliunas95}. In the end, both explanations may be simultaneously true, and stars may actually go through a phase of transient magnetic cycles as they age, before these cycles completely disappear when leaving the main sequence.

It must be noted that this discussion mainly concerns G dwarfs, since we find only one non-cycling star among K dwarfs ($T_{\mathrm{eff}} \lesssim 5300\, \mathrm{K}$). We have already noted that the distribution of mean $\log{R'_{\mathrm{HK}}}$ values varies markedly between G and K dwarfs (see Sect.~\ref{SectRHK}). Whereas $\log{R'_{\mathrm{HK}}}$ values show a clear peak at -5.0 for G dwarfs, they are much more spread between -4.7 and -5.0 for K dwarfs. The lack of non-cycling K dwarfs may be due to the fact that these stars are on average not old enough to have reached the $\log{R'_{\mathrm{HK}}}$ $\approx$ -5.0 level, which may be required to halt magnetic cycles.

\subsubsection{Subgiants}
\label{SectSubGiants}

It has long been known that stars evolving off the main sequence tend to show particularly low levels of chromospheric activity \citep[e.g.][]{wright04b}, and also less long-term variability than main-sequence stars. We confirm these findings here. As noted earlier, all 20 subgiants in the sample have a mean $\log{R'_{\mathrm{HK}}}$ value below -5.0. More importantly, only one of them shows a magnetic cycle (semi-amplitude $A$ = 0.055), while 5 stars have no cycle at all ($A < 0.04$). Moreover, many of the 14 stars not classified in either category have high probabilities to have no magnetic cycle. Summing these probabilities yields a global proportion of between 83-93\% of non-cycling stars among subgiants, to be compared to 30-47\% for main-sequence stars. We therefore conclude that most subgiants have indeed reached a stage of long-term stability in their chromospheric emission.

\subsubsection{The case of the Sun}

We can now try to consider the Sun within the broader context of our sample of solar-type stars. We show in Fig.~\ref{FigPlotTeffC} the position of the Sun at $T_{\mathrm{eff}}$ = 5770 K and $\log{R'_{\mathrm{HK}}}$ = -4.91. It is mainly surrounded by cycling stars in this figure, but is actually not far from the boundary between mainly cycling and mainly constant G dwarfs, located at $\log{R'_{\mathrm{HK}}}$ $\approx$ -4.95. One could therefore conjecture that the Sun will soon transition to a permanently quiet state, perhaps within $\sim$1-2 Gyr. On the other hand, the present-day solar cycle semi-amplitude of 0.26 appears rather large compared to other stars at the same activity level and effective temperature (see Fig.~\ref{FigPlotCK} and \ref{FigPlotTeffK}). However, one should keep in mind that there are significant variations in amplitude from one solar cycle to the next (up to 50\% and more), and that solar activity has been rather high since 1950 compared to previous periods, as indicated by sunspot data. On average, the Sun is therefore likely to have a cycle amplitude that is typical of stars in its activity and temperature range. It is also conceivable that the Sun is presently in an unstable state with activity cycles that are highly variable in amplitude, and that this could be the prelude to a transition to a permanently quiet state.

\subsection{Short-term variability}

\begin{figure}
\centering
\includegraphics[width=\columnwidth]{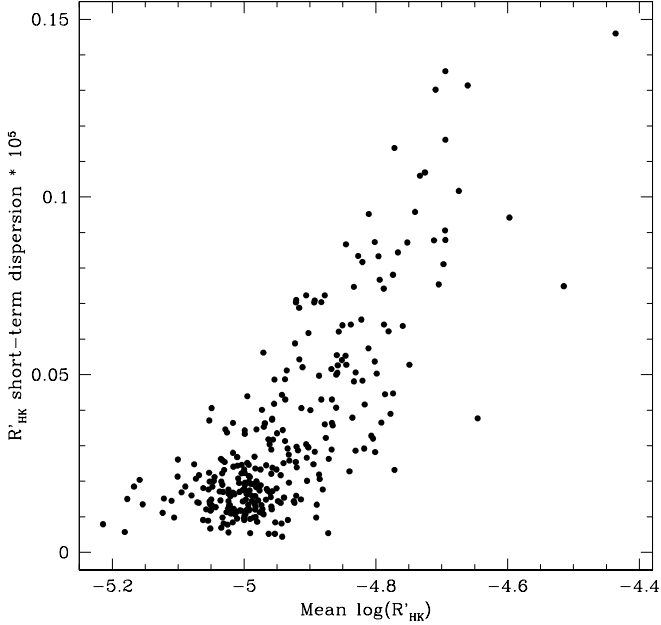}
\caption{$R'_{\mathrm{HK}}$ short-term dispersion as a function of mean activity level $\log{R'_{\mathrm{HK}}}$.}
\label{FigPlotRHKStdev}
\end{figure}

Besides the long-term variations in chromospheric emission caused by magnetic cycles, solar-type stars also show (sometimes prominent) short-term variability caused by active regions rotating with the star. For all stars in the sample, we characterize this short-term variability by computing the $R'_{\mathrm{HK}}$ standard deviation, after subtracting the best-fitting long-period signal with $P$ $>$ 2 yr (whether significant or not). We then search for correlations between $R'_{\mathrm{HK}}$ variability and stellar parameters. The only clear dependence we could find is with the mean activity level of the stars. Fig.~\ref{FigPlotRHKStdev} shows the dispersion in $R'_{\mathrm{HK}}$ as a function of $\log{R'_{\mathrm{HK}}}$ itself. As can be seen, there is a dense clump of low-variability stars with $\log{R'_{\mathrm{HK}}}$ between -5.1 and -4.9, and a marked trend towards high variability at higher activity levels. However, there are many variable stars at $\log{R'_{\mathrm{HK}}}$ = -4.90 already. Actually, the upper envelope of the points increases sharply from -5.00 towards higher activity levels. The lower envelope appears flat until about -4.95, where it starts increasing steadily with increasing activity level. The flattening at low activity may be due either to a genuine stellar variability floor or, more probably, to residual instrumental systematics. Incidentally, we note that the Sun reaches an activity level of about -4.97 at cycle minimum, when almost no sunspots are visible at the solar surface. It may be that this level represents a universal threshold that marks the appearance of significant active regions on solar-type stars.

We also note that the scatter around the overall trend in Fig.~\ref{FigPlotRHKStdev} may be caused at least partly by inclination effects, i.e. by the different space orientations of stellar rotation axes. Indeed, stars seen pole-on are expected to show less rotationally-induced variability since active regions maintain the same projected area as seen from the observer's standpoint.

In summary, it appears that the mean activity level, expressed by $\log{R'_{\mathrm{HK}}}$, is a good indicator of magnetically-induced variability on the short-term as well as for long-term activity cycles, since these also tend to increase in amplitude with the mean activity level.

In the context of radial velocity searches for exoplanets, more in-depth studies of short-term variability are fundamental to understand radial velocity jitter induced by stellar activity. It is clear that the degree of $R'_{\mathrm{HK}}$ short-term variability is linked to the degree of RV, FWHM, contrast and BIS variability. However, the exact relationships between all these quantities on the short term are complex and beyond the scope of this paper. Several well-sampled time series are available among the stars in this sample and deserve individual investigations. These will be the topic of a subsequent paper (Dumusque et al., in prep.).

\section{Impact of magnetic cycles on radial velocities and line shape parameters}

\begin{figure}
\centering
\includegraphics[width=\columnwidth]{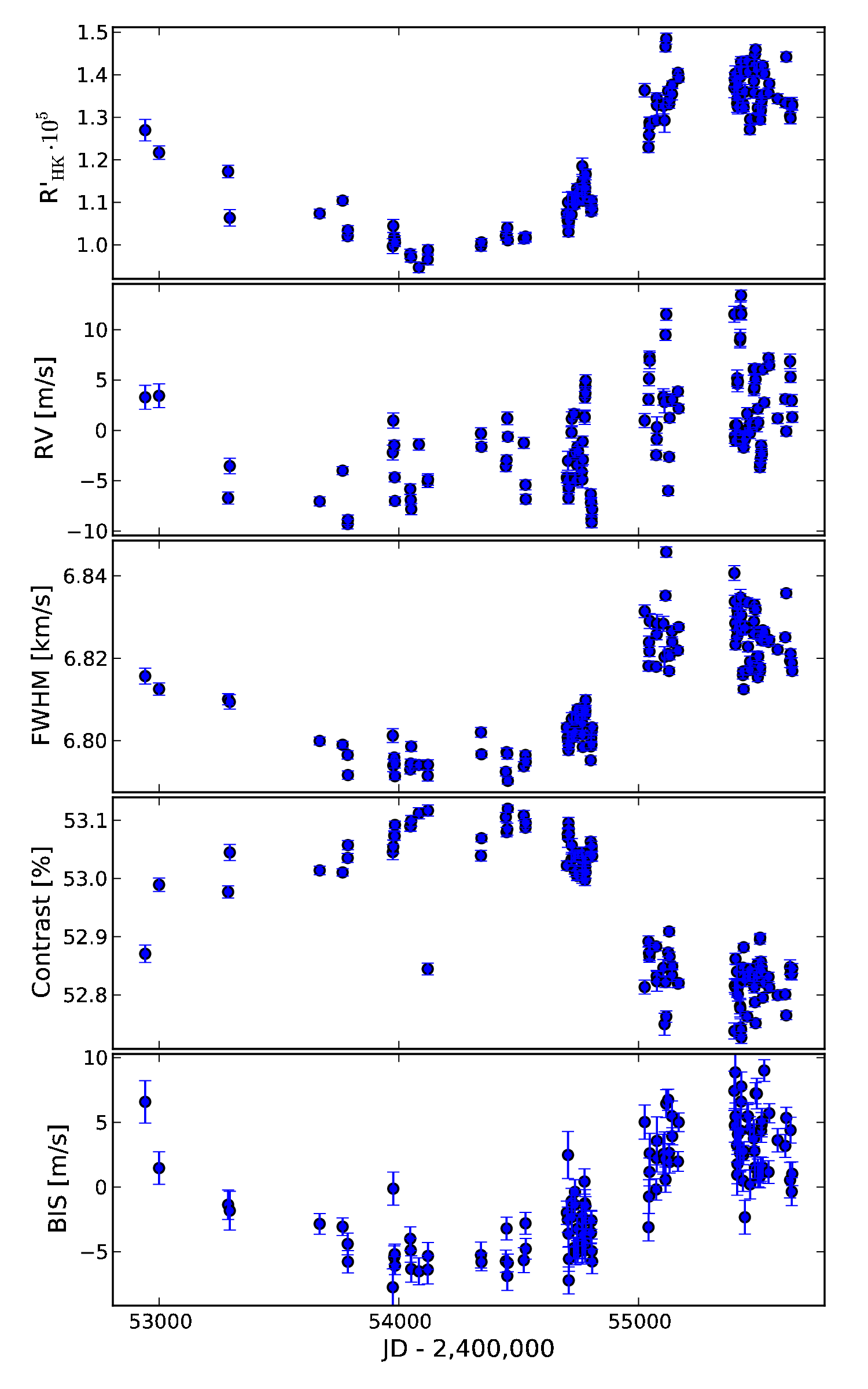}
\caption{Time series of $R'_{\mathrm{HK}}$, RV, FWHM, contrast and BIS measurements for the star HD\,21693, showing clear correlations between all quantities.}
\label{FigHD21693}
\end{figure}

We now want to study the possible impact of magnetic cycles on radial velocities, CCF FWHM, CCF contrast, and CCF bisector span. We consider the 99 main-sequence stars that have been found to have a magnetic cycle and try to find correlations between $R'_{\mathrm{HK}}$ and the other quantities. In several cases, correlations are actually obvious to the eye. An example is shown in Fig.~\ref{FigHD21693}, where the time series of the five quantities are displayed for the G8V star HD\,21693. As can be seen, the RV, FWHM, contrast and BIS measurements clearly vary in phase with $R'_{\mathrm{HK}}$. Similar behaviors are apparent on many other stars, but often less clearly so because of poorer sampling. We therefore need robust methods to study correlations in the whole sample of stars. Two of them are described below.

\subsection{Blind fitting of the $R'_{\mathrm{HK}}$-derived model}

One possible method is to fix the period and the phase of the putative RV, FWHM, contrast or BIS signal to the values found when fitting the magnetic cycle, and then compute via linear least-squares the best-fitting mean levels and semi-amplitudes on the RV, FWHM, contrast and BIS data. As for magnetic cycles, this assumes that we are looking for sinusoidal variations in these data, and that these variations are in phase with the $R'_{\mathrm{HK}}$ data. The advantage of this method is its robustness, since only two free parameters are fitted on the whole time series, and the long-period sinusoidal model is relatively insensitive to short-term scatter related to rotational modulations. In this way, effects from magnetic cycles on the one hand, and rotational modulations on the other hand can often be distinguished even if sampling is rather poor.

\begin{figure}
\centering
\includegraphics[width=\columnwidth,bb=390 185 567 690,clip]{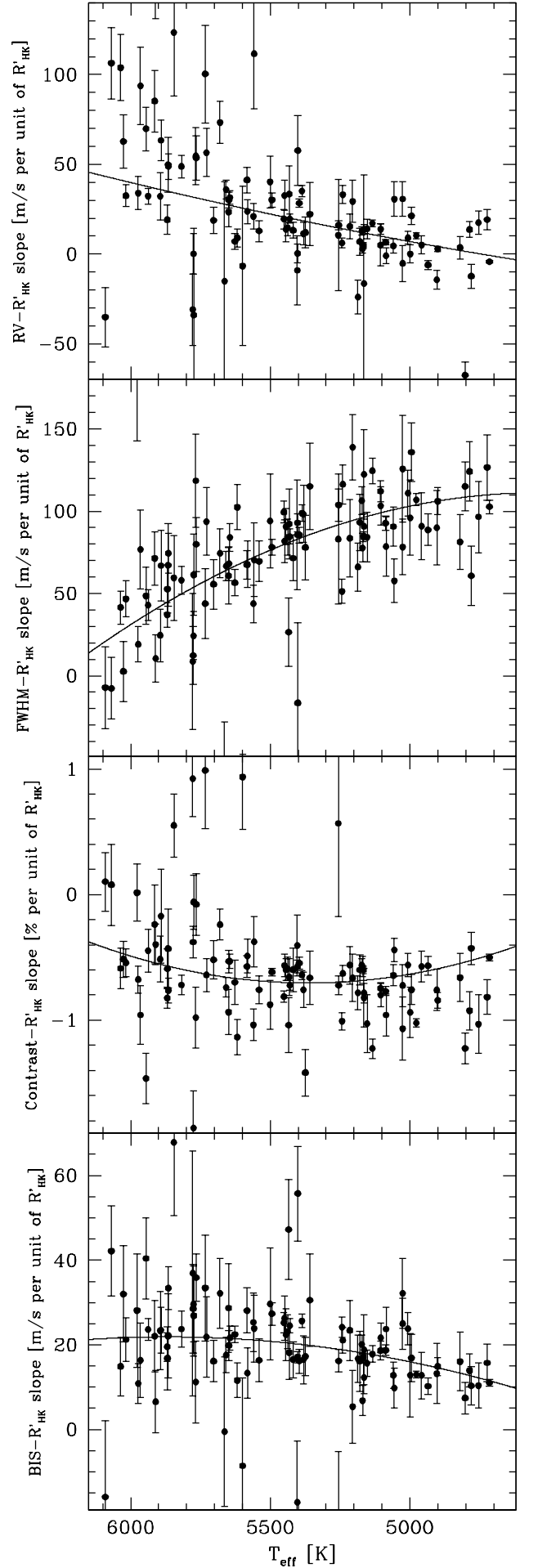}
\caption{Sensitivity of RV, FWHM, contrast and BIS measurements to magnetic cycles (expressed by $R'_{\mathrm{HK}}$ variations), as a function of $T_{\mathrm{eff}}$.}
\label{FigPlotCorr}
\end{figure}

Following this scheme, for each star with a magnetic cycle we obtain a best-fit mean value and semi-amplitude for the RV, FWHM, contrast and BIS data, at the period and phase given by the $R'_{\mathrm{HK}}$ data. Then the sensitivity of each of these quantities to magnetic cycles is obtained by simply dividing the RV, FWHM, contrast and BIS semi-amplitudes by the $R'_{\mathrm{HK}}$ semi-amplitude. Uncertainties in the sensitivities are derived from the covariance matrix of the least-squares fits. In this process we artificially inflate error bars on individual measurements to bring the reduced $\chi^2$ to 1, so as to "model" the short-term scatter in each quantity as a random Gaussian noise source. This approximation may be risky on a case-by-case basis since rotational modulations induce signals that are time-correlated and non-Gaussian. However, these aspects have only little effect in our case since we are fitting only the mean level and amplitude of a long-period signal, and we are mainly interested in the global results over the whole sample rather than to individual stars.

Fig.~\ref{FigPlotCorr} shows the obtained sensitivities of the RV, FWHM, contrast and BIS, plotted as a function of effective temperature. We emphasize that all stars are included in the procedure, with no attempt to filter out objects that show variability manifestly unrelated to magnetic cycles, including RV signals caused by orbiting companions, short-term activity or instrumental systematics. In spite of this, a clear dependence on $T_{\mathrm{eff}}$ can be seen for all quantities. Besides $T_{\mathrm{eff}}$, the other fundamental stellar parameter that is likely to play a role is metallicity. We proceed to fit linear models in $T_{\mathrm{eff}}$ and [Fe/H] to the sensitivities. Models combining a dependence on both parameters are found to yield the best reduced $\chi^2$, and second-order polynomials are found to be sufficient for both parameters (going beyond would increase the reduced $\chi^2$). We finally obtain the following empirical models:

\begin{eqnarray}
\label{EqSensitivities}
\nonumber C_{\mathrm{RV}} &=& (19.04 \pm 0.62) + (3.20 \pm 0.20)\cdot10^{-2} \cdot \tilde{T}_{\mathrm{eff}} \\
\nonumber && +\; (4.5 \pm 3.1)\cdot10^{-6} \cdot \tilde{T}_{\mathrm{eff}}^2 + (17.8 \pm 3.1) \cdot \mathrm{[Fe/H]} \\
&& +\; (24.3 \pm 5.2) \cdot \mathrm{[Fe/H]}^2 \\
\nonumber \\
\nonumber C_{\mathrm{FW}} &=& (83.2 \pm 1.4) - (6.46 \pm 0.36)\cdot10^{-2} \cdot \tilde{T}_{\mathrm{eff}} \\
\nonumber && -\; (3.71 \pm 0.79)\cdot10^{-5} \cdot \tilde{T}_{\mathrm{eff}}^2 + (33.7 \pm 6.8) \cdot \mathrm{[Fe/H]} \\
&& +\; (38 \pm 17) \cdot \mathrm{[Fe/H]}^2 \\
\nonumber \\
\nonumber C_{\mathrm{CO}} &=& (-0.703 \pm 0.012) + (3.9 \pm 3.4)\cdot10^{-5} \cdot \tilde{T}_{\mathrm{eff}} \\
\nonumber && +\; (5.28 \pm 0.67)\cdot10^{-7} \cdot \tilde{T}_{\mathrm{eff}}^2 + (0.333 \pm 0.055) \cdot \mathrm{[Fe/H]} \\
&& -\; (0.15 \pm 0.13) \cdot \mathrm{[Fe/H]}^2 \\
\nonumber \\
\nonumber C_{\mathrm{BIS}} &=& (20.18 \pm 0.44) + (7.3 \pm 1.3)\cdot10^{-3} \cdot \tilde{T}_{\mathrm{eff}} \\
\nonumber && -\; (7.8 \pm 2.4)\cdot10^{-6} \cdot \tilde{T}_{\mathrm{eff}}^2 + (8.2 \pm 2.3) \cdot \mathrm{[Fe/H]} \\
&& +\; (13.7 \pm 5.1) \cdot \mathrm{[Fe/H]}^2
\end{eqnarray}

In these equations, $C_{\mathrm{RV}}$, $C_{\mathrm{FW}}$, $C_{\mathrm{CO}}$ and $C_{\mathrm{BIS}}$ are the sensitivities expressed in m\,s$^{-1}$ (or percent for the contrast) per unit of $R'_{\mathrm{HK}}$, and $\tilde{T}_{\mathrm{eff}}$ = $T_{\mathrm{eff}}$\,--\,5400\,K (chosen to minimize covariances between polynomial coefficients). Fig.~\ref{FigPlotCorr} shows these models overplotted on the data for zero metallicity. Some of the scatter seen in the figure is in fact due to the effects of metallicity.

A dependence on stellar properties, in particular $T_{\mathrm{eff}}$ and [Fe/H], is expected since photospheric lines will respond differently to magnetic cycles depending on physical conditions and the details of the atmospheric structure. An important issue arising here is whether the simple dependencies derived above are able to capture the full effects of magnetic cycles on spectral lines, or whether other factors play an important role as well. A way to estimate this is to look at the reduced $\chi^2$ of the best-fit models to check if some variability remains unaccounted for. A related question is whether we should filter out some stars before performing the fit if they show additional variability unrelated to magnetic cycles. Taking all stars into account, we obtain reduced $\chi^2$ values of 6.79, 2.51, 5.64 and 1.91 for RV, FWHM, contrast and BIS, respectively. Clearly, additional variability is present in radial velocities and contrast, while FWHM and BIS data are closer to a reasonably good fit. The RV result is not surprising since we are considering here raw measurements, which in particular still contain the signals of many planets. We have no explanation for the rather poor $\chi^2$ value for the contrast, except that this parameter is the most sensitive to instrumental systematics.

We examine below the possibility to remove the RV signals of known planets before deriving RV sensitivities. But as a first step, we try to estimate the impact these signals might have on our fitted sensitivity curve (Eq.~\ref{EqSensitivities}). To do this, we artificially remove from the data the three most significant outliers as well as all stars with $R'_{\mathrm{HK}}$ semi-amplitudes smaller than 0.04 (which generally have less precise sensitivities), and we refit this filtered dataset. The obtained coefficients are indistinguishable from the original ones in Eq.~\ref{EqSensitivities}. This shows that the exact choice of stars to include in the fit, and the presence of planet-induced RV signals, have a negligible effect on the final RV sensitivity curve. It is mainly due to the fact that error bars on individual sensitivities usually take into account the mismatch between RV and $R'_{\mathrm{HK}}$ data if signals unrelated to magnetic cycles are present. In other words, a significant outlier can only arise if a planet induces a strong RV signal {\em exactly in phase} with the magnetic cycle of its host star, which is obviously quite unlikely.

We conclude that the obtained sensitivity curves in Eq.~\ref{EqSensitivities} are valid, even if they were derived based on uncorrected RV data. The relatively good reduced $\chi^2$ values for the FWHM and BIS data also indicate that the simple $T_{\mathrm{eff}}$ and [Fe/H] dependence is able to capture most of the variability seen in the sensitivities, and that other stellar properties probably have a much smaller effect.

\subsection{Data binning and subtraction of planetary signals}

In parallel, we also try an alternative approach for the RV sensitivities. We search for correlations directly through linear regression between RV and $R'_{\mathrm{HK}}$ data, using the Pearson correlation coefficient to estimate the quality of the correlation. Before that however, considering that RV data are affected by the presence of orbiting companions, we first remove the signals of planetary origin from the RV data to obtain a more reliable RV - $R'_{\mathrm{HK}}$ correlation. We also restrict ourselves to stars showing a magnetic cycle with $A > 0.04$ (83 stars out of 99). Moreover, to have sufficient sampling and minimize the impact of short-term variability we further refine the selection as follows:

\begin{itemize}
\item Stars must have more than 3 years of measurements
\item Measurements are binned over 3 months, with a required minimum of at least 3 measurements per bin
\item A total of at least 5 bins must be available
\end{itemize}

\begin{figure}
\centering
\includegraphics[width=\columnwidth]{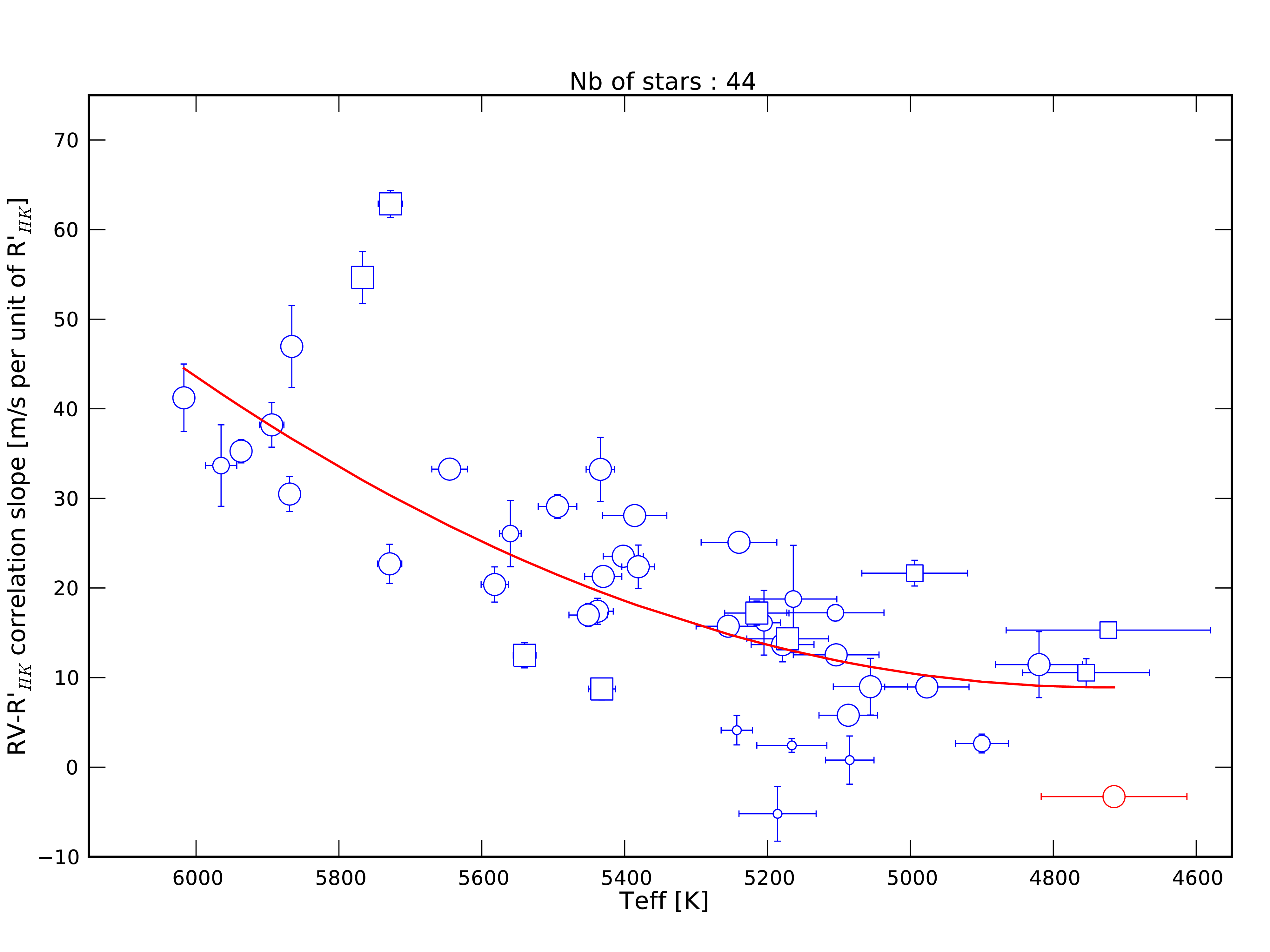}
\includegraphics[width=\columnwidth]{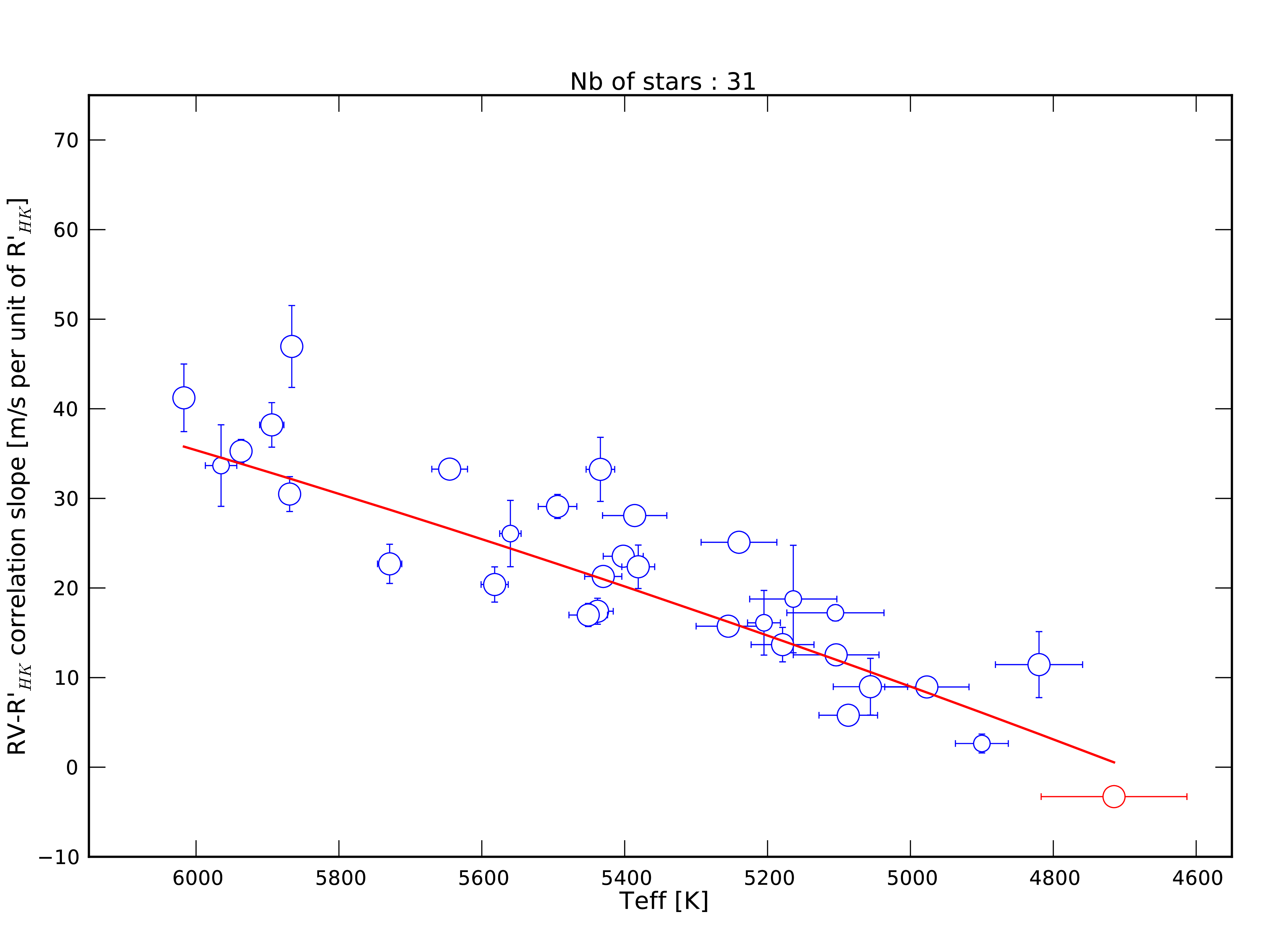}
\caption{{\em Top:} Correlation slope between RV and $R'_{\mathrm{HK}}$, as a function of effective temperature. Circles and squares are used to indicate inactive stars (max($\log{R'_{\mathrm{HK}}}$) $<$ -4.75) and active stars (max($\log{R'_{\mathrm{HK}}}$) $>$ -4.75), respectively (the maximum is calculated on the 3-month bins). In addition, three different symbol sizes are used to show the Pearson correlation coefficient $R$ of the RV - $R'_{\mathrm{HK}}$ correlation: small for $R < 0.5$, medium for $0.5 < R < 0.75$ and large for $R > 0.75$. {\em Bottom:} Same figure but with inactive stars only.}
\label{FigXavier}
\end{figure}

After this selection we are left with 44 stars, which all show clear magnetic cycles and offer a rather good sampling over more than 3 years. The following step consists of subtracting the RV signals of known planets, without erasing a possible signal coming from magnetic cycles. For the already known planets we use the orbital parameters found in the literature. There are also several new objects in this HARPS sample, for which the publication is in preparation \citep[][Udry et al., in prep; Queloz et al., in prep.]{pepe11,dumusque11c}. In these cases we take the preliminary orbital elements from these unpublished works. Usually, the magnetic cycle is first fitted with a sinusoid on the activity index, and the obtained model is then fitted to the RV data together with the Keplerians, just leaving the amplitude as a free parameter \citep[see][for an example]{dumusque11c}. The RV - $R'_{\mathrm{HK}}$ correlation slope is then obtained using the RV residuals (i.e. all planets subtracted). We plot in Fig.~\ref{FigXavier} the RV - $R'_{\mathrm{HK}}$ correlation slope as a function of effective temperature for the 44 selected stars.

As can be seen in Fig. \ref{FigXavier}, the RV - $R'_{\mathrm{HK}}$ correlation slope is well correlated to effective temperature. However, some stars fall quite far from the fitted quadratic model. Most stars which seem to be outliers happen to have a square symbol in Fig.~\ref{FigXavier}, which represents stars with a maximum activity value higher than $\log{R'_{\mathrm{HK}}}$ = -4.75. This corresponds to active stars according to the limit given by the Vaughan-Preston gap \citep{vaughan80}. The significant variability of active stars is produced by large magnetic features rotating with the star, that also induce large short-term RV variations. Indeed, all stars classified as active exhibit a minimum short-term RV variability of 10\,m\,s$^{-1}$. Combined with imperfect sampling, this does not allow us to properly estimate the RV - $R'_{\mathrm{HK}}$ correlation slope with the method described here.

There are also four inactive stars that seem to be outliers: HD\,85390, HD\,192310, HD\,72673 and HD\,144628, which present no RV - $R'_{\mathrm{HK}}$ correlation (small circles near 5200 K in Fig.~\ref{FigXavier}). The first two stars host multi-planet systems with long-period planets \citep[see][]{mordasini11,pepe11}. The long-period planets are not fully covered yet in phase and it is likely that the orbital solutions, and thus the RV residuals, will change when more points become available. Therefore, this can explain the observed lack of correlation. The third star presents a large peak in the periodogram at the rotational period of the star. We are therefore again confronted to short-term activity induced by magnetic features rotating with the star. In the case of an important signal related to short-term activity, even if we select only 3-month bins with more than 3 measurements, the sampling must be very dense to reduce the effect of activity. This is not the case for HD\,72673, which can explain the unexpected value of the RV - $R'_{\mathrm{HK}}$ correlation slope.

Once active stars are removed from the plot (as well as the four stars mentioned above), we obtain a much cleaner relation between the RV - $R'_{\mathrm{HK}}$ correlation slope and effective temperature, as shown in Fig.~\ref{FigXavier} (bottom panel). The fitted linear model is given by:

\begin{eqnarray}
\nonumber C_{\mathrm{RV}} &=& (20.17 \pm 0.27) + (2.689 \pm 0.097)\cdot10^{-2} \cdot \tilde{T}_{\mathrm{eff}} \\
\nonumber && -\; (2.6 \pm 1.8)\cdot10^{-6} \cdot \tilde{T}_{\mathrm{eff}}^2 + (20.2 \pm 1.3) \cdot \mathrm{[Fe/H]} \\
&& +\; (26.3 \pm 2.5) \cdot \mathrm{[Fe/H]}^2
\end{eqnarray}

The different variables are the same as in Eq.~\ref{EqSensitivities} above. Comparing the two equations, we see that this second method, which focuses on a smaller sample of carefully-selected stars, provides a $T_{\mathrm{eff}}$ dependence that is very similar to the previous one. Indeed, all the obtained polynomial coefficients are compatible with each other at the 2$\sigma$ level. We conclude that both methods confirm that 1) magnetic cycles do induce RV variations in solar-type stars, and 2) the sensitivity of RV measurements to a given magnetic cycle amplitude decreases towards cooler stars.

\section{Discussion and conclusion}

\subsection{Calibration of $R'_{\mathrm{HK}}$ taking into account metallicity}

In this paper we provide an updated calibration of the $R'_{\mathrm{HK}}$ index, which takes into account the effects of metallicity on stellar bolometric luminosity. Based on a calibration of effective temperature as a function of $B-V$ color and metallicity, a correction factor to the usual $R'_{\mathrm{HK}}$ definition can be applied and removes the observed trend of increasing $R'_{\mathrm{HK}}$ values with decreasing metallicity. The correction yields a sharper distribution of mean $R'_{\mathrm{HK}}$ values for our sample stars, confirming that stars of different metallicities were exhibiting an artificial spread in $R'_{\mathrm{HK}}$ with the uncorrected calibration.

\subsection{Statistics of magnetic cycles}

The present survey has allowed us to revisit some of the properties of magnetic cycles in solar-type stars. To the best of our knowledge, most of the information previously available on this subject comes from the Mt Wilson project, and in particular the global analysis of \citet{baliunas95} \citep[see however][for a recent study]{saar09}. Two main difficulties arise when aiming at characterizing magnetic cycles: the required long-term precision on Ca II H\&K measurements, and the necessity to monitor many stars over decadal timescales. It turns out that large radial-velocity planet-search surveys, which started in the 1990s, can actually provide both, since they are based on high-resolution echelle spectrographs and regularly monitor hundreds of solar-type stars over many years. Several surveys have already yielded mean activity levels for many stars, but {\em time series} of activity measurements have been rarely discussed.

The HARPS data presented here are of excellent quality in terms of precision on $R'_{\mathrm{HK}}$, but they span "only" 7-8 years, compared to 25 years for the published Mt Wilson data \citep{baliunas95}. However, as shown in Sect.~\ref{SectMagneticCycles}, this did not prevent us from detecting magnetic cycles with estimated periods up to $\sim$15 yr, and long-term drifts of undetermined periodicity. Obviously, the precision on fitted periods quickly decreases when going beyond the time span of observations, but even uncertainties of 25-50\% on the period still provide useful information on cycle properties.

An important simplification adopted in this work is the assumption of perfectly sinusoidal magnetic cycles \citep[as in][]{baliunas95}. We know from the Sun that they are in fact not really sinusoidal for any given cycle number, and moreover their amplitudes and shapes may vary significantly from one cycle to the next. However, our survey usually covers only one or two cycles, so that we are not able to study the {\em evolution} of the magnetic cycle in any given star. We rather obtain a random snapshot of cycles in many stars at the same time. With these limitations in mind, we can summarize the main results of this work as follows:

\begin{itemize}

\item The global occurrence of magnetic cycles with semi-amplitude $A$ $>$ 0.04 is 61$\pm$8\% among older solar-type stars. Correspondingly, 39$\pm$8\% of older solar-type stars show a flat activity record. These numbers are roughly compatible with the results of \citet{baliunas95}, although these authors do not explicitly provide comparable estimates and use a different classification scheme. The picture is markedly different for subgiants, for which the occurrence of magnetic cycles drops to only 12$\pm$5\%.

\item The distribution of magnetic cycle periods shows a broad maximum between 2-10 yr, and a sharp decrease towards longer periods. This may be due in part to the limited duration of the survey, but the number of observed long-term trends does not seem to be sufficient to produce a large number of long-period cycles. The decrease towards longer periods therefore seems to be real. There may be some discrepancies between our period distribution and the one of \citet{baliunas95}, in the sense that we detect more short-period cycles than they do. It must be noted, however, that short-period cycles are often less well-defined (more dispersion and evolution with time) than longer-period ones, and the discrepancy may be explained by a different classification scheme combined with a longer duration of observations (\citet{baliunas95} classify as "variable" the stars with poorly defined periodicities). On the other hand, both studies seem to agree on the low number of cycles with $P$ $\gtrsim$ 15 yr. This confirms that the limited duration of our survey does not significantly affect the derived overall properties of magnetic cycles.

\item The distribution of magnetic cycle semi-amplitudes shows a broad peak between zero and 0.2 in $R'_{\mathrm{HK}}$. Given the observational bias against low amplitudes, it seems that magnetic cycles significantly less pronounced than the solar one are the rule rather than the exception.

\item We find no clear correlations between cycle periods and amplitudes, nor between cycle periods and mean activity levels. The sometimes large error bars on cycle periods prevent us from going into further details. The occurrence and stability in time of short-period cycles ($P$ $\lesssim$ 5 yr) should be further investigated, together with methods to disentangle them from evolution of active regions on rotational timescales.

\item We find a clear trend of increasing cycle amplitudes with increasing mean activity levels. In particular, no stars with mean $\log{R'_{\mathrm{HK}}}$ $\approx$ -5.0 show cycles as pronounced as the Sun. We also find that the upper envelope of cycle amplitudes shows a peak in the temperature range 5000-5600\,K, while hotter stars do not exhibit very large amplitudes.

\item The distribution of mean $\log{R'_{\mathrm{HK}}}$ values is markedly different between G and K dwarfs, in the sense that G dwarfs tend to accumulate around -5.0, while K dwarfs are more spread between -4.7 and -5.0. A possible explanation is the slower decrease in activity with age in K dwarfs, which need more than $\sim$6-8 Gyr to reach a mean level of -5.0.

\item Non-cycling stars are only found at mean activity levels lower than about -4.95, and almost only among G dwarfs. Cycling stars are preferentially found at higher activity levels, although some of them are also found around -5.0. The number of well-observed K dwarfs below -4.95 is too low to properly study the occurrence of non-cycling stars in this region of parameter space.

\item A possible interpretation of these results is that magnetic cycle amplitudes decrease with age, and tend to disappear when stars reach a mean activity level of about -5.0. This is also supported by the quasi-absence of cycles in subgiants.

\item The Sun, with a mean $\log{R'_{\mathrm{HK}}}$ of -4.91, is still in the activity range where most stars exhibit magnetic cycles. One can conjecture that solar cycles will slowly disappear as the mean activity decreases to -5.0, possibly within a few Gyr. It is difficult to estimate from our data how frequent periods of extended minima like the Maunder minimum are for any given star, since we only have instantaneous snapshots of the state of many stars. What can be said is that the activity level of the Sun was almost certainly below -4.95 during the Maunder minimum since we do not find any non-cycling stars above that threshold. Considering the well-established decrease in activity with age in solar-type stars, it is more likely that the "clump" of G dwarfs around -5.0 is mainly the signature of an age effect, and not a population of stars in temporary Maunder-minimum states. However, an unknown proportion of these stars are probably in such a state. This also supports the view that Maunder-minimum stars cannot be found on the basis of their mean activity level only \citep{wright04b}, and should rather be identified by the demonstrated absence of a magnetic cycle.

\end{itemize}

\subsection{Effects of magnetic cycles on spectral lines and radial velocities}

It has long been hypothesized that magnetic cycles have an impact on photospheric spectral lines \citep[e.g.][]{dravins85,gray92,lindegren03}. Here we provide a global assessment of magnetically-induced variability in four quantities derived from the HARPS cross-correlation function (RV, FWHM, contrast and BIS), which is constructed as a "master" spectral line built from thousands of mainly weak to moderate individual Fe lines.

\begin{figure}
\centering
\includegraphics[width=\columnwidth]{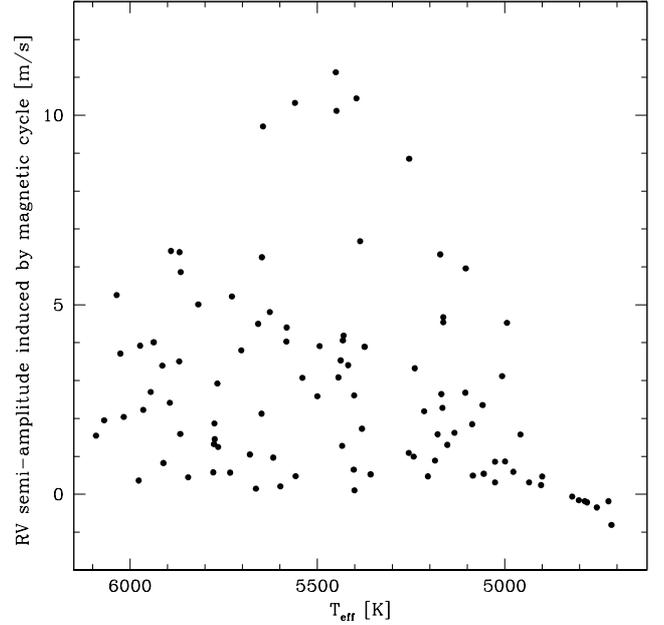}
\caption{Effective impact of magnetic cycles on RV for all sample stars with a detected cycle, shown as a function of $T_{\mathrm{eff}}$.}
\label{FigPlotEffectiveRVImpact}
\end{figure}

We observe long-term correlations between $R'_{\mathrm{HK}}$, RV, FWHM, contrast and BIS in many stars of our sample, with varying sensitivities between stars of different spectral types and various behaviors for different quantities. Globally speaking, for a given variation in $R'_{\mathrm{HK}}$, we find that RV and BIS are more sensitive in G dwarfs than in K dwarfs, whereas the opposite is true for the FWHM. The contrast sensitivity is roughly constant with spectral type, but varies significantly with metallicity. Overall, sensitivity variations with temperature are more important than with metallicity for RV, FWHM and BIS. The sensitivity of RV to magnetic cycles is almost zero at $T_{\mathrm{eff}}$ $\approx$ 4800 K, while the same is true for FWHM at $T_{\mathrm{eff}}$ $\approx$ 6200 K. However, the effective impact of a magnetic cycle for any given star is obtained by multiplying these sensitivities with the actual $R'_{\mathrm{HK}}$ semi-amplitude, which may be large and therefore have a non-negligible effect even if the sensitivity is low. We show in Fig.~\ref{FigPlotEffectiveRVImpact} the actual impact of magnetic cycles on the RV data for all stars in the sample with a detected cycle. For each star the RV sensitivity is computed from the model given in Eq.~\ref{EqSensitivities}, and the result is multiplied by the measured $R'_{\mathrm{HK}}$ semi-amplitude $A$. We see that induced RV semi-amplitudes can reach up to 11 m\,s$^{-1}$ in the worst case. More typical values are between 0 and 3 m\,s$^{-1}$ (there are unaffected stars at all spectral types). Interestingly, the model predicts an {\em anti-correlation} between $R'_{\mathrm{HK}}$ and RV (i.e. a negative RV semi-amplitude in Fig.~\ref{FigPlotEffectiveRVImpact}) for stars cooler than $\sim$4800 K. At least in one case this surprising result seems to be confirmed by an in-depth analysis: the star HD\,85512 \citep{pepe11}. Looking at Fig.~\ref{FigPlotEffectiveRVImpact}, we also see that despite the drop in sensitivity towards cool stars, one can still have RV effects of several m\,s$^{-1}$ because of the occurrence of strong magnetic cycles in K dwarfs.

The previous study that is most directly comparable to the results presented here is the one by \citet{santos10}. These authors obtained long-term HARPS data on eight stars known to have clear magnetic cycles, and studied the behavior of CCF parameters in the same way as we do here. Clear correlations were found between the S index, FWHM and BIS data, while the contrast was anti-correlated to the S index. These findings are in excellent agreement with the present results. On the other hand, attempts to find correlations with the RV data remained inconclusive, or actually RV effects were too small to be detected. The sensitivity model derived in this paper provides a natural explanation for this lack of correlations: it turns out that the five cycling stars actually studied by \citet{santos10} span a very narrow range of effective temperatures (200 K), centered at 5000 K. In this regime RV sensitivities are close to zero, and the S index variations were not large enough to induce a measurable RV effect. The only hot star in the sample, HD\,216385, is a magnetically-constant star that was chosen as a standard. Two other, higher-temperature stars in the sample could not be studied because of strong short-term variability and binarity. We conclude that it is only bad luck that prevented \citet{santos10} from finding correlations between RV and S index.

In general, the RV sensitivities to magnetic cycles are such that the induced RV effects can be of the same order of magnitude and period as the signal of Jupiter-like planets orbiting at several AUs from the star. Therefore, it is critical to check the behavior of the chromospheric activity index before announcing any long-period planet with a RV semi-amplitude lower than $\sim$15 m\,s$^{-1}$. It would be extremely useful for all planet-search teams to measure chromospheric activity (even at lower precision) in all stars exhibiting such low-amplitude, long-period RV signals, and to check whether any correlations are found with the magnetic cycle. This obviously also applies to already published planets, e.g. the Jupiter-analog candidate around HD\,154345 \citep{wright08}. If simultaneous activity measurements are available, then the present work shows that they can be easily used to correct RV data from magnetic cycle effects. Two approaches can be used: either one includes an additional sinusoidal component in the Keplerian model, fixing its period and phase to the ones of the magnetic cycle, or one can use the sensitivity model presented here to {\em a priori} correct the RV data based on a fit of the magnetic cycle on the activity data. Several examples of these correction procedures and their accuracies can be found in \citet{dumusque11c}.

The sensitivity model also allows us to estimate the RV effect that is expected in the Sun due to the solar magnetic cycle. With $T_{\mathrm{eff}}$ = 5770 K, [Fe/H] = 0 and $A$ = 0.26, we obtain a RV semi-amplitude of 8.2 m\,s$^{-1}$ and thus a peak-to-peak effect of about 16 m\,s$^{-1}$. This is in qualitative agreement with both theoretical estimates \citep[e.g.][]{lindegren03} and a previous attempt to actually measure it \citep{deming94}. We note that the effect is likely to depend on the particular set of spectral lines being considered, so that the comparison with literature values is not straightforward. This could potentially explain the null result of \citet{mcmillan93}. A more recent attempt to predict this same effect and its impact on exoplanet searches was published by \citet{meunier10}. They obtain a full amplitude over the solar cycle of about 8 m\,s$^{-1}$, which is lower than our result but can also be considered as qualitatively in agreement given the differences in the details between our observations and their simulations. We note in passing that \citet{meunier10} went on to simulate planet detection limits in the presence of stellar noise, and concluded that the effects of magnetic cycles would make the detection of Earth-like planets impossible. We emphasize here that these long-term effects are actually rather easy to model and correct using simultaneous $R'_{\mathrm{HK}}$ measurements. Their planet detection limits are therefore likely far too pessimistic.

Finally, we briefly discuss theoretical explanations for the observed effects of magnetic cycles on spectral lines. As already argued by several authors, the main phenomenon at play here is likely to be the "freezing" effect of magnetic fields on granular motions in the outer convective zone of solar-type stars \citep[e.g.][]{livingston82,brandt90}. The stronger the magnetic field, the slower the convective motions. As a consequence, the usual convective blueshift of spectral lines, caused by the temperature-velocity correlation in granules \citep{dravins82,kaisig82}, is decreased in active regions compared to the quiet photosphere. Therefore, stellar lines will appear redshifted when the average (or local) magnetic field strength is higher. This may explain the observed RV - $R'_{\mathrm{HK}}$ positive correlation over the magnetic cycle. However, other effects may also play a role, such as changes in large-scale surface flows over the magnetic cycle \citep{makarov10}, or perhaps changes in global mean effective temperature as the total spot/plage coverage varies (sensitivity of spectral lines to temperature).

Differential effects as a function of effective temperature or metallicity are more difficult to explain. Convective motions are slower in cooler stars, diminishing the amount of convective blueshift and therefore also reducing possible variations with magnetic field strength. The strong increase in FWHM sensitivity towards cooler temperatures should also be considered to help us understand the effect. The direct impact of magnetic fields on spectral lines, through Zeeman broadening, should be investigated. 3D hydrodynamical simulations of stellar atmospheres including magnetic fields are probably necessary to better understand all these observational findings. On the observational side, the lists of spectral lines that are used to build the cross-correlation function for precise RV measurements should be carefully checked and optimized to either minimize undesirable RV effects, or actually maximize them in order to better understand the underlying physics affecting line shapes over magnetic cycles.

In conclusion, the present study has allowed us to make important progress in the knowledge of magnetic cycles in solar-type stars in general, and in their effects on spectral lines in particular. Not only are these results hopefully useful in their own right for stellar physics, but they are also of great help for the search for exoplanets with the RV method. Understanding stellar RV "jitter" is the single most important challenge in the quest for Earth-like planets in the habitable zones of solar-type stars. The excellent quality of the HARPS data, and the different diagnostics available from the spectra, make it possible to monitor stars with unprecedented scrutiny. Further studies of the impact of magnetic activity at all timescales are ongoing and will allow us to disentangle even better stellar activity and planetary signals.

\begin{acknowledgements}
We are grateful to all technical and scientific collaborators of the HARPS Consortium, ESO Headquarters and ESO La Silla who have contributed with their extraordinary passion and valuable work to the success of the HARPS project. We would like to thank the Swiss National Science Foundation for its continuous support. We acknowledge support from French PNP. NCS and XD acknowledge the support by the European Research Council/European Community under the FP7 through Starting Grant agreement number 239953, and by Funda\c{c}\~ao para a Ci\^encia e a Tecnologia (FCT) and in the form of grants reference PTDC/CTE-AST/098528/2008 and PTDC/CTE-AST/098604/2008. NCS also acknowledges the support from FCT through program Ci\^encia\,2007 funded by FCT/MCTES (Portugal) and POPH/FSE (EC).
\end{acknowledgements}

\bibliographystyle{aa}
\bibliography{biblio}

\longtab{2}{
\begin{longtable}{lcccccclcccc}
\caption{Results for stars with a significant magnetic cycle.}
\label{TableResultsCycle} \\
\hline\hline
Name & $N_{\mathrm{meas}}$ & Span & $\log{R'_{\mathrm{HK}}}$ & $\log{R'_{\mathrm{HK}}}$ & $\sigma(R'_{\mathrm{HK}})$ & $\sigma(R'_{\mathrm{HK}})$ & $P_{\mathrm{cycle}}$ & $\gamma$ & $A$ & $P_{\mathrm{rot}}$ & Age \\
 & & [days] & mean & median & raw & short-term & [days] & & & [days] & [Gyr] \\
\hline
\endfirsthead
\caption{Continued.} \\
\hline\hline
Name & $N_{\mathrm{meas}}$ & Span & $\log{R'_{\mathrm{HK}}}$ & $\log{R'_{\mathrm{HK}}}$ & $\sigma(R'_{\mathrm{HK}})$ & $\sigma(R'_{\mathrm{HK}})$ & $P_{\mathrm{cycle}}$ & $\gamma$ & $A$ & $P_{\mathrm{rot}}$ & Age \\
 & & [days] & mean & median & raw & short-term & [days] & & & [days] & [Gyr] \\
\hline
\endhead
\hline
\endfoot
HD1461   	& 164 	& 2652 	& -4.996 	& -4.998 	& 0.0291 	& 0.0215 	& 3754$_{-565}^{+807}$ 	& -4.993 	& 0.0351 	& 29.0 $\pm$ 3.3 	& 5.12 $\pm$ 0.59 \\
HD1581   	& 127 	& 2625 	& -4.954 	& -4.954 	& 0.0099 	& 0.0084 	& 1018$_{-47}^{+51}$ 	& -4.953 	& 0.0099 	& 16.7 $\pm$ 2.6 	& 2.92 $\pm$ 0.34 \\
HD4915   	& 40 	& 1822 	& -4.818 	& -4.812 	& 0.1287 	& 0.0483 	& 1863$_{-211}^{+1951}$ 	& -4.820 	& 0.1804 	& 21.3 $\pm$ 3.2 	& 3.07 $\pm$ 0.35 \\
HD7199   	& 87 	& 2633 	& -4.938 	& -4.978 	& 0.2255 	& 0.0688 	& 2760$_{-319}^{+416}$ 	& -4.916 	& 0.2620 	& 41.4 $\pm$ 4.7 	& 6.38 $\pm$ 0.74 \\
HD8828   	& 46 	& 2224 	& -5.014 	& -5.012 	& 0.0212 	& 0.0144 	& 2163$_{-234}^{+300}$ 	& -5.024 	& 0.0383 	& 38.8 $\pm$ 3.8 	& 7.05 $\pm$ 0.81 \\
HD9246   	& 18 	& 2092 	& -4.858 	& -4.863 	& 0.0984 	& 0.0430 	& 2747$_{-387}^{+540}$ 	& -4.883 	& 0.1998 	& 39.8 $\pm$ 4.7 	& 5.85 $\pm$ 0.67 \\
HD10180   	& 218 	& 2601 	& -4.996 	& -4.995 	& 0.0270 	& 0.0223 	& 2737$_{-318}^{+414}$ 	& -4.999 	& 0.0216 	& 24.1 $\pm$ 3.0 	& 4.28 $\pm$ 0.49 \\
HD13060   	& 63 	& 2581 	& -4.830 	& -4.828 	& 0.1923 	& 0.0844 	& 5508$_{-4092}^{+\infty}$ 	& -4.766 	& 0.5957 	& 29.8 $\pm$ 4.3 	& 3.90 $\pm$ 0.45 \\
HD13724   	& 28 	& 2134 	& -4.736 	& -4.731 	& 0.1117 	& 0.0637 	& 2393$_{-1335}^{+390}$ 	& -4.759 	& 0.1582 	& 19.9 $\pm$ 3.3 	& 2.63 $\pm$ 0.30 \\
HD13808   	& 128 	& 2601 	& -4.910 	& -4.954 	& 0.2397 	& 0.0704 	& 3715$_{-562}^{+807}$ 	& -4.883 	& 0.2688 	& 40.2 $\pm$ 4.8 	& 5.87 $\pm$ 0.68 \\
HD15337   	& 43 	& 2258 	& -4.905 	& -4.916 	& 0.1004 	& 0.0617 	& 1111$_{-396}^{+72}$ 	& -4.902 	& 0.1185 	& 41.8 $\pm$ 4.9 	& 6.20 $\pm$ 0.71 \\
HD20003   	& 98 	& 2596 	& -4.969 	& -4.994 	& 0.1156 	& 0.0292 	& 3149$_{-414}^{+562}$ 	& -4.934 	& 0.1711 	& 37.1 $\pm$ 4.1 	& 6.08 $\pm$ 0.70 \\
HD20807   	& 38 	& 2309 	& -4.905 	& -4.906 	& 0.0393 	& 0.0266 	& 1133$_{-65}^{+1090}$ 	& -4.905 	& 0.0495 	& 17.7 $\pm$ 2.8 	& 2.87 $\pm$ 0.33 \\
HD20619   	& 26 	& 2138 	& -4.829 	& -4.817 	& 0.1071 	& 0.0553 	& 1687$_{-151}^{+184}$ 	& -4.846 	& 0.1438 	& 22.3 $\pm$ 3.2 	& 3.32 $\pm$ 0.38 \\
HD20794   	& 197 	& 2694 	& -5.025 	& -5.025 	& 0.0093 	& 0.0078 	& 751$_{-25}^{+290}$ 	& -5.025 	& 0.0065 	& 35.4 $\pm$ 3.6 	& 6.45 $\pm$ 0.74 \\
HD20781   	& 95 	& 2648 	& -5.045 	& -5.046 	& 0.0199 	& 0.0160 	& 8000$_{-4249}^{+\infty}$ 	& -5.080 	& 0.0848 	& 49.5 $\pm$ 4.4 	& 9.19 $\pm$ 1.06 \\
HD20782   	& 51 	& 2687 	& -4.924 	& -4.930 	& 0.0410 	& 0.0297 	& 1150$_{-58}^{+65}$ 	& -4.920 	& 0.0474 	& 21.6 $\pm$ 3.0 	& 3.52 $\pm$ 0.41 \\
HD21749   	& 52 	& 2225 	& -4.721 	& -4.719 	& 0.1896 	& 0.1060 	& 2164$_{-235}^{+299}$ 	& -4.733 	& 0.2037 	& 34.5 $\pm$ 6.8 	& 3.32 $\pm$ 0.38 \\
HD21693   	& 124 	& 2700 	& -4.920 	& -4.892 	& 0.1562 	& 0.0512 	& 2483$_{-256}^{+322}$ 	& -4.935 	& 0.2094 	& 36.3 $\pm$ 4.0 	& 5.98 $\pm$ 0.69 \\
HD26965A   	& 16 	& 2163 	& -4.993 	& -4.992 	& 0.0856 	& 0.0121 	& 3352$_{-543}^{+805}$ 	& -4.958 	& 0.1585 	& 42.2 $\pm$ 4.4 	& 6.95 $\pm$ 0.80 \\
HD27063   	& 40 	& 1576 	& -4.749 	& -4.750 	& 0.0774 	& 0.0528 	& 1316$_{-124}^{+153}$ 	& -4.749 	& 0.0903 	& 19.5 $\pm$ 3.3 	& 2.54 $\pm$ 0.29 \\
HD34688   	& 17 	& 2714 	& -4.924 	& -4.948 	& 0.1834 	& 0.0275 	& 2026$_{-172}^{+208}$ 	& -4.932 	& 0.2842 	& 41.3 $\pm$ 4.5 	& 6.57 $\pm$ 0.76 \\
HD39194   	& 133 	& 2718 	& -5.026 	& -5.025 	& 0.0269 	& 0.0170 	& 8000$_{-2151}^{+\infty}$ 	& -5.032 	& 0.0424 	& 41.4 $\pm$ 4.0 	& 7.54 $\pm$ 0.87 \\
HD38858   	& 51 	& 2965 	& -4.948 	& -4.948 	& 0.0184 	& 0.0139 	& 3406$_{-428}^{+570}$ 	& -4.945 	& 0.0208 	& 23.6 $\pm$ 3.1 	& 3.95 $\pm$ 0.45 \\
HD40307   	& 183 	& 2720 	& -4.995 	& -5.021 	& 0.1252 	& 0.0373 	& 3804$_{-566}^{+806}$ 	& -4.957 	& 0.1894 	& 47.2 $\pm$ 5.3 	& 7.08 $\pm$ 0.81 \\
HD45184   	& 81 	& 2739 	& -4.896 	& -4.895 	& 0.0728 	& 0.0406 	& 1595$_{-108}^{+125}$ 	& -4.913 	& 0.0979 	& 21.4 $\pm$ 3.0 	& 3.45 $\pm$ 0.40 \\
HD45364   	& 64 	& 2697 	& -4.978 	& -4.979 	& 0.0448 	& 0.0353 	& 8000$_{-6958}^{+\infty}$ 	& -4.970 	& 0.0717 	& 34.0 $\pm$ 3.7 	& 5.86 $\pm$ 0.67 \\
HD51608   	& 116 	& 2698 	& -4.986 	& -4.986 	& 0.0408 	& 0.0346 	& 2481$_{-256}^{+322}$ 	& -4.981 	& 0.0317 	& 39.6 $\pm$ 4.1 	& 6.84 $\pm$ 0.79 \\
HD55693   	& 22 	& 3006 	& -4.963 	& -4.958 	& 0.0451 	& 0.0200 	& 2403$_{-218}^{+266}$ 	& -4.981 	& 0.0773 	& 27.4 $\pm$ 3.2 	& 4.76 $\pm$ 0.55 \\
HD59468   	& 141 	& 2755 	& -4.993 	& -4.991 	& 0.0251 	& 0.0178 	& 8000$_{-2131}^{+\infty}$ 	& -5.000 	& 0.0361 	& 31.8 $\pm$ 3.5 	& 5.66 $\pm$ 0.65 \\
HD63765   	& 48 	& 2303 	& -4.760 	& -4.763 	& 0.1537 	& 0.1138 	& 2219$_{-1015}^{+304}$ 	& -4.772 	& 0.2275 	& 26.2 $\pm$ 3.8 	& 3.52 $\pm$ 0.41 \\
HD65907A   	& 62 	& 2609 	& -4.948 	& -4.950 	& 0.0209 	& 0.0139 	& 8000$_{-2216}^{+\infty}$ 	& -4.923 	& 0.0779 	& 14.9 $\pm$ 2.6 	& 2.57 $\pm$ 0.30 \\
HD65277   	& 15 	& 2703 	& -5.039 	& -5.058 	& 0.0833 	& 0.0171 	& 3791$_{-565}^{+2048}$ 	& -5.037 	& 0.0949 	& 54.9 $\pm$ 6.2 	& 8.05 $\pm$ 0.93 \\
HD67458   	& 25 	& 3009 	& -4.927 	& -4.926 	& 0.0544 	& 0.0177 	& 8000$_{-3194}^{+\infty}$ 	& -4.881 	& 0.1910 	& 17.1 $\pm$ 2.8 	& 2.69 $\pm$ 0.31 \\
HD68607   	& 40 	& 1210 	& -4.722 	& -4.713 	& 0.1623 	& 0.1069 	& 863$_{-130}^{+\infty}$ 	& -4.725 	& 0.1495 	& 29.9 $\pm$ 4.7 	& 3.54 $\pm$ 0.41 \\
HD68978A   	& 61 	& 2340 	& -4.875 	& -4.876 	& 0.0459 	& 0.0364 	& 1021$_{-310}^{+58}$ 	& -4.867 	& 0.0566 	& 18.8 $\pm$ 2.9 	& 2.88 $\pm$ 0.33 \\
HD69830   	& 205 	& 2747 	& -4.986 	& -4.996 	& 0.0565 	& 0.0217 	& 5865$_{-1235}^{+\infty}$ 	& -4.944 	& 0.1439 	& 35.8 $\pm$ 3.9 	& 5.98 $\pm$ 0.69 \\
HD71479   	& 16 	& 2355 	& -5.015 	& -5.016 	& 0.0382 	& 0.0149 	& 1525$_{-114}^{+1993}$ 	& -4.997 	& 0.0798 	& 26.6 $\pm$ 3.2 	& 4.72 $\pm$ 0.54 \\
HD71835   	& 72 	& 2698 	& -4.898 	& -4.907 	& 0.1052 	& 0.0588 	& 3222$_{-418}^{+565}$ 	& -4.923 	& 0.1803 	& 36.5 $\pm$ 4.1 	& 5.91 $\pm$ 0.68 \\
HD72673   	& 65 	& 2464 	& -4.968 	& -4.961 	& 0.0668 	& 0.0233 	& 3050$_{-408}^{+558}$ 	& -4.975 	& 0.0910 	& 40.2 $\pm$ 4.1 	& 6.86 $\pm$ 0.79 \\
HD73121   	& 20 	& 2709 	& -5.048 	& -5.048 	& 0.0219 	& 0.0121 	& 1581$_{-108}^{+3021}$ 	& -5.057 	& 0.0343 	& 19.2 $\pm$ 2.6 	& 3.72 $\pm$ 0.43 \\
HD73524   	& 60 	& 2694 	& -4.981 	& -4.981 	& 0.0385 	& 0.0197 	& 2801$_{-322}^{+984}$ 	& -4.981 	& 0.0464 	& 19.8 $\pm$ 2.8 	& 3.49 $\pm$ 0.40 \\
HD78747   	& 43 	& 2554 	& -4.991 	& -4.992 	& 0.0167 	& 0.0121 	& 2389$_{-250}^{+316}$ 	& -4.994 	& 0.0188 	& 16.5 $\pm$ 2.6 	& 3.06 $\pm$ 0.35 \\
HD82342   	& 31 	& 3009 	& -5.012 	& -5.010 	& 0.0753 	& 0.0333 	& 3434$_{-428}^{+571}$ 	& -4.999 	& 0.1184 	& 50.9 $\pm$ 5.6 	& 7.68 $\pm$ 0.88 \\
HD82516   	& 56 	& 2320 	& -4.954 	& -4.963 	& 0.1222 	& 0.0537 	& 6065$_{-4074}^{+\infty}$ 	& -4.802 	& 0.5874 	& 36.7 $\pm$ 5.3 	& 4.59 $\pm$ 0.53 \\
HD85390   	& 65 	& 2689 	& -4.969 	& -4.967 	& 0.0631 	& 0.0401 	& 2476$_{-255}^{+322}$ 	& -4.973 	& 0.0790 	& 44.9 $\pm$ 4.7 	& 7.37 $\pm$ 0.85 \\
HD85512   	& 175 	& 2705 	& -4.936 	& -4.937 	& 0.1143 	& 0.0317 	& 3793$_{-566}^{+806}$ 	& -4.956 	& 0.2035 	& 50.9 $\pm$ 7.0 	& 6.41 $\pm$ 0.74 \\
HD86140   	& 17 	& 2690 	& -4.838 	& -4.844 	& 0.1002 	& 0.0506 	& 2798$_{-1427}^{+418}$ 	& -4.831 	& 0.1836 	& 39.0 $\pm$ 5.4 	& 4.98 $\pm$ 0.57 \\
HD86065   	& 24 	& 3008 	& -4.872 	& -4.877 	& 0.1245 	& 0.0621 	& 891$_{-32}^{+1780}$ 	& -4.856 	& 0.1275 	& 40.7 $\pm$ 5.4 	& 5.37 $\pm$ 0.62 \\
HD89454   	& 48 	& 1468 	& -4.685 	& -4.688 	& 0.1353 	& 0.0906 	& 1573$_{-333}^{+575}$ 	& -4.695 	& 0.1606 	& 21.1 $\pm$ 3.7 	& 2.51 $\pm$ 0.29 \\
HD90156   	& 79 	& 2665 	& -4.973 	& -4.973 	& 0.0131 	& 0.0111 	& 1279$_{-72}^{+81}$ 	& -4.973 	& 0.0092 	& 27.0 $\pm$ 3.2 	& 4.67 $\pm$ 0.54 \\
HD90711   	& 18 	& 2306 	& -4.974 	& -4.972 	& 0.0837 	& 0.0269 	& 2925$_{-400}^{+551}$ 	& -4.984 	& 0.1180 	& 43.1 $\pm$ 4.4 	& 7.33 $\pm$ 0.84 \\
HD90812   	& 22 	& 2307 	& -4.981 	& -4.994 	& 0.1170 	& 0.0286 	& 8000$_{-2419}^{+\infty}$ 	& -4.831 	& 0.5507 	& 34.1 $\pm$ 4.4 	& 4.85 $\pm$ 0.56 \\
HD93083   	& 66 	& 2279 	& -4.941 	& -4.933 	& 0.1627 	& 0.0443 	& 3453$_{-550}^{+806}$ 	& -4.942 	& 0.2273 	& 46.2 $\pm$ 5.3 	& 6.81 $\pm$ 0.78 \\
HD94151   	& 20 	& 2703 	& -4.960 	& -4.981 	& 0.0915 	& 0.0233 	& 3792$_{-566}^{+806}$ 	& -4.951 	& 0.1561 	& 33.1 $\pm$ 3.7 	& 5.58 $\pm$ 0.64 \\
HD96700   	& 152 	& 3011 	& -4.967 	& -4.968 	& 0.0199 	& 0.0178 	& 830$_{-27}^{+\infty}$ 	& -4.965 	& 0.0141 	& 20.6 $\pm$ 2.9 	& 3.55 $\pm$ 0.41 \\
HD98281   	& 53 	& 2227 	& -4.914 	& -4.911 	& 0.0670 	& 0.0487 	& 2861$_{-396}^{+1354}$ 	& -4.938 	& 0.1120 	& 33.5 $\pm$ 3.8 	& 5.57 $\pm$ 0.64 \\
HD100508   	& 32 	& 2284 	& -5.001 	& -5.008 	& 0.0776 	& 0.0297 	& 8000$_{-3735}^{+\infty}$ 	& -4.903 	& 0.3236 	& 39.5 $\pm$ 4.5 	& 6.09 $\pm$ 0.70 \\
HD101930   	& 33 	& 2261 	& -5.016 	& -5.019 	& 0.0497 	& 0.0304 	& 8000$_{-2453}^{+\infty}$ 	& -4.906 	& 0.3137 	& 43.0 $\pm$ 5.1 	& 6.25 $\pm$ 0.72 \\
HD106116   	& 103 	& 2658 	& -5.005 	& -5.007 	& 0.0243 	& 0.0143 	& 3194$_{-417}^{+564}$ 	& -5.002 	& 0.0335 	& 33.1 $\pm$ 3.5 	& 5.90 $\pm$ 0.68 \\
HD106275   	& 17 	& 2649 	& -4.870 	& -4.891 	& 0.1956 	& 0.0709 	& 3187$_{-417}^{+564}$ 	& -4.893 	& 0.3246 	& 41.6 $\pm$ 5.0 	& 6.05 $\pm$ 0.70 \\
HD108309   	& 20 	& 2647 	& -5.003 	& -5.000 	& 0.0341 	& 0.0163 	& 1552$_{-574}^{+3005}$ 	& -5.014 	& 0.0547 	& 29.9 $\pm$ 3.3 	& 5.39 $\pm$ 0.62 \\
HD109200   	& 118 	& 2866 	& -4.976 	& -4.981 	& 0.0771 	& 0.0335 	& 8000$_{-2069}^{+\infty}$ 	& -4.950 	& 0.2118 	& 43.0 $\pm$ 4.6 	& 6.92 $\pm$ 0.80 \\
HD114613   	& 96 	& 1768 	& -5.059 	& -5.063 	& 0.0545 	& 0.0371 	& 897$_{-53}^{+61}$ 	& -5.053 	& 0.0553 	& 34.1 $\pm$ 3.5 	& 6.38 $\pm$ 0.73 \\
HD114747   	& 38 	& 1473 	& -4.829 	& -4.859 	& 0.2596 	& 0.0767 	& 1820$_{-243}^{+333}$ 	& -4.794 	& 0.3770 	& 36.0 $\pm$ 5.2 	& 4.49 $\pm$ 0.52 \\
HD115617   	& 129 	& 2639 	& -4.992 	& -4.995 	& 0.0241 	& 0.0198 	& 1548$_{-811}^{+266}$ 	& -4.993 	& 0.0200 	& 33.9 $\pm$ 3.6 	& 5.99 $\pm$ 0.69 \\
HD115674   	& 39 	& 2252 	& -4.919 	& -4.924 	& 0.0484 	& 0.0295 	& 1947$_{-588}^{+236}$ 	& -4.902 	& 0.0852 	& 25.5 $\pm$ 3.3 	& 4.08 $\pm$ 0.47 \\
HD119638   	& 24 	& 3000 	& -4.934 	& -4.932 	& 0.0417 	& 0.0196 	& 6000$_{-2000}^{+\infty}$ 	& -4.932 	& 0.0484 	& 12.0 $\pm$ 2.3 	& 2.35 $\pm$ 0.27 \\
HD125072   	& 24 	& 2174 	& -4.941 	& -4.952 	& 0.1343 	& 0.0555 	& 1146$_{-70}^{+982}$ 	& -4.859 	& 0.2797 	& 42.0 $\pm$ 5.9 	& 5.23 $\pm$ 0.60 \\
HD125881   	& 26 	& 3003 	& -4.872 	& -4.867 	& 0.0795 	& 0.0289 	& 1044$_{-43}^{+47}$ 	& -4.867 	& 0.1241 	& 15.7 $\pm$ 2.7 	& 2.45 $\pm$ 0.28 \\
HD129642   	& 54 	& 1510 	& -4.971 	& -4.973 	& 0.0488 	& 0.0363 	& 1419$_{-149}^{+436}$ 	& -4.968 	& 0.0464 	& 48.0 $\pm$ 5.3 	& 7.27 $\pm$ 0.84 \\
HD132648   	& 28 	& 2624 	& -4.874 	& -4.890 	& 0.1187 	& 0.0528 	& 5792$_{-2060}^{+\infty}$ 	& -4.845 	& 0.2082 	& 28.1 $\pm$ 3.7 	& 4.18 $\pm$ 0.48 \\
HD134664   	& 28 	& 1066 	& -4.871 	& -4.858 	& 0.0823 	& 0.0263 	& 2097$_{-691}^{+\infty}$ 	& -4.871 	& 0.1588 	& 23.2 $\pm$ 3.2 	& 3.58 $\pm$ 0.41 \\
HD136352   	& 123 	& 2543 	& -4.986 	& -4.986 	& 0.0107 	& 0.0098 	& 1041$_{-97}^{+581}$ 	& -4.986 	& 0.0060 	& 25.0 $\pm$ 3.1 	& 4.37 $\pm$ 0.50 \\
HD136713   	& 41 	& 2203 	& -4.787 	& -4.771 	& 0.1429 	& 0.0710 	& 5706$_{-3791}^{+\infty}$ 	& -4.921 	& 0.5547 	& 45.4 $\pm$ 5.6 	& 6.36 $\pm$ 0.73 \\
HD138549   	& 21 	& 2622 	& -4.834 	& -4.819 	& 0.1370 	& 0.0639 	& 2180$_{-205}^{+253}$ 	& -4.851 	& 0.1759 	& 28.4 $\pm$ 3.7 	& 4.26 $\pm$ 0.49 \\
HD137388   	& 64 	& 2118 	& -4.879 	& -4.867 	& 0.1461 	& 0.0723 	& 2770$_{-389}^{+541}$ 	& -4.877 	& 0.1844 	& 40.6 $\pm$ 5.0 	& 5.79 $\pm$ 0.67 \\
HD144628   	& 53 	& 2105 	& -5.004 	& -5.003 	& 0.0508 	& 0.0280 	& 4103$_{-804}^{+1321}$ 	& -5.017 	& 0.0797 	& 48.2 $\pm$ 4.7 	& 8.21 $\pm$ 0.95 \\
HD146233   	& 38 	& 2157 	& -4.923 	& -4.923 	& 0.0793 	& 0.0287 	& 2803$_{-392}^{+2663}$ 	& -4.918 	& 0.1476 	& 23.8 $\pm$ 3.2 	& 3.87 $\pm$ 0.45 \\
HD148303   	& 31 	& 2164 	& -4.667 	& -4.657 	& 0.2140 	& 0.1017 	& 2416$_{-1274}^{+936}$ 	& -4.674 	& 0.3000 	& 28.9 $\pm$ 5.6 	& 2.82 $\pm$ 0.33 \\
HD154088   	& 107 	& 1866 	& -5.034 	& -5.037 	& 0.0267 	& 0.0174 	& 8000$_{-6487}^{+\infty}$ 	& -4.977 	& 0.1550 	& 42.6 $\pm$ 4.4 	& 7.20 $\pm$ 0.83 \\
HD154577   	& 118 	& 2550 	& -4.974 	& -4.981 	& 0.0750 	& 0.0289 	& 2702$_{-316}^{+412}$ 	& -4.959 	& 0.1294 	& 45.8 $\pm$ 5.0 	& 7.18 $\pm$ 0.83 \\
HD157172   	& 81 	& 2099 	& -4.990 	& -5.001 	& 0.0798 	& 0.0228 	& 8000$_{-2581}^{+\infty}$ 	& -4.840 	& 0.4855 	& 32.7 $\pm$ 4.1 	& 4.78 $\pm$ 0.55 \\
HD157830   	& 55 	& 2526 	& -4.822 	& -4.824 	& 0.1164 	& 0.0655 	& 1613$_{-119}^{+140}$ 	& -4.822 	& 0.1487 	& 24.3 $\pm$ 3.5 	& 3.51 $\pm$ 0.40 \\
HD161098   	& 65 	& 1866 	& -4.946 	& -4.952 	& 0.0500 	& 0.0282 	& 8000$_{-2791}^{+\infty}$ 	& -4.801 	& 0.4861 	& 22.5 $\pm$ 3.4 	& 3.16 $\pm$ 0.36 \\
HD172513   	& 41 	& 1081 	& -4.777 	& -4.777 	& 0.0751 	& 0.0503 	& 1221$_{-151}^{+481}$ 	& -4.799 	& 0.1205 	& 28.3 $\pm$ 3.9 	& 3.95 $\pm$ 0.45 \\
HD177565   	& 26 	& 1684 	& -4.927 	& -4.949 	& 0.1218 	& 0.0248 	& 2016$_{-262}^{+355}$ 	& -4.895 	& 0.1710 	& 28.4 $\pm$ 3.5 	& 4.50 $\pm$ 0.52 \\
HD188559   	& 35 	& 1153 	& -4.811 	& -4.826 	& 0.1981 	& 0.0952 	& 2221$_{-431}^{+704}$ 	& -4.811 	& 0.3193 	& 39.1 $\pm$ 6.2 	& 4.45 $\pm$ 0.51 \\
HD192310   	& 152 	& 2704 	& -4.996 	& -5.009 	& 0.1060 	& 0.0258 	& 3792$_{-566}^{+806}$ 	& -4.931 	& 0.2049 	& 43.7 $\pm$ 4.9 	& 6.69 $\pm$ 0.77 \\
HD197823   	& 19 	& 1056 	& -4.719 	& -4.751 	& 0.2345 	& 0.0377 	& 764$_{-63}^{+2001}$ 	& -4.645 	& 0.4882 	& 21.0 $\pm$ 4.0 	& 2.27 $\pm$ 0.26 \\
HD199960   	& 30 	& 1864 	& -4.982 	& -4.978 	& 0.0488 	& 0.0174 	& 1076$_{-72}^{+83}$ 	& -4.992 	& 0.0857 	& 25.2 $\pm$ 3.1 	& 4.43 $\pm$ 0.51 \\
HD203432   	& 36 	& 1405 	& -4.816 	& -4.838 	& 0.1904 	& 0.0641 	& 2080$_{-324}^{+472}$ 	& -4.838 	& 0.2825 	& 29.5 $\pm$ 3.8 	& 4.35 $\pm$ 0.50 \\
HD204313   	& 66 	& 1548 	& -4.993 	& -4.997 	& 0.0323 	& 0.0233 	& 987$_{-73}^{+\infty}$ 	& -5.000 	& 0.0371 	& 33.1 $\pm$ 3.5 	& 5.88 $\pm$ 0.68 \\
HD204941   	& 34 	& 2180 	& -4.962 	& -4.971 	& 0.0695 	& 0.0238 	& 2132$_{-232}^{+296}$ 	& -4.957 	& 0.0892 	& 45.3 $\pm$ 4.9 	& 7.15 $\pm$ 0.82 \\
HD207129   	& 79 	& 1876 	& -4.898 	& -4.907 	& 0.0890 	& 0.0400 	& 1520$_{-139}^{+171}$ 	& -4.899 	& 0.1069 	& 17.6 $\pm$ 2.8 	& 2.83 $\pm$ 0.33 \\
HD209100   	& 46 	& 1916 	& -4.806 	& -4.806 	& 0.1025 	& 0.0641 	& 1719$_{-315}^{+217}$ 	& -4.788 	& 0.1486 	& 37.6 $\pm$ 6.2 	& 4.12 $\pm$ 0.47 \\
HD215152   	& 167 	& 2740 	& -4.871 	& -4.878 	& 0.0837 	& 0.0516 	& 713$_{-22}^{+1795}$ 	& -4.867 	& 0.0876 	& 41.8 $\pm$ 5.6 	& 5.49 $\pm$ 0.63 \\
HD222237   	& 18 	& 1115 	& -5.015 	& -5.025 	& 0.0575 	& 0.0246 	& 2168$_{-424}^{+2049}$ 	& -4.993 	& 0.0918 	& 50.7 $\pm$ 5.7 	& 7.54 $\pm$ 0.87 \\
HD222595   	& 26 	& 1053 	& -4.806 	& -4.833 	& 0.2024 	& 0.0833 	& 1667$_{-275}^{+411}$ 	& -4.796 	& 0.2280 	& 24.1 $\pm$ 3.5 	& 3.36 $\pm$ 0.39 \\
HD222669   	& 47 	& 1403 	& -4.858 	& -4.860 	& 0.0715 	& 0.0526 	& 1336$_{-257}^{+180}$ 	& -4.858 	& 0.0654 	& 17.5 $\pm$ 2.9 	& 2.66 $\pm$ 0.31 \\
\end{longtable}
}

\longtab{3}{
\begin{longtable}{lcccccccccc}
\caption{Results for stars with no significant magnetic cycle.}
\label{TableResultsNoCycle} \\
\hline\hline
Name & $N_{\mathrm{meas}}$ & Span & $\log{R'_{\mathrm{HK}}}$ & $\log{R'_{\mathrm{HK}}}$ & $\sigma(R'_{\mathrm{HK}})$ & $\sigma(R'_{\mathrm{HK}})$ & $A_{lim}$ & $Prob$ & $P_{\mathrm{rot}}$ & Age \\
 & & [days] & mean & median & raw & short-term & & $(A < 0.04)$ & [days] & [Gyr] \\
\hline
\endfirsthead
\caption{Continued.} \\
\hline\hline
Name & $N_{\mathrm{meas}}$ & Span & $\log{R'_{\mathrm{HK}}}$ & $\log{R'_{\mathrm{HK}}}$ & $\sigma(R'_{\mathrm{HK}})$ & $\sigma(R'_{\mathrm{HK}})$ & $A_{lim}$ & $Prob$ & $P_{\mathrm{rot}}$ & Age \\
 & & [days] & mean & median & raw & short-term & & $(A < 0.04)$ & [days] & [Gyr] \\
\hline
\endhead
\hline
\endfoot
HD224789    	& 32 	& 1328 	& -4.436 	& -4.438 	& 0.1638 	& 0.1460 	& 0.5968 	& 0.000 	& 11.8 $\pm$ 4.7 	& 0.68 $\pm$ 0.08 \\
HD55    	& 10 	& 1311 	& -4.893 	& -4.891 	& 0.0965 	& 0.0283 	& 0.6010 	& 0.000 	& 45.2 $\pm$ 6.4 	& 5.59 $\pm$ 0.64 \\
HD361    	& 16 	& 2134 	& -4.712 	& -4.702 	& 0.1259 	& 0.0878 	& 0.5226 	& 0.000 	& 14.1 $\pm$ 2.9 	& 1.73 $\pm$ 0.20 \\
HD967    	& 38 	& 2556 	& -4.987 	& -4.986 	& 0.0218 	& 0.0192 	& 0.0468 	& 0.833 	& 26.3 $\pm$ 3.2 	& 4.60 $\pm$ 0.53 \\
HD1320    	& 6 	& 2510 	& -4.904 	& -4.879 	& 0.0710 	& 0.0201 	& 0.6010 	& 0.000 	& 23.4 $\pm$ 3.2 	& 3.74 $\pm$ 0.43 \\
HD1388    	& 57 	& 2647 	& -4.979 	& -4.979 	& 0.0160 	& 0.0146 	& 0.0517 	& 0.917 	& 19.7 $\pm$ 2.8 	& 3.47 $\pm$ 0.40 \\
HD3569    	& 7 	& 2081 	& -4.938 	& -4.965 	& 0.1808 	& 0.0430 	& 0.6010 	& 0.000 	& 42.7 $\pm$ 4.7 	& 6.75 $\pm$ 0.78 \\
HD3823    	& 33 	& 2266 	& -5.017 	& -5.015 	& 0.0135 	& 0.0118 	& 0.0375 	& 1.000 	& 15.8 $\pm$ 2.5 	& 3.08 $\pm$ 0.35 \\
HD4308    	& 63 	& 2178 	& -4.999 	& -4.997 	& 0.0230 	& 0.0190 	& 0.0364 	& 1.000 	& 28.0 $\pm$ 3.2 	& 4.96 $\pm$ 0.57 \\
HD4307    	& 31 	& 1856 	& -5.069 	& -5.068 	& 0.0147 	& 0.0139 	& 0.0443 	& 0.917 	& 22.2 $\pm$ 2.8 	& 4.26 $\pm$ 0.49 \\
HD6348    	& 8 	& 2559 	& -4.983 	& -4.990 	& 0.0389 	& 0.0236 	& 0.5749 	& 0.000 	& 42.3 $\pm$ 4.3 	& 7.22 $\pm$ 0.83 \\
HD6673    	& 9 	& 1142 	& -4.818 	& -4.815 	& 0.0804 	& 0.0292 	& 0.6010 	& 0.000 	& 37.7 $\pm$ 5.3 	& 4.82 $\pm$ 0.56 \\
HD6735    	& 9 	& 2560 	& -4.890 	& -4.891 	& 0.0260 	& 0.0134 	& 0.3863 	& 0.000 	& 14.2 $\pm$ 2.6 	& 2.35 $\pm$ 0.27 \\
HD7134    	& 16 	& 1767 	& -4.980 	& -4.980 	& 0.0133 	& 0.0086 	& 0.0637 	& 0.750 	& 18.6 $\pm$ 2.7 	& 3.29 $\pm$ 0.38 \\
HD7449    	& 82 	& 2633 	& -4.860 	& -4.861 	& 0.0477 	& 0.0407 	& 0.0642 	& 0.667 	& 13.5 $\pm$ 2.6 	& 2.16 $\pm$ 0.25 \\
HD8638    	& 35 	& 1827 	& -4.996 	& -4.996 	& 0.0216 	& 0.0165 	& 0.0726 	& 0.750 	& 30.4 $\pm$ 3.4 	& 5.39 $\pm$ 0.62 \\
HD9796    	& 12 	& 2112 	& -4.920 	& -4.919 	& 0.0658 	& 0.0239 	& 0.4264 	& 0.000 	& 37.1 $\pm$ 4.1 	& 5.98 $\pm$ 0.69 \\
HD9782    	& 27 	& 2217 	& -4.964 	& -4.963 	& 0.0179 	& 0.0173 	& 0.1053 	& 0.000 	& 18.1 $\pm$ 2.7 	& 3.16 $\pm$ 0.36 \\
HD10002    	& 8 	& 2088 	& -5.051 	& -5.049 	& 0.0230 	& 0.0117 	& 0.6010 	& 0.000 	& 48.5 $\pm$ 4.5 	& 8.72 $\pm$ 1.00 \\
HD10700    	& 153 	& 2650 	& -5.017 	& -5.017 	& 0.0089 	& 0.0084 	& 0.0086 	& 1.000 	& 37.4 $\pm$ 3.8 	& 6.75 $\pm$ 0.78 \\
HD11226    	& 28 	& 1766 	& -4.998 	& -4.998 	& 0.0114 	& 0.0108 	& 0.0597 	& 0.750 	& 16.6 $\pm$ 2.6 	& 3.09 $\pm$ 0.36 \\
HD11505    	& 12 	& 1392 	& -5.024 	& -5.026 	& 0.0136 	& 0.0123 	& 0.1396 	& 0.000 	& 26.3 $\pm$ 3.1 	& 4.78 $\pm$ 0.55 \\
HD11964A    	& 60 	& 2514 	& -5.154 	& -5.154 	& 0.0152 	& 0.0135 	& 0.0184 	& 1.000 	& 53.9 $\pm$ 4.4 	& 10.69 $\pm$ 1.23 \\
HD12387    	& 14 	& 1092 	& -5.017 	& -5.004 	& 0.0544 	& 0.0364 	& 0.5131 	& 0.000 	& 27.4 $\pm$ 3.2 	& 4.94 $\pm$ 0.57 \\
HD12345    	& 14 	& 1729 	& -5.017 	& -5.013 	& 0.0289 	& 0.0168 	& 0.1242 	& 0.167 	& 39.5 $\pm$ 3.9 	& 7.12 $\pm$ 0.82 \\
HD12617    	& 12 	& 1544 	& -4.694 	& -4.717 	& 0.2616 	& 0.1161 	& 0.6010 	& 0.000 	& 30.7 $\pm$ 5.9 	& 3.00 $\pm$ 0.35 \\
HD14374    	& 17 	& 1805 	& -4.660 	& -4.671 	& 0.1760 	& 0.1314 	& 0.6010 	& 0.000 	& 20.6 $\pm$ 3.8 	& 2.31 $\pm$ 0.27 \\
HD14747    	& 12 	& 2092 	& -5.004 	& -5.001 	& 0.0275 	& 0.0246 	& 0.2822 	& 0.000 	& 30.8 $\pm$ 3.4 	& 5.48 $\pm$ 0.63 \\
HD16297    	& 10 	& 2052 	& -4.705 	& -4.710 	& 0.1561 	& 0.0754 	& 0.6000 	& 0.000 	& 24.7 $\pm$ 4.1 	& 2.96 $\pm$ 0.34 \\
HD16417    	& 76 	& 2623 	& -5.029 	& -5.031 	& 0.0307 	& 0.0254 	& 0.0458 	& 0.833 	& 27.8 $\pm$ 3.2 	& 5.08 $\pm$ 0.58 \\
HD16714    	& 26 	& 2089 	& -4.988 	& -4.987 	& 0.0188 	& 0.0159 	& 0.0737 	& 0.417 	& 33.7 $\pm$ 3.6 	& 5.92 $\pm$ 0.68 \\
HD17970    	& 8 	& 1739 	& -5.071 	& -5.065 	& 0.0225 	& 0.0141 	& 0.5599 	& 0.000 	& 49.7 $\pm$ 4.5 	& 9.10 $\pm$ 1.05 \\
HD19034    	& 8 	& 2136 	& -4.957 	& -4.951 	& 0.0327 	& 0.0172 	& 0.5048 	& 0.000 	& 28.9 $\pm$ 3.4 	& 4.90 $\pm$ 0.56 \\
HD19467    	& 15 	& 2170 	& -5.015 	& -5.015 	& 0.0149 	& 0.0122 	& 0.0697 	& 0.583 	& 27.3 $\pm$ 3.2 	& 4.92 $\pm$ 0.57 \\
HD20407    	& 8 	& 2524 	& -4.943 	& -4.942 	& 0.0189 	& 0.0081 	& 0.4619 	& 0.000 	& 17.6 $\pm$ 2.7 	& 3.00 $\pm$ 0.35 \\
HD21209A    	& 9 	& 2707 	& -4.876 	& -4.883 	& 0.0872 	& 0.0322 	& 0.5814 	& 0.000 	& 43.9 $\pm$ 6.3 	& 5.36 $\pm$ 0.62 \\
HD21019    	& 13 	& 1800 	& -5.122 	& -5.122 	& 0.0166 	& 0.0151 	& 0.0735 	& 0.333 	& 38.4 $\pm$ 3.5 	& 7.66 $\pm$ 0.88 \\
HD21938    	& 12 	& 2702 	& -4.990 	& -4.986 	& 0.0241 	& 0.0193 	& 0.2066 	& 0.000 	& 18.8 $\pm$ 2.7 	& 3.37 $\pm$ 0.39 \\
HD22879    	& 54 	& 2687 	& -4.996 	& -4.997 	& 0.0140 	& 0.0119 	& 0.0259 	& 1.000 	& 14.2 $\pm$ 2.4 	& 2.83 $\pm$ 0.33 \\
HD23456    	& 20 	& 2201 	& -4.939 	& -4.937 	& 0.0216 	& 0.0154 	& 0.0930 	& 0.417 	& 9.4 $\pm$ 2.1 	& 2.85 $\pm$ 0.33 \\
HD23249    	& 54 	& 1877 	& -5.181 	& -5.181 	& 0.0061 	& 0.0057 	& 0.0117 	& 1.000 	& 61.4 $\pm$ 5.2 	& 11.56 $\pm$ 1.33 \\
HD23356    	& 13 	& 2228 	& -4.774 	& -4.769 	& 0.0748 	& 0.0447 	& 0.3493 	& 0.000 	& 34.8 $\pm$ 5.3 	& 4.19 $\pm$ 0.48 \\
HD24331    	& 11 	& 2702 	& -4.820 	& -4.808 	& 0.1848 	& 0.0817 	& 0.6010 	& 0.000 	& 37.5 $\pm$ 5.1 	& 4.89 $\pm$ 0.56 \\
HD24892    	& 13 	& 2749 	& -5.124 	& -5.123 	& 0.0146 	& 0.0111 	& 0.1421 	& 0.250 	& 45.0 $\pm$ 3.9 	& 8.95 $\pm$ 1.03 \\
HD25673    	& 13 	& 2123 	& -4.971 	& -4.995 	& 0.1436 	& 0.0562 	& 0.5434 	& 0.000 	& 43.0 $\pm$ 4.4 	& 7.17 $\pm$ 0.83 \\
HD28471    	& 13 	& 1194 	& -5.010 	& -5.011 	& 0.0432 	& 0.0268 	& 0.3656 	& 0.000 	& 27.1 $\pm$ 3.2 	& 4.87 $\pm$ 0.56 \\
HD28701    	& 19 	& 1330 	& -5.021 	& -5.017 	& 0.0230 	& 0.0123 	& 0.1494 	& 0.083 	& 27.5 $\pm$ 3.2 	& 4.98 $\pm$ 0.57 \\
HD28821    	& 12 	& 2706 	& -4.984 	& -4.985 	& 0.0314 	& 0.0140 	& 0.4125 	& 0.000 	& 29.9 $\pm$ 3.4 	& 5.24 $\pm$ 0.60 \\
HD30306    	& 15 	& 1454 	& -5.052 	& -5.053 	& 0.0262 	& 0.0199 	& 0.1436 	& 0.250 	& 41.4 $\pm$ 3.9 	& 7.71 $\pm$ 0.89 \\
HD30278    	& 21 	& 1514 	& -5.024 	& -5.021 	& 0.0235 	& 0.0165 	& 0.0892 	& 0.667 	& 39.9 $\pm$ 3.9 	& 7.23 $\pm$ 0.83 \\
HIP22059    	& 10 	& 1357 	& -4.515 	& -4.508 	& 0.0909 	& 0.0749 	& 0.6010 	& 0.000 	& 17.9 $\pm$ 5.4 	& 1.25 $\pm$ 0.14 \\
HD34449    	& 12 	& 2664 	& -4.886 	& -4.888 	& 0.0257 	& 0.0219 	& 0.3564 	& 0.000 	& 18.3 $\pm$ 2.9 	& 2.88 $\pm$ 0.33 \\
HD31527    	& 166 	& 2720 	& -4.972 	& -4.972 	& 0.0127 	& 0.0122 	& 0.0140 		& 1.000 	& 20.8 $\pm$ 2.9 	& 3.61 $\pm$ 0.42 \\
HD31822    	& 45 	& 2353 	& -4.885 	& -4.886 	& 0.0216 	& 0.0206 	& 0.0415 	& 0.917 	& 15.1 $\pm$ 2.6 	& 2.44 $\pm$ 0.28 \\
HD32724    	& 13 	& 2704 	& -5.045 	& -5.044 	& 0.0228 	& 0.0212 	& 0.2045 	& 0.000 	& 24.2 $\pm$ 2.9 	& 4.51 $\pm$ 0.52 \\
HD33725    	& 10 	& 2720 	& -4.970 	& -5.003 	& 0.1040 	& 0.0107 	& 0.5416 	& 0.167 	& 41.5 $\pm$ 4.3 	& 7.00 $\pm$ 0.81 \\
HD35854    	& 12 	& 2717 	& -4.804 	& -4.784 	& 0.1396 	& 0.0320 	& 0.5847 	& 0.000 	& 37.2 $\pm$ 5.4 	& 4.59 $\pm$ 0.53 \\
HD36108    	& 13 	& 2971 	& -5.011 	& -5.013 	& 0.0148 	& 0.0114 	& 0.2170 	& 0.000 	& 19.3 $\pm$ 2.7 	& 3.54 $\pm$ 0.41 \\
HD36003    	& 61 	& 1495 	& -4.894 	& -4.901 	& 0.0984 	& 0.0703 	& 0.1893 	& 0.000 	& 45.7 $\pm$ 6.6 	& 5.54 $\pm$ 0.64 \\
HD36379    	& 46 	& 2342 	& -4.994 	& -4.994 	& 0.0153 	& 0.0134 	& 0.0267 	& 1.000 	& 15.3 $\pm$ 2.5 	& 2.92 $\pm$ 0.34 \\
HIP26542    	& 6 	& 1135 	& -4.833 	& -4.817 	& 0.1217 	& 0.0481 	& 0.6010 	& 0.000 	& 40.5 $\pm$ 6.1 	& 4.80 $\pm$ 0.55 \\
HD37986    	& 16 	& 2156 	& -5.050 	& -5.047 	& 0.0225 	& 0.0133 	& 0.0677 	& 0.750 	& 46.1 $\pm$ 4.3 	& 8.43 $\pm$ 0.97 \\
HD38277    	& 6 	& 2714 	& -5.024 	& -5.025 	& 0.0147 	& 0.0056 	& 0.6010 	& 0.000 	& 26.2 $\pm$ 3.1 	& 4.77 $\pm$ 0.55 \\
HD38973    	& 24 	& 2354 	& -4.966 	& -4.969 	& 0.0278 	& 0.0245 	& 0.0732 	& 0.333 	& 18.2 $\pm$ 2.7 	& 3.18 $\pm$ 0.37 \\
HD40105    	& 19 	& 1794 	& -5.177 	& -5.176 	& 0.0198 	& 0.0150 	& 0.0716 	& 0.500 	& 59.9 $\pm$ 5.0 	& 11.53 $\pm$ 1.33 \\
HD40397    	& 11 	& 2756 	& -5.024 	& -5.023 	& 0.0190 	& 0.0177 	& 0.3235 	& 0.000 	& 36.5 $\pm$ 3.7 	& 6.65 $\pm$ 0.77 \\
HD44447    	& 19 	& 2745 	& -4.995 	& -4.992 	& 0.0319 	& 0.0252 	& 0.1234 	& 0.333 	& 15.3 $\pm$ 2.5 	& 2.93 $\pm$ 0.34 \\
HD44120    	& 11 	& 2745 	& -5.050 	& -5.052 	& 0.0174 	& 0.0119 	& 0.1082 	& 0.250 	& 20.3 $\pm$ 2.7 	& 3.87 $\pm$ 0.45 \\
HD44594    	& 18 	& 2360 	& -4.991 	& -4.995 	& 0.0424 	& 0.0212 	& 0.1359 	& 0.417 	& 27.7 $\pm$ 3.2 	& 4.88 $\pm$ 0.56 \\
HD44573    	& 26 	& 2354 	& -4.597 	& -4.586 	& 0.1664 	& 0.0942 	& 0.5963 	& 0.000 	& 23.0 $\pm$ 5.2 	& 2.04 $\pm$ 0.23 \\
HD44420    	& 14 	& 2356 	& -4.999 	& -4.990 	& 0.0443 	& 0.0196 	& 0.2541 	& 0.000 	& 31.8 $\pm$ 3.5 	& 5.64 $\pm$ 0.65 \\
HD45289    	& 12 	& 2749 	& -5.033 	& -5.034 	& 0.0201 	& 0.0190 	& 0.0936 	& 0.250 	& 30.6 $\pm$ 3.3 	& 5.62 $\pm$ 0.65 \\
HD47186    	& 98 	& 2635 	& -5.023 	& -5.021 	& 0.0146 	& 0.0132 	& 0.0143 	& 1.000 	& 35.3 $\pm$ 3.6 	& 6.42 $\pm$ 0.74 \\
HD48611    	& 11 	& 1189 	& -4.817 	& -4.821 	& 0.0984 	& 0.0416 	& 0.5928 	& 0.000 	& 28.5 $\pm$ 3.8 	& 4.08 $\pm$ 0.47 \\
HD50590    	& 15 	& 2333 	& -4.999 	& -5.001 	& 0.0797 	& 0.0343 	& 0.3572 	& 0.250 	& 50.9 $\pm$ 5.6 	& 7.68 $\pm$ 0.88 \\
HD50806    	& 45 	& 2084 	& -5.095 	& -5.093 	& 0.0199 	& 0.0169 	& 0.0318 	& 1.000 	& 38.6 $\pm$ 3.6 	& 7.51 $\pm$ 0.87 \\
HD52919    	& 7 	& 1067 	& -4.772 	& -4.770 	& 0.0346 	& 0.0232 	& 0.6010 	& 0.000 	& 36.7 $\pm$ 6.4 	& 3.87 $\pm$ 0.45 \\
HD59711A    	& 15 	& 2691 	& -4.959 	& -4.961 	& 0.0301 	& 0.0221 	& 0.1274 	& 0.250 	& 24.1 $\pm$ 3.1 	& 4.09 $\pm$ 0.47 \\
HD65562    	& 18 	& 2332 	& -4.954 	& -4.953 	& 0.0884 	& 0.0486 	& 0.2262 	& 0.000 	& 44.7 $\pm$ 4.8 	& 7.08 $\pm$ 0.82 \\
HD66221    	& 19 	& 2351 	& -5.034 	& -5.036 	& 0.0263 	& 0.0199 	& 0.1183 	& 0.250 	& 37.0 $\pm$ 3.7 	& 6.81 $\pm$ 0.78 \\
HD69655    	& 18 	& 2753 	& -4.959 	& -4.959 	& 0.0175 	& 0.0155 	& 0.0447 	& 0.917 	& 16.9 $\pm$ 2.6 	& 2.97 $\pm$ 0.34 \\
HD70889    	& 17 	& 1580 	& -4.781 	& -4.788 	& 0.1065 	& 0.0622 	& 0.3947 	& 0.333 	& 14.4 $\pm$ 2.8 	& 1.98 $\pm$ 0.23 \\
HD71334    	& 16 	& 2355 	& -4.995 	& -4.993 	& 0.0188 	& 0.0164 	& 0.0935 	& 0.000 	& 25.3 $\pm$ 3.1 	& 4.46 $\pm$ 0.51 \\
HD72579    	& 18 	& 2707 	& -5.062 	& -5.062 	& 0.0229 	& 0.0181 	& 0.0734 	& 0.583 	& 45.9 $\pm$ 4.2 	& 8.52 $\pm$ 0.98 \\
HD72769    	& 22 	& 2704 	& -5.052 	& -5.052 	& 0.0210 	& 0.0143 	& 0.0574 	& 0.500 	& 40.2 $\pm$ 3.8 	& 7.51 $\pm$ 0.86 \\
HD74014    	& 15 	& 2702 	& -5.043 	& -5.042 	& 0.0179 	& 0.0127 	& 0.1276 	& 0.333 	& 42.0 $\pm$ 4.0 	& 7.73 $\pm$ 0.89 \\
HD78429    	& 59 	& 1961 	& -4.921 	& -4.926 	& 0.0760 	& 0.0703 	& 0.1244 	& 0.000 	& 25.1 $\pm$ 3.2 	& 4.09 $\pm$ 0.47 \\
HD78558    	& 15 	& 2354 	& -5.020 	& -5.020 	& 0.0158 	& 0.0109 	& 0.0499 	& 0.833 	& 23.5 $\pm$ 2.9 	& 4.26 $\pm$ 0.49 \\
HD78612    	& 19 	& 2703 	& -5.030 	& -5.030 	& 0.0256 	& 0.0232 	& 0.0733 	& 0.583 	& 22.4 $\pm$ 2.9 	& 4.13 $\pm$ 0.48 \\
HD81639    	& 19 	& 2707 	& -5.007 	& -5.006 	& 0.0284 	& 0.0195 	& 0.1272 	& 0.167 	& 35.8 $\pm$ 3.7 	& 6.40 $\pm$ 0.74 \\
HD83529    	& 23 	& 2707 	& -4.993 	& -4.993 	& 0.0148 	& 0.0122 	& 0.0405 	& 0.917 	& 17.6 $\pm$ 2.6 	& 3.21 $\pm$ 0.37 \\
HD86171    	& 8 	& 1209 	& -4.807 	& -4.807 	& 0.1104 	& 0.0328 	& 0.6010 	& 0.000 	& 28.8 $\pm$ 3.9 	& 4.05 $\pm$ 0.47 \\
HD88084    	& 18 	& 2693 	& -4.982 	& -4.981 	& 0.0157 	& 0.0099 	& 0.0918 	& 0.833 	& 26.1 $\pm$ 3.2 	& 4.55 $\pm$ 0.52 \\
HD88218    	& 19 	& 2701 	& -5.055 	& -5.056 	& 0.0123 	& 0.0089 	& 0.0909 	& 0.833 	& 23.1 $\pm$ 2.9 	& 4.36 $\pm$ 0.50 \\
HD88742    	& 23 	& 1869 	& -4.694 	& -4.696 	& 0.2041 	& 0.1354 	& 0.5632 	& 0.000 	& 11.3 $\pm$ 2.7 	& 1.37 $\pm$ 0.16 \\
HD92588    	& 33 	& 1883 	& -5.167 	& -5.167 	& 0.0216 	& 0.0185 	& 0.0424 	& 0.917 	& 58.7 $\pm$ 4.9 	& 11.30 $\pm$ 1.30 \\
HD92719    	& 15 	& 2324 	& -4.860 	& -4.861 	& 0.0668 	& 0.0500 	& 0.4446 	& 0.000 	& 18.6 $\pm$ 2.9 	& 2.83 $\pm$ 0.33 \\
HD93385    	& 121 	& 2694 	& -4.984 	& -4.984 	& 0.0147 	& 0.0135 	& 0.0187 	& 1.000 	& 18.6 $\pm$ 2.7 	& 3.32 $\pm$ 0.38 \\
HD95456    	& 77 	& 2339 	& -4.963 	& -4.965 	& 0.0447 	& 0.0318 	& 0.0607 	& 0.417 	& 11.6 $\pm$ 2.3 	& 2.54 $\pm$ 0.29 \\
HD95521    	& 15 	& 2349 	& -4.886 	& -4.905 	& 0.1364 	& 0.0497 	& 0.5493 	& 0.000 	& 21.6 $\pm$ 3.1 	& 3.38 $\pm$ 0.39 \\
HD96423    	& 21 	& 2703 	& -5.021 	& -5.023 	& 0.0189 	& 0.0158 	& 0.0619 	& 0.750 	& 31.4 $\pm$ 3.4 	& 5.70 $\pm$ 0.66 \\
HD97037    	& 18 	& 2669 	& -5.004 	& -5.003 	& 0.0139 	& 0.0117 	& 0.1007 	& 0.583 	& 21.7 $\pm$ 2.9 	& 3.89 $\pm$ 0.45 \\
HD97343    	& 20 	& 2677 	& -5.021 	& -5.022 	& 0.0169 	& 0.0133 	& 0.0605 	& 0.750 	& 40.8 $\pm$ 4.0 	& 7.36 $\pm$ 0.85 \\
HD97998    	& 17 	& 2352 	& -4.948 	& -4.948 	& 0.0164 	& 0.0144 	& 0.0958 	& 0.583 	& 22.5 $\pm$ 3.0 	& 3.79 $\pm$ 0.44 \\
HD98356    	& 8 	& 2326 	& -4.786 	& -4.790 	& 0.0905 	& 0.0445 	& 0.6010 	& 0.000 	& 32.5 $\pm$ 4.5 	& 4.30 $\pm$ 0.49 \\
HD102117    	& 32 	& 2282 	& -5.053 	& -5.053 	& 0.0240 	& 0.0224 	& 0.0440 	& 0.833 	& 37.9 $\pm$ 3.7 	& 7.10 $\pm$ 0.82 \\
HD102438    	& 39 	& 2244 	& -4.982 	& -4.982 	& 0.0134 	& 0.0127 	& 0.0211 	& 1.000 	& 29.9 $\pm$ 3.4 	& 5.21 $\pm$ 0.60 \\
HD104006    	& 26 	& 2233 	& -5.054 	& -5.053 	& 0.0208 	& 0.0177 	& 0.0572 	& 0.583 	& 44.5 $\pm$ 4.1 	& 8.24 $\pm$ 0.95 \\
HD104067    	& 83 	& 2271 	& -4.752 	& -4.753 	& 0.0975 	& 0.0872 	& 0.1218 	& 0.000 	& 34.0 $\pm$ 5.6 	& 3.82 $\pm$ 0.44 \\
HD104263    	& 29 	& 2245 	& -5.035 	& -5.033 	& 0.0284 	& 0.0263 	& 0.0725 	& 0.500 	& 40.5 $\pm$ 3.9 	& 7.42 $\pm$ 0.85 \\
HD104982    	& 39 	& 2271 	& -4.974 	& -4.973 	& 0.0140 	& 0.0133 	& 0.0316 	& 1.000 	& 25.8 $\pm$ 3.2 	& 4.46 $\pm$ 0.51 \\
HD105837    	& 21 	& 2652 	& -4.877 	& -4.879 	& 0.0558 	& 0.0360 	& 0.1903 	& 0.417 	& 13.9 $\pm$ 2.6 	& 2.27 $\pm$ 0.26 \\
HD109409    	& 22 	& 2649 	& -5.110 	& -5.111 	& 0.0207 	& 0.0145 	& 0.0570 	& 0.833 	& 35.1 $\pm$ 3.4 	& 6.92 $\pm$ 0.80 \\
HD110619    	& 17 	& 2648 	& -4.922 	& -4.925 	& 0.0527 	& 0.0254 	& 0.2233 	& 0.250 	& 25.1 $\pm$ 3.2 	& 4.10 $\pm$ 0.47 \\
HD111031    	& 26 	& 2245 	& -5.036 	& -5.036 	& 0.0176 	& 0.0146 	& 0.0479 	& 0.833 	& 33.4 $\pm$ 3.5 	& 6.14 $\pm$ 0.71 \\
HD112540    	& 7 	& 2295 	& -4.774 	& -4.771 	& 0.2205 	& 0.0781 	& 0.6010 	& 0.000 	& 24.8 $\pm$ 3.7 	& 3.35 $\pm$ 0.39 \\
HD114853    	& 37 	& 3011 	& -4.957 	& -4.959 	& 0.0393 	& 0.0376 	& 0.0842 	& 0.083 	& 24.0 $\pm$ 3.1 	& 4.07 $\pm$ 0.47 \\
HD115585    	& 19 	& 2641 	& -5.076 	& -5.084 	& 0.0299 	& 0.0248 	& 0.1113 	& 0.250 	& 41.4 $\pm$ 3.8 	& 7.90 $\pm$ 0.91 \\
HD116920    	& 8 	& 2305 	& -4.827 	& -4.820 	& 0.1210 	& 0.0834 	& 0.6010 	& 0.000 	& 37.9 $\pm$ 5.1 	& 4.99 $\pm$ 0.57 \\
HD117105    	& 18 	& 2658 	& -4.976 	& -4.976 	& 0.0128 	& 0.0113 	& 0.0529 	& 0.750 	& 17.3 $\pm$ 2.6 	& 3.09 $\pm$ 0.36 \\
HD117207    	& 15 	& 2642 	& -5.035 	& -5.035 	& 0.0076 	& 0.0069 	& 0.0985 	& 0.667 	& 37.0 $\pm$ 3.7 	& 6.82 $\pm$ 0.78 \\
HD119782    	& 17 	& 2636 	& -4.697 	& -4.703 	& 0.0945 	& 0.0811 	& 0.4908 	& 0.000 	& 28.2 $\pm$ 4.7 	& 3.18 $\pm$ 0.37 \\
HD123265    	& 17 	& 2642 	& -5.073 	& -5.078 	& 0.0250 	& 0.0207 	& 0.1342 	& 0.000 	& 49.8 $\pm$ 4.5 	& 9.14 $\pm$ 1.05 \\
HD122862    	& 18 	& 2295 	& -5.027 	& -5.026 	& 0.0129 	& 0.0113 	& 0.0647 	& 0.667 	& 18.5 $\pm$ 2.6 	& 3.48 $\pm$ 0.40 \\
HD124292    	& 25 	& 3006 	& -5.009 	& -5.012 	& 0.0243 	& 0.0208 	& 0.0791 	& 0.083 	& 37.0 $\pm$ 3.8 	& 6.62 $\pm$ 0.76 \\
HD124364    	& 14 	& 2636 	& -4.865 	& -4.880 	& 0.1139 	& 0.0357 	& 0.5563 	& 0.000 	& 24.1 $\pm$ 3.3 	& 3.68 $\pm$ 0.42 \\
HD125184    	& 68 	& 2131 	& -5.029 	& -5.025 	& 0.0411 	& 0.0346 	& 0.0880 	& 0.500 	& 36.8 $\pm$ 3.7 	& 6.72 $\pm$ 0.77 \\
HD125455    	& 18 	& 2165 	& -4.916 	& -4.915 	& 0.1211 	& 0.0543 	& 0.5083 	& 0.250 	& 42.3 $\pm$ 4.8 	& 6.43 $\pm$ 0.74 \\
HD126525    	& 42 	& 2518 	& -4.993 	& -4.993 	& 0.0138 	& 0.0118 	& 0.0308 	& 1.000 	& 30.3 $\pm$ 3.4 	& 5.34 $\pm$ 0.62 \\
HD128674    	& 19 	& 2263 	& -4.958 	& -4.957 	& 0.0188 	& 0.0129 	& 0.0809 	& 0.500 	& 27.7 $\pm$ 3.3 	& 4.71 $\pm$ 0.54 \\
HD130992    	& 18 	& 1831 	& -4.858 	& -4.860 	& 0.0722 	& 0.0505 	& 0.3187 	& 0.000 	& 42.2 $\pm$ 6.1 	& 5.17 $\pm$ 0.59 \\
HD130930    	& 18 	& 2117 	& -5.016 	& -5.015 	& 0.0407 	& 0.0241 	& 0.1726 	& 0.000 	& 50.3 $\pm$ 5.1 	& 8.22 $\pm$ 0.95 \\
HD134060    	& 100 	& 2616 	& -4.983 	& -4.983 	& 0.0172 	& 0.0158 	& 0.0209 	& 1.000 	& 22.3 $\pm$ 2.9 	& 3.91 $\pm$ 0.45 \\
HD134985    	& 15 	& 2619 	& -5.068 	& -5.074 	& 0.0309 	& 0.0217 	& 0.3366 	& 0.000 	& 44.3 $\pm$ 4.1 	& 8.34 $\pm$ 0.96 \\
HD134606    	& 107 	& 2489 	& -5.049 	& -5.049 	& 0.0138 	& 0.0133 	& 0.0155 	& 1.000 	& 40.1 $\pm$ 3.8 	& 7.46 $\pm$ 0.86 \\
HD136894    	& 38 	& 2044 	& -5.006 	& -5.007 	& 0.0130 	& 0.0118 	& 0.0267 	& 1.000 	& 35.7 $\pm$ 3.7 	& 6.39 $\pm$ 0.74 \\
HD137303    	& 16 	& 2044 	& -4.811 	& -4.807 	& 0.1328 	& 0.0574 	& 0.4776 	& 0.000 	& 39.0 $\pm$ 6.1 	& 4.48 $\pm$ 0.52 \\
HD140901    	& 24 	& 1546 	& -4.694 	& -4.697 	& 0.1380 	& 0.0879 	& 0.5881 	& 0.000 	& 20.3 $\pm$ 3.6 	& 2.43 $\pm$ 0.28 \\
HD142709    	& 13 	& 2117 	& -5.049 	& -5.058 	& 0.1120 	& 0.0406 	& 0.4298 	& 0.000 	& 56.9 $\pm$ 6.7 	& 8.08 $\pm$ 0.93 \\
HD143114    	& 19 	& 1789 	& -4.991 	& -4.989 	& 0.0115 	& 0.0094 	& 0.0734 	& 0.667 	& 21.3 $\pm$ 2.9 	& 3.77 $\pm$ 0.43 \\
HD144411    	& 11 	& 1719 	& -4.851 	& -4.843 	& 0.0709 	& 0.0541 	& 0.5024 	& 0.000 	& 40.9 $\pm$ 5.6 	& 5.22 $\pm$ 0.60 \\
HD144585    	& 14 	& 2539 	& -5.033 	& -5.043 	& 0.0449 	& 0.0098 	& 0.4071 	& 0.000 	& 29.3 $\pm$ 3.2 	& 5.37 $\pm$ 0.62 \\
HD145809    	& 13 	& 2174 	& -5.062 	& -5.061 	& 0.0115 	& 0.0091 	& 0.0704 	& 0.750 	& 24.8 $\pm$ 2.9 	& 4.68 $\pm$ 0.54 \\
HD145598    	& 36 	& 2107 	& -5.004 	& -5.003 	& 0.0256 	& 0.0222 	& 0.0584 	& 0.833 	& 28.2 $\pm$ 3.2 	& 5.03 $\pm$ 0.58 \\
HD145666    	& 22 	& 1410 	& -4.778 	& -4.777 	& 0.0543 	& 0.0390 	& 0.2174 	& 0.083 	& 14.3 $\pm$ 2.8 	& 1.96 $\pm$ 0.23 \\
HD147512    	& 29 	& 1734 	& -4.999 	& -4.999 	& 0.0126 	& 0.0103 	& 0.0322 	& 1.000 	& 35.4 $\pm$ 3.7 	& 6.28 $\pm$ 0.72 \\
HD150433    	& 50 	& 2077 	& -5.000 	& -5.000 	& 0.0107 	& 0.0097 	& 0.0181 	& 1.000 	& 24.2 $\pm$ 3.0 	& 4.29 $\pm$ 0.49 \\
HD151504    	& 14 	& 1494 	& -5.031 	& -5.032 	& 0.0097 	& 0.0081 	& 0.0629 	& 0.667 	& 42.3 $\pm$ 4.1 	& 7.68 $\pm$ 0.88 \\
HD154363    	& 17 	& 2516 	& -4.906 	& -4.910 	& 0.1488 	& 0.0723 	& 0.4597 	& 0.000 	& 46.4 $\pm$ 6.5 	& 5.75 $\pm$ 0.66 \\
HD157347    	& 20 	& 2893 	& -5.011 	& -5.010 	& 0.0144 	& 0.0095 	& 0.0684 	& 0.667 	& 31.0 $\pm$ 3.4 	& 5.57 $\pm$ 0.64 \\
HD157338    	& 25 	& 1461 	& -4.975 	& -4.972 	& 0.0232 	& 0.0189 	& 0.0850 	& 0.333 	& 18.4 $\pm$ 2.7 	& 3.25 $\pm$ 0.37 \\
HD161612    	& 32 	& 2154 	& -5.014 	& -5.014 	& 0.0117 	& 0.0107 	& 0.0221 	& 1.000 	& 39.4 $\pm$ 3.9 	& 7.07 $\pm$ 0.81 \\
HD162236    	& 16 	& 1069 	& -4.709 	& -4.714 	& 0.1443 	& 0.1302 	& 0.6010 	& 0.000 	& 22.4 $\pm$ 3.8 	& 2.74 $\pm$ 0.32 \\
HD162396    	& 39 	& 1885 	& -5.007 	& -5.006 	& 0.0203 	& 0.0170 	& 0.0414 	& 0.917 	& 11.3 $\pm$ 2.2 	& 2.94 $\pm$ 0.34 \\
HD165920    	& 13 	& 1791 	& -5.048 	& -5.048 	& 0.0239 	& 0.0185 	& 0.1102 	& 0.000 	& 48.9 $\pm$ 4.6 	& 8.72 $\pm$ 1.00 \\
HD166724    	& 47 	& 2526 	& -4.740 	& -4.739 	& 0.1268 	& 0.0958 	& 0.2286 	& 0.000 	& 30.9 $\pm$ 4.7 	& 3.74 $\pm$ 0.43 \\
HD168871    	& 22 	& 2537 	& -4.988 	& -4.989 	& 0.0180 	& 0.0145 	& 0.0500 	& 0.833 	& 18.8 $\pm$ 2.7 	& 3.36 $\pm$ 0.39 \\
HD170493    	& 13 	& 2101 	& -4.802 	& -4.802 	& 0.2087 	& 0.0873 	& 0.6010 	& 0.000 	& 38.7 $\pm$ 6.3 	& 4.29 $\pm$ 0.49 \\
HD171665    	& 13 	& 1730 	& -4.914 	& -4.906 	& 0.0505 	& 0.0149 	& 0.2931 	& 0.000 	& 28.2 $\pm$ 3.5 	& 4.57 $\pm$ 0.53 \\
HD171990    	& 34 	& 1797 	& -5.101 	& -5.099 	& 0.0237 	& 0.0213 	& 0.0594 	& 0.667 	& 21.6 $\pm$ 2.7 	& 4.31 $\pm$ 0.50 \\
HD174545    	& 13 	& 2535 	& -4.911 	& -4.894 	& 0.1092 	& 0.0521 	& 0.5910 	& 0.000 	& 42.4 $\pm$ 4.9 	& 6.35 $\pm$ 0.73 \\
HD176986    	& 80 	& 2474 	& -4.833 	& -4.829 	& 0.0790 	& 0.0747 	& 0.1288 	& 0.250 	& 39.0 $\pm$ 5.3 	& 5.04 $\pm$ 0.58 \\
HD177758    	& 7 	& 2749 	& -4.991 	& -4.993 	& 0.0087 	& 0.0054 	& 0.5928 	& 0.000 	& 16.4 $\pm$ 2.6 	& 3.04 $\pm$ 0.35 \\
HD177409    	& 18 	& 1980 	& -4.867 	& -4.878 	& 0.0866 	& 0.0430 	& 0.2814 	& 0.083 	& 16.7 $\pm$ 2.8 	& 2.58 $\pm$ 0.30 \\
HD180409    	& 8 	& 2747 	& -4.942 	& -4.942 	& 0.0094 	& 0.0044 	& 0.3603 	& 0.000 	& 15.3 $\pm$ 2.6 	& 2.70 $\pm$ 0.31 \\
HD181433    	& 154 	& 2882 	& -5.100 	& -5.100 	& 0.0276 	& 0.0261 	& 0.0211 	& 1.000 	& 58.4 $\pm$ 5.8 	& 9.47 $\pm$ 1.09 \\
HD183658    	& 14 	& 1695 	& -4.983 	& -4.980 	& 0.0185 	& 0.0120 	& 0.0875 	& 0.750 	& 24.9 $\pm$ 3.1 	& 4.34 $\pm$ 0.50 \\
HD183783    	& 8 	& 2526 	& -4.954 	& -4.942 	& 0.1540 	& 0.0418 	& 0.6010 	& 0.000 	& 49.7 $\pm$ 6.5 	& 6.54 $\pm$ 0.75 \\
HD185615    	& 22 	& 2305 	& -5.033 	& -5.032 	& 0.0294 	& 0.0259 	& 0.0795 	& 0.000 	& 38.1 $\pm$ 3.8 	& 7.00 $\pm$ 0.81 \\
HD187456    	& 11 	& 2145 	& -4.925 	& -4.926 	& 0.0304 	& 0.0144 	& 0.2144 	& 0.417 	& 47.1 $\pm$ 6.2 	& 6.14 $\pm$ 0.71 \\
HD189625    	& 20 	& 1751 	& -4.788 	& -4.779 	& 0.1556 	& 0.0742 	& 0.4726 	& 0.250 	& 19.2 $\pm$ 3.2 	& 2.65 $\pm$ 0.30 \\
HD188748    	& 19 	& 2122 	& -4.981 	& -4.980 	& 0.0209 	& 0.0183 	& 0.0732 	& 0.333 	& 28.6 $\pm$ 3.3 	& 4.98 $\pm$ 0.57 \\
HD189567    	& 154 	& 2742 	& -4.941 	& -4.943 	& 0.0390 	& 0.0344 	& 0.0421 	& 0.917 	& 24.7 $\pm$ 3.2 	& 4.11 $\pm$ 0.47 \\
HD190248    	& 84 	& 2491 	& -5.052 	& -5.052 	& 0.0164 	& 0.0137 	& 0.0244 	& 1.000 	& 41.4 $\pm$ 3.9 	& 7.70 $\pm$ 0.89 \\
HD190954    	& 9 	& 2111 	& -5.015 	& -5.017 	& 0.0279 	& 0.0116 	& 0.4987 	& 0.000 	& 35.0 $\pm$ 3.6 	& 6.31 $\pm$ 0.73 \\
HD192031    	& 8 	& 2525 	& -5.051 	& -5.051 	& 0.0107 	& 0.0067 	& 0.4282 	& 0.000 	& 39.0 $\pm$ 3.8 	& 7.29 $\pm$ 0.84 \\
HD193193    	& 31 	& 1760 	& -4.938 	& -4.943 	& 0.0411 	& 0.0313 	& 0.1614 	& 0.000 	& 17.5 $\pm$ 2.7 	& 2.96 $\pm$ 0.34 \\
HD195564    	& 11 	& 2731 	& -5.106 	& -5.104 	& 0.0231 	& 0.0098 	& 0.3661 	& 0.000 	& 36.3 $\pm$ 3.5 	& 7.15 $\pm$ 0.82 \\
HD196761    	& 8 	& 2730 	& -4.951 	& -4.949 	& 0.0667 	& 0.0180 	& 0.5672 	& 0.000 	& 33.1 $\pm$ 3.7 	& 5.58 $\pm$ 0.64 \\
HD197210    	& 7 	& 2073 	& -4.890 	& -4.880 	& 0.0933 	& 0.0098 	& 0.5845 	& 0.000 	& 29.3 $\pm$ 3.6 	& 4.61 $\pm$ 0.53 \\
HD199288    	& 14 	& 2491 	& -4.963 	& -4.962 	& 0.0101 	& 0.0052 	& 0.0658 	& 0.667 	& 18.1 $\pm$ 2.7 	& 3.16 $\pm$ 0.36 \\
HD203384    	& 9 	& 2333 	& -5.002 	& -5.001 	& 0.0489 	& 0.0175 	& 0.5404 	& 0.000 	& 39.8 $\pm$ 4.0 	& 7.04 $\pm$ 0.81 \\
HD203850    	& 7 	& 2871 	& -4.921 	& -4.924 	& 0.0534 	& 0.0159 	& 0.6010 	& 0.000 	& 44.3 $\pm$ 5.2 	& 6.49 $\pm$ 0.75 \\
HD204385    	& 7 	& 2222 	& -4.967 	& -4.968 	& 0.0280 	& 0.0148 	& 0.6010 	& 0.000 	& 19.4 $\pm$ 2.8 	& 3.37 $\pm$ 0.39 \\
HD205536    	& 23 	& 2219 	& -5.026 	& -5.022 	& 0.0397 	& 0.0337 	& 0.2436 	& 0.000 	& 41.1 $\pm$ 4.0 	& 7.44 $\pm$ 0.86 \\
HD207700    	& 15 	& 2111 	& -4.999 	& -5.000 	& 0.0245 	& 0.0191 	& 0.0897 	& 0.000 	& 33.0 $\pm$ 3.5 	& 5.87 $\pm$ 0.68 \\
HD209742    	& 7 	& 1069 	& -4.872 	& -4.891 	& 0.1331 	& 0.0054 	& 0.6005 	& 0.000 	& 39.1 $\pm$ 4.7 	& 5.68 $\pm$ 0.65 \\
HD210752    	& 12 	& 1779 	& -4.934 	& -4.934 	& 0.0134 	& 0.0091 	& 0.1771 	& 0.167 	& 12.1 $\pm$ 2.3 	& 2.36 $\pm$ 0.27 \\
HD211038    	& 73 	& 1880 	& -5.214 	& -5.214 	& 0.0086 	& 0.0079 	& 0.0153 	& 1.000 	& 62.3 $\pm$ 5.0 	& 12.35 $\pm$ 1.42 \\
HD210918    	& 31 	& 1858 	& -5.010 	& -5.008 	& 0.0249 	& 0.0244 	& 0.0304 	& 1.000 	& 27.1 $\pm$ 3.2 	& 4.86 $\pm$ 0.56 \\
HD212708    	& 34 	& 1072 	& -5.046 	& -5.047 	& 0.0233 	& 0.0201 	& 0.0731 	& 0.583 	& 38.8 $\pm$ 3.8 	& 7.21 $\pm$ 0.83 \\
HD213628    	& 7 	& 2176 	& -4.953 	& -4.951 	& 0.0328 	& 0.0052 	& 0.6010 	& 0.000 	& 33.2 $\pm$ 3.7 	& 5.62 $\pm$ 0.65 \\
HD213941    	& 26 	& 2218 	& -4.961 	& -4.961 	& 0.0333 	& 0.0304 	& 0.1169 	& 0.333 	& 27.8 $\pm$ 3.3 	& 4.74 $\pm$ 0.55 \\
HD215456    	& 94 	& 2616 	& -5.089 	& -5.087 	& 0.0206 	& 0.0185 	& 0.0224 	& 1.000 	& 28.5 $\pm$ 3.1 	& 5.51 $\pm$ 0.63 \\
HD219249    	& 22 	& 1283 	& -4.960 	& -4.959 	& 0.0283 	& 0.0168 	& 0.1333 	& 0.000 	& 30.2 $\pm$ 3.5 	& 5.15 $\pm$ 0.59 \\
HD220256    	& 23 	& 2222 	& -5.033 	& -5.034 	& 0.0320 	& 0.0222 	& 0.1114 	& 0.500 	& 48.6 $\pm$ 4.7 	& 8.49 $\pm$ 0.98 \\
HD220339    	& 11 	& 2122 	& -4.836 	& -4.818 	& 0.1627 	& 0.0379 	& 0.5919 	& 0.000 	& 37.6 $\pm$ 4.9 	& 5.13 $\pm$ 0.59 \\
HD220507    	& 50 	& 2221 	& -5.050 	& -5.051 	& 0.0182 	& 0.0149 	& 0.0455 	& 0.917 	& 34.0 $\pm$ 3.5 	& 6.34 $\pm$ 0.73 \\
HD221356    	& 15 	& 2165 	& -4.940 	& -4.939 	& 0.0329 	& 0.0251 	& 0.2324 	& 0.000 	& 11.2 $\pm$ 2.3 	& 2.40 $\pm$ 0.28 \\
HD221420    	& 28 	& 1420 	& -5.158 	& -5.155 	& 0.0244 	& 0.0204 	& 0.0641 	& 0.667 	& 37.0 $\pm$ 3.4 	& 7.62 $\pm$ 0.88 \\
HD222422    	& 7 	& 2113 	& -4.792 	& -4.791 	& 0.0671 	& 0.0365 	& 0.6010 	& 0.000 	& 26.4 $\pm$ 3.8 	& 3.66 $\pm$ 0.42 \\
HD223121    	& 13 	& 2133 	& -4.845 	& -4.848 	& 0.1415 	& 0.0867 	& 0.6010 	& 0.000 	& 39.8 $\pm$ 5.3 	& 5.22 $\pm$ 0.60 \\
HD223171    	& 64 	& 1809 	& -4.995 	& -4.995 	& 0.0476 	& 0.0439 	& 0.0695 	& 0.333 	& 27.9 $\pm$ 3.2 	& 4.92 $\pm$ 0.57 \\
HD224619    	& 11 	& 2136 	& -5.000 	& -5.001 	& 0.0256 	& 0.0091 	& 0.1550 	& 0.417 	& 37.6 $\pm$ 3.8 	& 6.67 $\pm$ 0.77 \\
\end{longtable}
}

\end{document}